\documentclass[12pt]{article}
\usepackage{docmute}

\usepackage{cite}
\usepackage{graphicx}
\usepackage{dcolumn}
\usepackage{bm}

\usepackage[hidelinks]{hyperref}


\usepackage{epsf}
\usepackage{amsmath,amssymb}
\usepackage{slashed}
\usepackage{graphicx}
\usepackage{subfig}
\usepackage[usenames,dvipsnames]{xcolor}

\usepackage{tikz}
\usetikzlibrary{quantikz2}

\let\oldRe\Re
\let\oldIm\Im
\usepackage{physics}
\let\Re\oldRe
\let\Im\oldIm

\usepackage{dsfont}
\newcommand{\mathhyphen}{\mathchar"712D}

\usepackage{comment}

\usepackage{tikz}
\usetikzlibrary{quantikz2}

\begin{document}

\begin{titlepage}

\def\thefootnote{\fnsymbol{footnote}}

\begin{center}

\hfill August, 2026\\

\vskip 1in

{\Large \bf
  
  Qubits for Dark Matter Hunting\\
  
}

\vskip .5in

{\large
  Takeo Moroi
}

\vskip .5in

{\em Department of Physics, The University of Tokyo, Tokyo 113-0033, Japan}

\end{center}
\vskip .5in

\begin{abstract}

An introductory review is provided for those who are interested in exploring applications of qubits and other quantum excitations to the detection of dark matter (and any other physics beyond the Standard Model). Topics covered include the fundamental properties of qubits, the excitation mechanism of qubits due to the electric field induced by dark matter (with attention to the effects of the coherence of dark matter), the dynamics of coupled qubit-cavity systems modeled by the Jaynes-Cummings framework, and the influence of noise and decoherence (especially Markovian noise described by the Lindblad equation). In addition, the article introduces essential concepts in quantum sensing, including the operator-sum representation and positive operator-valued measures, the Cram\'er-Rao bound, the standard quantum limit and the Heisenberg limit, and the potential enhancement of sensitivity to dark matter achievable with entangled states. Throughout, these topics are discussed with particular emphasis on their application to the detection of wave-like dark matter.

\end{abstract}

\end{titlepage}

\newpage

\renewcommand{\thepage}{\roman{page}}
\setcounter{page}{1}

\tableofcontents

\newpage


\newcommand{\TMpreface}{

\section*{Preface}

This review article is based on the author's personal notes summarizing topics related to the use of quantum excitations for dark matter detection. As readers may notice, much of the material covered here is likely already familiar to experts. For this reason, the author initially hesitated to make these notes public, particularly at a time when powerful and helpful AI tools are so readily available. The author recalled, however, that when he first looked into the idea of using qubits for dark matter detection, he found it rather difficult to locate relevant information on qubits, cavities, quantum sensing, and so on. This article may therefore be of some help to others who find themselves in a similar situation. For newcomers to the field, especially students and postdocs, a review covering the basics may serve as a useful entry point.

This review is intended for readers, especially those in particle physics and cosmology, who are just beginning to explore the detection of dark matter (or anything else) using qubits and other quantum excitations. The author hopes that readers will find something useful, clarifying, or at least thought-provoking somewhere in these pages.

}

\TMpreface

\clearpage

\begin{table}
  \centering
  \caption*{\textbf{List of symbols}}
  \renewcommand{\arraystretch}{1.3}
  \begin{tabular}{p{0.2\textwidth} p{0.725\textwidth}}
    \hline\hline
    \textbf{Symbol} & \textbf{Meaning} \\
    \hline
    $t$ & Time (or total exposure time per cycle)\\
    \hline
    $\rho$ & Density matrix (state)\\
    $\rho_n$ & Density matrix of $n$-th subsystem (or $n$-th qubit)\\
    $\mathds{I}$ & Identity operator\\
    \hline
    $\ket{0},\,\ket{1}$ & Qubit ground and excited states\\
    $X,\, Y,\, Z$ & Pauli operators acting on a qubit; see Eqs.\ \eqref{PauliX} -- \eqref{PauliZ}\\
    $a,\, a^\dagger$ & Qubit lowering and raising operators; see Eqs.\ \eqref{op_a} and \eqref{op_adagger}\\
    $\vec{\rho} = (\rho_X, \rho_Y, \rho_Z)$ & Bloch vector components\\
    $T_1$ & Qubit energy relaxation (longitudinal) time\\
    $T_2$ & Qubit dephasing (transverse) time\\
    $T_0$ & Qubit spontaneous excitation (dark-count) time\\
    $\tau_{\rm Q}$ & Overall qubit coherence time\\
    \hline
    $\omega$ & Qubit transition angular frequency; $H_0=\omega\ketbra{1}{1}$ \\
    $m$ & Angular frequency of the drive field (DM mass)\\
    $\varphi$ & Drive phase (DM phase)\\
    $\eta$ & Effective drive strength; $H_1=-2\eta X\cos(mt-\varphi)$ \\
    $p_{\rm ge}(t)$ & Excitation probability from $\ket{0}$ to $\ket{1}$ after time $t$ \\
    $\Gamma_{\rm ge}(t)$ & Effective excitation rate; $\Gamma_{\rm ge}\equiv p_{\rm ge}(t)/t$ (for $p_{\rm ge}\ll 1$) \\
    $p_{\rm ge}^{\rm (dark\mathhyphen count)}$ & Dark-count probability \\
    \hline
    $\rho_{\rm DM}$ & DM energy density\\
    $v_{\rm DM}$ & Mean DM velocity; $v_{\rm DM}\sim 10^{-3}$\\
    $v$ & DM velocity\\
    $\mathcal{F}(v)$ & DM velocity distribution; see Eq.\ \eqref{F(v)}\\
    $\tau_{\rm DM}= 2\pi/mv_{\rm DM}^2$ & Coherence time of the DM field\\
    $\ell_{\rm DM}= 2\pi/mv_{\rm DM}$ & Spatial coherence length of the DM field\\
    $\mathcal{T}_{\rm DM}(v_\omega)$ & Effective DM coherence time; see Eq.~\eqref{T_DM(v)}\\
    \hline \hline
  \end{tabular}
\end{table}

\begin{table}
  \centering
  \caption*{\textbf{List of symbols (contd.)}}
  \renewcommand{\arraystretch}{1.3}
  \begin{tabular}{p{0.2\textwidth} p{0.725\textwidth}}
    \hline\hline
    \textbf{Symbol} & \textbf{Meaning} \\
    \hline \hline
    $c,c^\dagger$ & Cavity-mode annihilation and creation operators (selected mode)\\
    $c_n,c_n^\dagger$ & Cavity-mode annihilation and creation operators (mode $n$)\\
    $\omega_\gamma$, $\omega_n$ & Cavity angular frequency \\
    $\gamma_n$ & Cavity energy-loss rate (mode $n$)\\
    $Q_n$ & Cavity quality factor (mode $n$)\\
    $\kappa$ & Cavity (packaging) coefficient\\
    \hline
    $g$ & Qubit-cavity coupling in the Jaynes-Cummings model \\
    $\chi$ & Dispersive shift\\
    \hline
    $K_a$ & Kraus operators; CPTP map $\mathcal{E}[\rho]=\sum_a K_a\rho K_a^\dagger$ \\
    $E_a$ & POVM elements; $E_a=K_a^\dagger K_a$, $\sum_a E_a=\mathds{I}$ \\
    $L_a$ & Lindblad (jump) operators; $\partial_t\rho=-i[H,\rho]+\sum_a\mathcal{D}_{L_a}[\rho]$ \\
    $\mathcal{D}_L$ & Lindblad dissipator; $\mathcal{D}_L[\rho]=L\rho L^\dagger-\frac{1}{2}\{L^\dagger L,\rho\}$ \\
    $F_{\rm C}$ & Classical Fisher information; see Eq.\ \eqref{defCFI}\\
    $F_{\rm Q}$ & Quantum Fisher information; see Eq.\ \eqref{defQFI}\\
    \hline
    $N$ & Number of subsystems (or qubits) \\
    $T$ & Total experimental time\\
    $\nu$ & Number of independent measurement repetitions; often $\nu=T/t$ \\
    \hline
    $g_{a\gamma\gamma}$ & Axion-photon-photon coupling; see Eq.\ \eqref{L_agamma}\\
    $\epsilon$ & Kinetic mixing parameter for the dark photon; see Eq.\ \eqref{L_darkphoton}\\
    \hline\hline
  \end{tabular}
\end{table}

\clearpage

\renewcommand{\thepage}{\arabic{page}}
\setcounter{page}{1}
\def\thefootnote{\fnsymbol{footnote}}
\setcounter{footnote}{0}
\renewcommand{\theequation}{\thesection.\arabic{equation}}
\renewcommand{\thefigure}{\thesection.\arabic{figure}}

\section{Introduction}
\setcounter{equation}{0}
\setcounter{figure}{0}
\setcounter{footnote}{1}

Dark matter (DM) of our Universe is one of the most serious mysteries in particle physics and cosmology (for a review, see, e.g., Refs.\ \cite{Feng:2010gw, Cirelli:2024ssz}). Various astrophysical and cosmological observations have confirmed the existence of DM, but evidence for DM has so far been provided only through its gravitational interaction. Direct detection of DM through particle-physics processes is therefore an important challenge. Despite significant efforts so far, direct detection of DM has not yet been successful.

One of the difficulties of DM detection originates from our ignorance of its mass and interaction strength, as well as from the weakness of the DM interaction. It is therefore essential to explore every available detection strategy. If the mass of the DM is (much) heavier than $O(10)\ {\rm eV}$, the occupation number of the DM within its de Broglie wavelength becomes smaller than unity so that the DM is particle-like; weakly interacting massive particles (WIMPs) are popular candidates for such DM. Currently, scattering processes off nuclei and electrons are commonly used to detect particle-like DM; for recent progress in detecting heavy and light WIMP DM, see, e.g., Refs.\ \cite{LZ:2024zvo, XENON:2025vwd, PandaX:2024qfu} and \cite{SENSEI:2023zdf, CDEX:2023wfz, SuperCDMS:2024yiv, DAMIC-M:2025luv, PandaX:2025rrz}, respectively. In contrast, if the DM mass is (much) lighter, the occupation number becomes of order unity or larger, resulting in wave-like DM. For wave-like DM detection, cavity experiments are a popular option, as first proposed in Ref.\ \cite{Sikivie:1983ip} (see also Ref.\ \cite{Bradley:2003kg}); for the most recent efforts, see, e.g., Refs.\ \cite{ADMX:2024xbv, ADMX:2025vom, HAYSTAC:2024jch, Yang:2023yry, Kim:2023vpo, CAPP:2024dtx, Quiskamp:2024oet, QUAX:2024fut, Ahyoune:2024klt, Dixit:2020ymh, Cervantes:2022gtv, SHANHE:2023kxz, Schneemann:2023bqc, APEX:2024jxw, Chang:2025ahb, Quiskamp:2025wme, Zhao:2025thg, Nakazono:2025tak}. In this article, we focus on the detection of wave-like DM; basic properties of wave-like DM are summarized in Appendix~\ref{app:dm}.

Recently, there has been growing interest in utilizing quantum sensors, particularly qubits (quantum bits), for the detection of wave-like DM. A qubit is a two-level quantum system, originally developed mainly for quantum computation. Qubits have, however, many useful and interesting properties that may be utilized for DM detection, as we will see throughout this article. Indeed, qubits have already been used in actual DM search experiments as a single-photon counter \cite{Dixit:2020ymh} or as a shifter of the cavity frequency \cite{Zhao:2025thg, Nakazono:2025tak}. 

The main focus of this article is a more direct use of qubits for DM detection. It has been pointed out that wave-like DM can be detected via the direct excitation of superconducting qubits \cite{Chen:2022quj, Chen:2024aya}. Ideas of using other types of quantum sensors, such as diamond NV centers \cite{Chigusa:2023roq, Chigusa:2024psk}, ion traps \cite{Ito:2023zhp}, and Rydberg atoms \cite{Graham:2023sow, Engelhardt:2023qjf, Chigusa:2025rqs, Banerjee:2025jss}, have also been subsequently proposed and discussed. The advantages of using qubits for DM detection lie not only in the fact that they provide a simple setup for a DM detection experiment, but also in the possibility that their quantum properties may be used to enhance the DM signal \cite{Chen:2023swh, Chen:2025tgj, Bodas:2025vff, Fukuda:2025afi, He:2025ovo, Zheng:2025qgv, Bogorad:2026ggm}. (For the possibility of quantum enhancement in cavity experiments, see also Refs.\ \cite{HAYSTAC:2020kwv, Agrawal:2023umy, Freiman:2025tse}.) Thus, we regard qubits as a highly promising platform for DM detection.\footnote
{The use of qubits may not be limited to the detection of DM. For other applications, see Refs.\ \cite{Kanno:2023whr, Santoso:2025gmf, Fukuda:2025zcf, Dong:2025mdk, Ito:2025mgm}.
}

In particular, a new experiment using the direct excitation of superconducting qubits for DM detection has already been launched; the experiment is named the ``DarQ'' (Dark matter search using Qubits) experiment \cite{DarQ} (see also Ref.\ \cite{Kang:2025kaf}). A study of the possibility of using diamond NV centers for DM detection is also underway \cite{QUP}. We expect and hope that this trend will continue and that the importance of quantum technologies (including qubits) in the field of particle physics and cosmology will keep on growing.

The purpose of this review article is to provide a basic theoretical background primarily (but not exclusively) for those in the field of particle physics and cosmology who are trying to understand, design, and/or perform DM search experiments with qubits. The primary focus is on a type of experiment using the direct excitation of superconducting qubits, but much of the material discussed in this article is hopefully applicable to broader research areas using quantum sensors for detecting new phenomena. The subjects treated in this article range from basic features of qubits to an overview of quantum sensing and the effects of entanglement in multi-qubit systems, which may be utilized to enhance the sensitivity of DM detection.

This article is organized as follows. In the first half, an overview of the simple dynamics of qubits is presented. In Section~\ref{sec:101}, we introduce basic properties of qubits. In Section~\ref{sec:evolution}, we discuss how qubits evolve, particularly under the influence of wave-like DM. In Section~\ref{sec:environments}, we provide a theoretical framework for studying qubit evolution under the effects of decoherence and noise. In Section~\ref{sec:qubitwithnoise}, several examples of qubit evolution in the presence of noise are discussed by using the Lindblad equation. In Section~\ref{sec:cavity}, the effects of the cavity on the qubit dynamics are discussed. In the second half, we consider several issues related to quantum sensing and the effects of entanglement in qubit systems. In Section~\ref{sec:sensing}, the basics of quantum sensing are introduced. In Section~\ref{sec:HL}, the difficulty in achieving the Heisenberg limit is discussed. In Section~\ref{sec:ghz}, the possibility of using a highly entangled state for DM detection is considered. Then, at the end, using the information given throughout this article, a possible setup for a DM detection experiment using the direct excitation of superconducting qubits, which eventually grew into the DarQ project, is introduced in Section~\ref{sec:DMsearch}. Section~\ref{sec:conclusions} is devoted to conclusions and discussion. Throughout this article, natural units ($\hbar=c=1$) are used unless otherwise stated.

\section{Qubit 101}
\label{sec:101}
\setcounter{equation}{0}
\setcounter{figure}{0}
\setcounter{footnote}{1}

In this section, we summarize basic properties of qubits. We also introduce operators and quantities used throughout this article.  More complete and detailed discussions of qubits (as well as those of quantum sensing) can be found, e.g., in Refs.\ \cite{PreskillQCNotes, Nielsen:2012yss}.

\subsection{Notations}

A single qubit is the simplest quantum system, characterized as a two-level quantum system. The two basis states, typically referred to as the ground state and the excited state, are denoted by
\begin{align}
  \text{Ground State}:&\quad \ket{0} \text{~or~} \ket{\rm g},\\
  \text{Excited State}:&\quad \ket{1} \text{~or~} \ket{\rm e}.
\end{align}
In the following, we will discuss various operators that act on the single-qubit Hilbert space. Among these, several operators play fundamental roles in quantum information and computation.

First, the identity operator is given by
\begin{align}
  \mathds{I} \equiv \ketbra{0}{0} + \ketbra{1}{1}.
\end{align}
Next, following the convention adopted in Ref.\ \cite{Nielsen:2012yss}, we introduce the Pauli operators as
\begin{align}
  X &\equiv \ketbra{0}{1} + \ketbra{1}{0},
  \label{PauliX} 
  \\
  Y &\equiv -i\ketbra{0}{1} + i\ketbra{1}{0}, \\
  Z &\equiv \ketbra{0}{0} - \ketbra{1}{1},
  \label{PauliZ}
\end{align}
where $X$, $Y$, and $Z$ generate rotations on the Bloch sphere representation of the qubit. Throughout this article, we treat the basis states $\ket{0}$ and $\ket{1}$ as time-independent. Therefore, the Pauli operators are also time-independent.

These operators are related to the familiar Pauli matrices by representing them in the $\{\ket{0},\ket{1}\}$ basis:
\begin{align}
  \sigma_{X} = &\, \begin{pmatrix} 0 & 1 \\ 1 & 0 \end{pmatrix}
             = \begin{pmatrix}
                 \mel{0}{X}{0} & \mel{0}{X}{1} \\
                 \mel{1}{X}{0} & \mel{1}{X}{1}
               \end{pmatrix}, \\[2mm]
  \sigma_{Y} = &\, \begin{pmatrix} 0 & -i \\ i & 0 \end{pmatrix}
             = \begin{pmatrix}
                 \mel{0}{Y}{0} & \mel{0}{Y}{1} \\
                 \mel{1}{Y}{0} & \mel{1}{Y}{1}
               \end{pmatrix}, \\[2mm]
  \sigma_{Z} = &\, \begin{pmatrix} 1 & 0 \\ 0 & -1 \end{pmatrix}
             = \begin{pmatrix}
                 \mel{0}{Z}{0} & \mel{0}{Z}{1} \\
                 \mel{1}{Z}{0} & \mel{1}{Z}{1}
               \end{pmatrix}.
\end{align}

Additionally, we introduce the lowering (de-excitation) and raising (excitation) operators:
\begin{align}
  a \equiv &\, \ketbra{0}{1} = \frac{1}{2}\left(X + i Y\right),
  \label{op_a}
  \\
  a^\dagger \equiv &\, \ketbra{1}{0} = \frac{1}{2}\left(X - i Y\right),
  \label{op_adagger}
\end{align}
which are related to the $\sigma_+$ and $\sigma_-$ matrices as\footnote
{In the present convention, the lowering operator $a$ corresponds to the matrix $\sigma_+$; this choice may be unfamiliar to some readers. The author hopes that readers will get used to the convention adopted in this article.}
\begin{align}
  \sigma_+ = &\, \begin{pmatrix}
    0 & 1 \\ 0 & 0
  \end{pmatrix} = \begin{pmatrix}
    \mel{0}{a}{0} & \mel{0}{a}{1} \\
    \mel{1}{a}{0} & \mel{1}{a}{1}
  \end{pmatrix},
  \\
  \sigma_- = &\, \begin{pmatrix}
    0 & 0 \\ 1 & 0
  \end{pmatrix} = \begin{pmatrix}
    \mel{0}{a^\dagger}{0} & \mel{0}{a^\dagger}{1} \\
    \mel{1}{a^\dagger}{0} & \mel{1}{a^\dagger}{1}
  \end{pmatrix}.
\end{align}

To define the free Hamiltonian for the qubit, we set the energy of the ground state $\ket{0}$ to zero, and denote the energy gap between the two states by $\omega$. Thus, the free (unperturbed) Hamiltonian is
\begin{align}
  H_0 = \omega \ketbra{1}{1} = -\frac{\omega}{2}\left( Z - \mathds{I} \right).
\end{align}

\begin{figure}[t]
\begin{center}
  \includegraphics[width=0.6\textwidth]{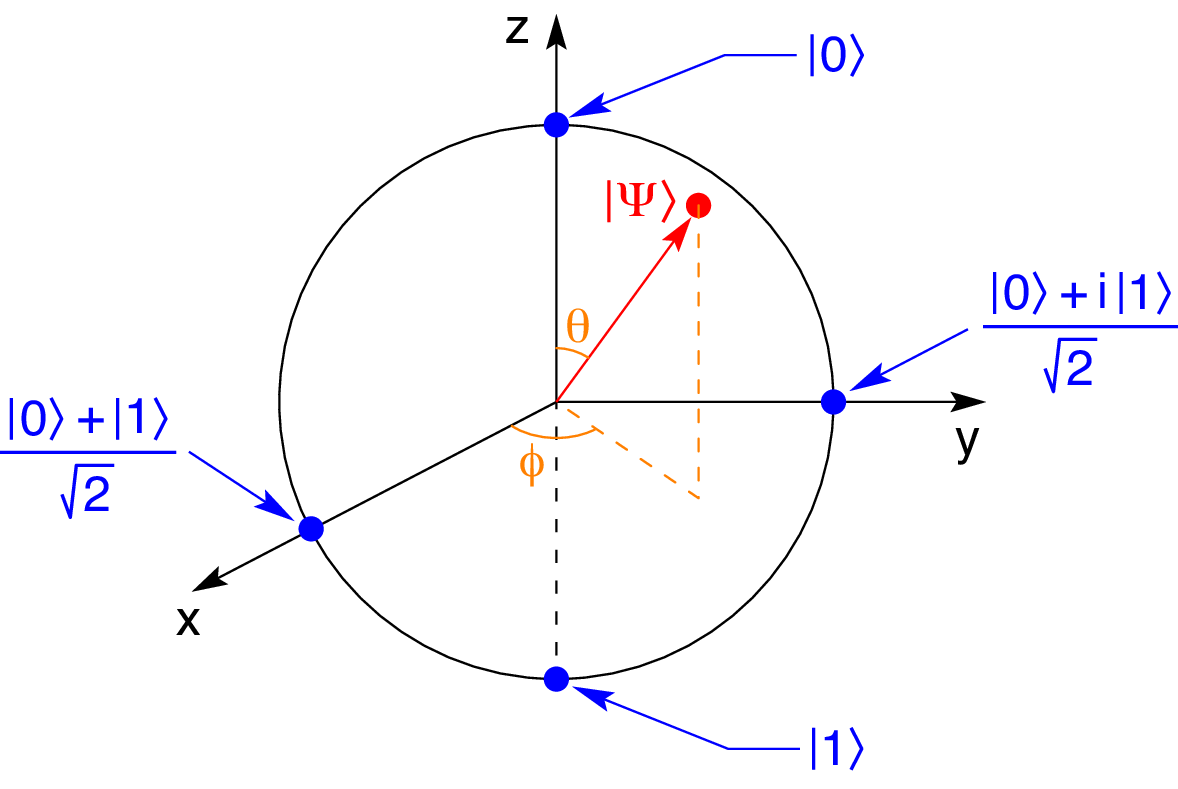}
\end{center}
\caption{Bloch sphere.}
\label{fig:blochsphere01}
\end{figure}

An arbitrary pure state of a single qubit can be written as a superposition of $\ket{0}$ and $\ket{1}$. Since the overall (global) phase of the state is physically irrelevant, it is conventional to parameterize the pure state as
\begin{align}
  \ket{\Psi} = \cos\frac{\theta}{2}\,\ket{0} + e^{i\phi}\sin\frac{\theta}{2}\,\ket{1},
  \label{blochsph}
\end{align}
where $\theta$ and $\phi$ are real parameters in the ranges
\begin{align}
  0 \leq \theta \leq \pi, \qquad 0 \leq \phi < 2\pi.
\end{align}
This parameterization corresponds to the standard spherical coordinate representation on the surface of a unit sphere, known as the Bloch sphere. Each point on the surface of the Bloch sphere represents a distinct pure qubit state (see Fig.~\ref{fig:blochsphere01}).

\subsection{Density Matrix of a Qubit}

In many practical situations, a qubit is not perfectly isolated from its environment. As a consequence, noise and other uncontrolled degrees of freedom can render the qubit state mixed rather than pure. A convenient and general framework for describing both pure and mixed states is given by the density matrix, denoted by $\rho$.

The density-matrix formalism applies not only to a qubit but also to general quantum systems. Suppose that the system is prepared in one of the (normalized) states $\{ \ket{\varphi_i} \}$ with probability $p_i \ge 0$, where the index $i$ labels different preparation outcomes. The density matrix is then defined as
\begin{align}
  \rho \equiv \sum_i p_i \ketbra{\varphi_i}{\varphi_i}.
  \label{defrho}
\end{align}
With the normalization condition $\sum_i p_i = 1$, we have
\begin{align}
  \mathrm{tr}(\rho) = 1.
\end{align}
Note that $\rho$ is Hermitian and positive semidefinite by construction. Using the density matrix, the expectation value of an operator $\mathcal{O}$ is given by
\begin{align}
  \ev{\mathcal{O}} = \mathrm{tr}(\rho\, \mathcal{O}).
\end{align}

For a single qubit, any density matrix can be expanded in the operator basis formed by the identity and the Pauli operators:
\begin{align}
  \rho = \frac{1}{2}\left(\mathds{I} + \rho_X X + \rho_Y Y + \rho_Z Z\right),
  \label{bloch_rep_rho}
\end{align}
where the coefficients $\rho_X$, $\rho_Y$, and $\rho_Z$ are real. The physicality of the state (i.e., $\rho$ being positive semidefinite) implies the constraint
\begin{align}
  \rho_X^2 + \rho_Y^2 + \rho_Z^2 \leq 1.
\end{align}
The state is pure if and only if the equality holds. In particular, if a pure state is parameterized as in Eq.~\eqref{blochsph}, then the coefficients in Eq.\ \eqref{bloch_rep_rho} are given by
\begin{align}
  \rho_X = &\, \sin\theta\cos\phi,
  \\
  \rho_Y = &\, \sin\theta\sin\phi,
  \\
  \rho_Z = &\, \cos\theta.
\end{align}
On the other hand, a mixed state satisfies $\rho_X^2 + \rho_Y^2 + \rho_Z^2 < 1$. The vector
\begin{align}
  \vec{\rho} \equiv (\rho_X,\rho_Y,\rho_Z)
\end{align}
is called the Bloch vector. 

One may also represent the single-qubit density matrix in a different form; in the computational basis $\{\ket{0},\ket{1}\}$, 
\begin{align}
  \rho = \sum_{I,J=0,1} \rho_{IJ}\ketbra{I}{J},
  \label{rho_IJ}
\end{align}
i.e., in the matrix form, 
\begin{align}
  \rho \equiv
  \begin{pmatrix}
    \rho_{00} & \rho_{01}\\
    \rho_{10} & \rho_{11}
  \end{pmatrix}
  =
  \begin{pmatrix}
    \mel{0}{\rho}{0} & \mel{0}{\rho}{1}\\
    \mel{1}{\rho}{0} & \mel{1}{\rho}{1}
  \end{pmatrix}.
\end{align}
The normalization $\mbox{tr} ( \rho) =1$ yields
\begin{align}
  \rho_{00} + \rho_{11} = 1.
\end{align}
Moreover, since $\rho$ is Hermitian, the diagonal elements $\rho_{00}$ and $\rho_{11}$ are real, while the off-diagonal elements satisfy $\rho_{10} = \rho_{01}^*$. The off-diagonal elements quantify the coherence between $\ket{0}$ and $\ket{1}$, and they are typically reduced by dephasing noise.

Through a procedure known as purification, the density matrix of a mixed state can always be expressed as that of a pure state defined on an enlarged Hilbert space that includes an auxiliary environment. Because the density matrix is Hermitian and positive semidefinite, it can be decomposed as
\begin{align}
  \rho = \sum_i \lambda_i \ketbra{\psi_i}{\psi_i},
\end{align}
where $\lambda_i\geq 0$ are the eigenvalues of $\rho$, and $\ket{\psi_i}$ are the corresponding normalized, mutually orthogonal eigenstates. Then, by introducing
\begin{align}
  \tilde{\rho} \equiv \ketbra{\Psi}{\Psi},
\end{align}
where
\begin{align}
  \ket{\Psi} \equiv \sum_i \sqrt{\lambda_i} \ket{\psi_i} \otimes \ket{e_i},
\end{align}
with $\ket{e_i}$ being auxiliary environment states that are orthogonal to each other, i.e., $\braket{e_i}{e_j} = \delta_{i,j}$, the density matrix of the system is recovered as
\begin{align}
  \rho = \mbox{tr}_{\rm E} ( \tilde{\rho} ),
\end{align}
where $\mbox{tr}_{\rm E}$ denotes the partial trace over the environment. In this way, any density matrix describing a mixed state can always be ``purified,'' that is, expressed as a pure state on the enlarged system that includes the auxiliary environment.

\subsection{Coherence Time of a Qubit}

In discussing the coherence time of a single qubit, there are two commonly used time scales. The first is the so-called ``energy relaxation time'' (or ``longitudinal relaxation time''), which characterizes how rapidly an excited qubit relaxes to the ground state due to interactions with its environment. This time scale is typically denoted by $T_1$, in terms of which the excited-state population (the $(1,1)$ element of the density matrix) decays as
\begin{align}
  \rho_{11}(t) \sim \rho_{11}(0)\, e^{-t/T_1}.
\end{align}

The other important time scale is the so-called ``dephasing time'' (or ``transverse relaxation time''), which characterizes the loss of phase coherence. This time scale is usually denoted by $T_2$ and describes the decay of the off-diagonal elements of the density matrix as
\begin{align}
  |\rho_{01}(t)| \sim |\rho_{01}(0)|\, e^{-t/T_2}.
\end{align}

The $T_1$ and $T_2$ parameters are commonly used to quantify the coherence quality of a qubit; a qubit with longer $T_1$ and longer $T_2$ has higher quality. A discussion of the $T_1$ and $T_2$ parameters in the context of specific noise models can be found in Section~\ref{sec:qubitwithnoise}.

\section{Unitary Evolution of a Single Qubit}
\label{sec:evolution}
\setcounter{equation}{0}
\setcounter{figure}{0}
\setcounter{footnote}{1}

To perform DM detection using qubits, the first step is to understand the excitation process of a single qubit under the influence of a DM-induced electromagnetic (EM) field. In this article, we mainly study the excitation of the qubit due to an oscillating electric field (which may arise from DM oscillations). For this purpose, as we explain below, we employ an effective Hamiltonian that is applicable to a wide class of qubits, including superconducting qubits. (For an overview of superconducting qubits, one of the most popular laboratory implementations of qubits, see Appendix~\ref{app:superconducting}.) We also discuss the effects of DM decoherence, which are not captured by this Hamiltonian, as well as the estimation of the DM coherence time in terms of the DM velocity distribution.

\subsection{Qubit-DM Interaction}

In this article, we consider wave-like DM, namely a coherently oscillating bosonic field whose oscillation frequency is (almost) equal to its mass. The basic properties of wave-like DM are summarized in Appendix \ref{app:dm}. The evolution of the bosonic field at the location of a qubit can be approximated as
\begin{align}
  \sigma (t) \simeq \bar{\sigma} \cos (m t - \varphi),
  \label{phi(Minkowski)}
\end{align}
where $\bar{\sigma}$ is the amplitude, $m$ is the DM mass, and $\varphi$ is an unknown phase. Owing to the DM velocity distribution, the phase $\varphi$ is effectively randomized and can be regarded as being reset after each DM coherence time (see the discussion below).

In what follows, we focus on the case where the DM field induces an AC electric field and thereby drives excitation and de-excitation processes of qubits. The effective single-qubit Hamiltonian we employ is
\begin{align}
  H = H_0 + H_1,
  \label{Hamiltonian_tot}
\end{align}
where
\begin{align}
  H_0 &= \omega \ketbra{1}{1}
  = -\frac{1}{2}\omega\left(Z-\mathds{I}\right),
  \label{Hamiltonian_0}
  \\
  H_1 &= -2\eta\left(\ketbra{0}{1}+\ketbra{1}{0}\right)\cos(mt-\varphi)
  = -2\eta\,X\cos(mt-\varphi).
  \label{Hamiltonian_1}
\end{align}
Here, $H_0$ is the free Hamiltonian, whereas $H_1$ describes the interaction with an oscillating external field. This effective Hamiltonian applies to a wide range of qubit platforms. The drive strength $\eta$ depends on both the DM properties and the specific qubit realization. For superconducting qubits, the corresponding expression for $\eta$ is given in Appendix \ref{app:superconducting}. In general, the coupling encoded in $\eta$ is extremely small; the goal of a qubit-based DM detection experiment is to establish a nonzero value of $\eta$.

Throughout this article we consider the case in which the interaction term is proportional to the Pauli-$X$ operator. As discussed in Appendix \ref{app:superconducting}, such an effective interaction of the qubit arises in particular when the qubit couples to an electric field, as is the case for superconducting qubits interacting with a DM-induced electric field. Such an effective interaction is not, however, the most general one. Another possibility is that the effective interaction of the qubit is proportional to the Pauli-$Z$ operator. Such an interaction becomes important when the qubit interacts with an external magnetic field. Because the main focus of this article is to consider the use of the direct excitation of superconducting qubits for DM detection, we only consider the interaction proportional to the Pauli-$X$ operator.

\subsection{Evolution of a Single Qubit}

First, we study the evolution of a single qubit using the Hamiltonian given in Eq.\ \eqref{Hamiltonian_tot}. For this purpose, it is convenient to move to the interaction picture with $H_0$ as the free Hamiltonian. The interaction-picture Hamiltonian is then
\begin{align}
  H_{\rm I}
  \equiv &\, e^{iH_0 t} H_1 e^{-iH_0 t}
  = -2\eta \cos(mt-\varphi)\left(X\cos\omega t + Y\sin\omega t\right).
\end{align}
Accordingly, the interaction-picture state $\ket{\Psi_{\rm I}}$ evolves as
\begin{align}
  i\,\partial_t \ket{\Psi_{\rm I}(t)} = H_{\rm I}(t)\ket{\Psi_{\rm I}(t)}.
\end{align}
We expand the state by using the states $\ket{0}$ and $\ket{1}$ as
\begin{align}
  \ket{\Psi_{\rm I}(t)} = \psi_0 (t)\ket{0} + \psi_1(t) \ket{1}.
\end{align}
Notice that, with the same wave function, the state in the Schr\"odinger picture is described as 
\begin{align}
  \ket{\Psi_{\rm S}(t)} = \psi_0 (t)\ket{0} + \psi_1(t) e^{-i\omega t}\ket{1}.
\end{align}
The evolution equation of the wave function is then given by
\begin{align}
  i \partial_t
  \begin{pmatrix}
    \psi_0 \\ \psi_1
  \end{pmatrix}
  = -2\eta \cos(mt-\varphi)
  \begin{pmatrix}
    0 & e^{-i\omega t} \\
    e^{i\omega t} & 0
  \end{pmatrix}
  \begin{pmatrix}
    \psi_0 \\ \psi_1
  \end{pmatrix}.
  \label{Schrodinger}
\end{align}

We first perturbatively treat the interaction term to analyze the transition probability. In the context of DM detection, $\eta$ is expected to be highly suppressed. We therefore expand the wave function in powers of $\eta$ as
\begin{align}
  \psi_0 (t) = \sum_{p=0}^\infty \psi_0^{(p)} (t),~~~
  \psi_1 (t) = \sum_{p=0}^\infty \psi_1^{(p)} (t),
  \label{fetap}
\end{align}
where $\psi_0^{(p)}$ and $\psi_1^{(p)}$ are of $O(\eta^p)$.

\begin{figure}[t]
  \centering
  \includegraphics[width=0.45\textwidth]{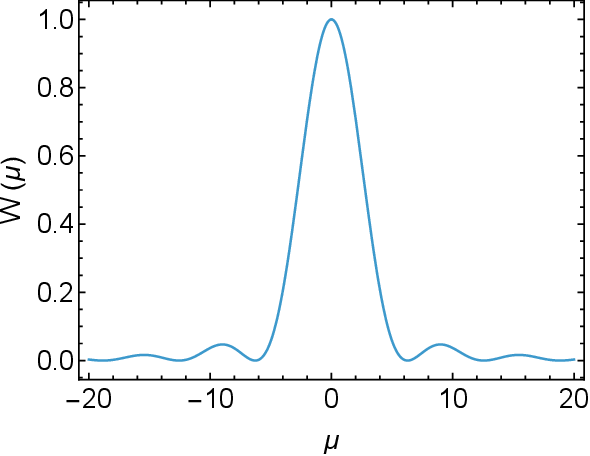}
  \caption{The shape of the function $W(\mu)$.}
  \label{fig:fnW}
\end{figure}

Here, in order to understand the evolution of the qubit, we consider the excitation starting from the ground state. The initial conditions are then
\begin{align}
  \psi_0^{(0)} (0)=1, ~~~\psi_0^{(p\geq 1)} (0)=\psi_1^{(p)} (0)=0.
\end{align}
Substituting Eq.\ \eqref{fetap} into Eq.\ \eqref{Schrodinger}, we obtain
\begin{align}
  \psi_0 (t) = &\, 1 + O(\eta^2),~~
  \label{fgeta}
  \\
  \psi_1 (t) = &\, \eta
  \left(
    \frac{e^{i(\omega -m)t}-1}{\omega -m} e^{i\varphi}
    + \frac{e^{i(\omega +m)t}-1}{\omega +m} e^{-i\varphi}
    \right) + O(\eta^3).
    \label{feeta}
\end{align}
The excitation probability from the ground state to the excited state is given by $p_{\rm ge}(t)=|\psi_1 (t)|^2$. When $|\omega-m|\ll \omega+m$, the first term in the parentheses on the right-hand side of Eq.\ \eqref{feeta} dominates. In this regime, to leading order in $\eta$, $p_{\rm ge}(t)$ is estimated as
\begin{align}
  p_{\rm ge}(t) \simeq (\eta t)^2 \, W ((\omega-m) t),
\end{align}
where
\begin{align}
  W (\mu) = \frac{2(1-\cos\mu)}{\mu^2}.
\end{align}
In Fig.\ \ref{fig:fnW}, we plot $W (\mu)$. The function $W (\mu)$ is peaked at $\mu=0$ and satisfies
\begin{align}
  \int_{-\infty}^\infty d\mu W(\mu) = 2 \pi.
  \label{IntW=2pi}
\end{align}
It also admits the expansion
\begin{align}
  W (\mu) = 1 - \frac{1}{12} \mu^2 + O(\mu^4),
\end{align}
while $W(\mu)\sim O(\mu^{-2})$ as $\mu\rightarrow\infty$. Thus, $p_{\rm ge}(t)$ grows quadratically in time as long as $t\lesssim |\omega-m|^{-1}$, and is bounded by $\sim O(\eta^2/|\omega-m|^2)$. In the resonance limit, where the qubit frequency matches the oscillation frequency of the external field (i.e., $\omega\rightarrow m$), this growth persists until $t\sim \eta^{-1}$, and the excitation probability can be substantially enhanced. In DM detection experiments, the advantages of operating near resonance are often exploited to the fullest extent.

Next, let us consider the resonant limit, i.e., $\omega\rightarrow m$, in which we can analyze the evolution of the qubit without perturbative treatment of the interaction.

In the resonance limit, one can often employ the rotating-wave approximation (RWA), in which rapidly oscillating terms in the Hamiltonian are neglected. Under the RWA, we drop the terms oscillating at frequencies of $O(m)$, and the interaction-picture Hamiltonian in the resonant limit reduces to
\begin{align}
  H_{\rm I} \simeq -\eta (X \cos\varphi + Y \sin\varphi).
  \label{H_I(resonance)}
\end{align}
The unitary operator governing the time evolution is then
\begin{align}
  U_{\rm I} \equiv e^{-i H_{\rm I}t} \simeq
  \cos \eta t + i (X \cos\varphi + Y \sin\varphi) \sin \eta t.
\end{align}
Here and in what follows, we omit the identity operator unless it is needed explicitly.
Accordingly, the wave function evolves as
\begin{align}
  \begin{pmatrix}
    \psi_0 (t) \\ \psi_1(t)
  \end{pmatrix}
  \simeq &\,
  \left[ \cos \eta t + i (\sigma_X \cos\varphi + \sigma_Y \sin\varphi) \sin \eta t \right]
  \begin{pmatrix}
    \psi_0 (0) \\ \psi_1(0)
  \end{pmatrix}
  \nonumber \\ = &\,
  \begin{pmatrix}
    \cos \eta t & i e^{-i\varphi}\sin \eta t \\
    i e^{i\varphi} \sin \eta t & \cos \eta t
  \end{pmatrix}
  \begin{pmatrix}
    \psi_0 (0) \\ \psi_1(0)
  \end{pmatrix}.
\end{align}
We can see, for example, that for $\ket{\Psi_{\rm I}(0)}=\ket{0}$ the state traces out a great circle on the Bloch sphere, called Rabi oscillation.

\begin{figure}[t]
  \centering
  \includegraphics[width=0.45\textwidth]{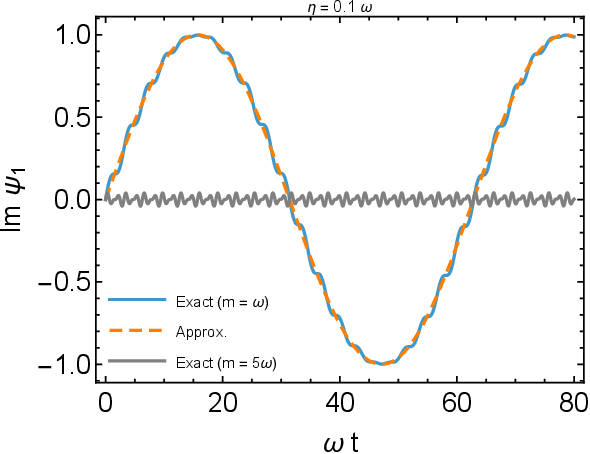}
  \caption{Evolution of $\mbox{Im} \psi_1$ as a function of time, taking $\varphi=0$. The initial condition is set as $(\psi_0(0),\psi_1(0))=(1,0)$. The results for the resonance limit (i.e., $\omega\rightarrow m$) are shown in blue-solid (exact result) and magenta-dashed (RWA result) lines. The gray-solid line shows the exact result for the non-resonance case with $\omega=\frac{1}{5}m$.}
  \label{fig:precession}
\end{figure}

In the discussion above, we have adopted the RWA. However, even if we solve the exact Schr\"odinger equation, we can observe an approximate circular motion of the state on a great circle of the Bloch sphere. In Fig.\ \ref{fig:precession}, we plot $\mbox{Im}\psi_1$ as a function of time, taking $\varphi=0$ and $\omega=m$. Comparing the exact and RWA results, we can see that the RWA provides a good approximation for the analysis of the evolution of the state.  In the same figure, we also show the exact result for the non-resonance case, taking $\omega=\frac{1}{5}m$. In the non-resonance case, excitation of the qubit is highly suppressed, as expected.

\subsection{Coherence of DM}

So far, we have studied the evolution of a single qubit coupled to an oscillating external field with a fixed frequency and a fixed phase $\varphi$. If, however, the field originates from a coupling to a wave-like DM background, this description is generally insufficient; the DM wave in the Solar neighborhood is composed of many velocity components and therefore exhibits a corresponding distribution of frequencies. In this section, we discuss the consequences of such a frequency distribution (see also Appendix~\ref{subsec:decoherence}). A nontrivial velocity distribution is known to affect the event rates in haloscope DM detection experiments \cite{Krauss:1985ub, Turner:1990qx, Foster:2017hbq, Foster:2020fln, Gramolin:2021mqv, Dror:2022xpi}. In particular, an important consequence is the phase decoherence of the DM oscillation, as we show below. We therefore analyze the impact of the DM velocity distribution in the context of DM detection with qubits.

To make the discussion concrete, we focus on wave-like DM and denote the corresponding real scalar field by $\sigma$. Assuming that the DM field can be described as a superposition of modes with different velocities, the field amplitude at the location $\vec{x}$ can be written as \cite{Foster:2017hbq, Foster:2020fln}
\begin{align}
  \sigma (t, \vec{x}) =
  \sum_{\vec{v}}
  \tilde{\sigma}_{\vec{v}} \cos \left(
  E_v t - m \vec{v} \vec{x} - \varphi_{\vec{v}}
  \right),
  \label{sigma_DM}
\end{align}
where the sum runs over the velocity components. Here, $\tilde{\sigma}_{\vec{v}}$ is the (real) amplitude of the mode with velocity $\vec{v}$, and $\varphi_{\vec{v}}$ is the corresponding random phase. The mode energy is
\begin{align}
  E_v \equiv \frac{m}{\sqrt{1-v^2}} \simeq m + \frac{1}{2} m v^2,
\end{align}
where $v \equiv |\vec{v}|$, and in the second equality we have used the fact that the relevant DM components are non-relativistic. 

Hereafter, we fix the location of the experimental apparatus at $\vec{x}=0$. Then, for a qubit coupled to such a DM field $\sigma$, the interaction Hamiltonian is expected to take the form\footnote
{Here, for simplicity, we consider the case of a scalar field. For the case of vector-field DM, we may have to take into account a possible mismatch between the direction of the DM polarization and the sensitivity axis of the qubit (see Eq.\ \eqref{eta_DM}). Even in such a case, the following arguments remain unchanged.}
\begin{align}
  H_1 = -2 \eta \left(\ketbra{0}{1}+\ketbra{1}{0}\right)
  \sum_{\vec{v}}
  f_{\vec{v}} \cos \left( E_v t - \varphi_{\vec{v}} \right),
  \label{Hamiltonian_1'}
\end{align}
where
\begin{align}
  f_{\vec{v}} \equiv \frac{\tilde{\sigma}_{\vec{v}}}{\bar{\sigma}},
\end{align}
with
\begin{align}
  \bar{\sigma} \equiv \sqrt{ \sum_{\vec{v}} \tilde{\sigma}_{\vec{v}}^2 }.
\end{align}

Before proceeding, we comment on the spatial coherence length of the DM field. According to Eq.\ \eqref{sigma_DM}, the DM amplitude is position dependent; the characteristic wavelength is
\begin{align}
  \ell_{\rm DM} \equiv \frac{2\pi}{m v_{\rm DM}}
  \simeq 1.2\ {\rm km} \times
  \left( \frac{m}{1\, \mu{\rm eV}} \right)^{-1}
  \left( \frac{v_{\rm DM}}{10^{-3}} \right)^{-1},
\end{align}
where $v_{\rm DM}\sim 10^{-3}$ is the mean DM velocity. The quantity $\ell_{\rm DM}$ can also be interpreted as the coherence length of the DM field, within which the spatial variation of the DM amplitude is negligible. In what follows, we consider setups involving multiple qubits. In this case, we assume that all qubits are located within a distance much shorter than $\ell_{\rm DM}$, so that they effectively experience the same DM field.

We now consider qubit evolution under the interaction Hamiltonian in Eq.~\eqref{Hamiltonian_1'}. Treating the interaction perturbatively, we can straightforwardly obtain the dynamics. For the initial condition $\ket{\Psi(0)}=\ket{0}$, we can use Eqs.~\eqref{fgeta} and \eqref{feeta} (neglecting irrelevant terms); we find the state at time $t$ (in the Schr\"odinger picture) as
\begin{align}
  \ket{\Psi (t)} \simeq \ket{0} +
  \eta
  \left( \sum_{\vec{v}} \frac{e^{i(\omega -E_v)t}-1}{\omega -E_v} e^{i\varphi_{\vec{v}}} f_{\vec{v}} \right) e^{-i\omega t}
  \ket{1}.
\end{align}
The transition probability from the ground state to the excited state is then
\begin{align}
  p_{\rm ge} (t) = \left| \braket{1}{\Psi (t)} \right|^2 =
  \eta^2 \sum_{\vec{v},\vec{v}'}
  \left( \frac{e^{i(\omega -E_v)t}-1}{\omega -E_v} e^{i\varphi_{\vec{v}}} f_{\vec{v}} \right)
  \left( \frac{e^{-i(\omega -E_{v'})t}-1}{\omega -E_{v'}} e^{-i\varphi_{\vec{v}'}} f_{\vec{v}'} \right).
\end{align}
The terms with $\vec{v}\neq\vec{v}'$ involve random phase factors $e^{i(\varphi_{\vec{v}}-\varphi_{\vec{v}'})}$, in addition to phases that vary in time. As a result, these cross terms are expected to cancel through destructive interference when performing an ensemble (or volume) average,\footnote
{
We adopt the ergodic hypothesis and assume the DM field is statistically homogeneous. Then, the ensemble average over the random phases $\varphi_{\vec{v}}$ can be equivalently evaluated as a spatial average over a sufficiently large volume. Here, we consider box regularization in a cubic region of volume $L^3$. In this setting, the velocity takes discrete values $\vec{v}=\frac{2\pi}{mL}(n_x,n_y,n_z)$, where $n_{x,y,z}$ are integers. As can be seen from Eq.\ \eqref{sigma_DM}, the effective phase parameter at position $\vec{x}$ is given by $\varphi_{\vec{v}}^{\rm (eff)}(\vec{x})=m\vec{v}\vec{x}+\varphi_{\vec{v}}$. The volume average is then
\begin{align*}
  \ev{e^{i\varphi_{\vec{v}}^{\rm (eff)}} e^{-i\varphi_{\vec{v}'}^{\rm (eff)}}}_{\rm volume} =
  \frac{1}{L^3} \int d^3 x
  e^{i(m\vec{v}\vec{x}+\varphi_{\vec{v}})} e^{-i(m\vec{v}'\vec{x}+\varphi_{\vec{v}'})}
  = \delta_{\vec{v},\vec{v}'},
\end{align*}
which implies that only the diagonal terms survive.
}
as well as under a time average. 

This observation motivates us to focus on the diagonal contributions with $\vec{v}=\vec{v}'$, and to assume that the net contribution from the $\vec{v}\neq\vec{v}'$ terms is subdominant after ensemble averaging. By retaining only the $\vec{v}=\vec{v}'$ terms, we obtain
\begin{align}
  p_{\rm ge} (t) \simeq
  \eta^2 t^2 \sum_{\vec{v}} W((\omega -E_v)t) f_{\vec{v}}^2 .
  \label{pge_stochastic}
\end{align}

We assume that the state density in velocity space is dense enough so that $f_{\vec{v}}^2$ can be treated as a smooth function of $\vec{v}$. Then, we introduce the DM velocity distribution function (with directional information being integrated out) as
\begin{align}
  \mathcal{F} (v) \equiv \frac{1}{\Delta v}
  \sum_{v\leq |\vec{v}'|\leq v+\Delta v} f_{\vec{v}'}^2,
  \label{F(v)}
\end{align}
where $v\equiv |\vec{v}|$, and the sum runs over $\vec{v}'$ with $v\leq |\vec{v}'|\leq v+\Delta v$. Here, $\Delta v$ is a parameter chosen to be much smaller than the characteristic DM speed; the distribution function is normalized as
\begin{align}
  \int_0^\infty dv \mathcal{F} (v) = 1.
\end{align}
Notice that the velocity distribution of the local DM energy density in the Solar system is expressed as 
\begin{align}
  \frac{d \rho_{\rm DM}}{dv} \simeq
  \rho_{\rm DM}^{\rm (Solar)} \mathcal{F} (v),
\end{align}
where $\rho_{\rm DM}^{\rm (Solar)}$ denotes the total DM energy density in the Solar system (see Appendix \ref{app:dm}).

In the standard halo model \cite{Drukier:1986tm, Lewin:1995rx}, the velocity distribution in the Solar rest frame is
\begin{align}
  \mathcal{F}^{\rm (SHM)} (v) = \mathcal{N}^{-1} \frac{v}{\sqrt{\pi}v_0 v_\odot} 
  e^{-(v^2+v_\odot^2)/v_0^2} 
  \left( e^{2vv_\odot/v_0^2} - e^{-2c_\star vv_\odot/v_0^2} \right),
\end{align}
where $v_{\rm esc}$ is the local escape velocity, $v_0$ is the local circular velocity, $v_\odot$ is the Solar velocity relative to the Galaxy, 
\begin{align}
  \mathcal{N} = \left[ \erf (v_{\rm esc}/v_0) - \frac{2v_{\rm esc}}{\sqrt{\pi}v_0} e^{-v_{\rm esc}^2/v_0^2} \right],
\end{align}
and\footnote
{The clip function is defined as
\begin{align*}
  \mbox{clip}_{[-1,1]} (x) = 
  \left\{
  \begin{array}{ll}
    -1 & ~:~ x<-1 \\
    x & ~:~ -1 \leq x \leq 1 \\
    1 & ~:~ x>1
  \end{array}
  \right. .
\end{align*}
}
\begin{align}
  c_\star \equiv \mbox{clip}_{[-1,1]} \left( \frac{v_{\rm esc}^2 - v^2 - v_\odot^2}{2vv_\odot} \right).
\end{align}

With the velocity distribution function, the transition probability can be rewritten as \cite{Fukuda:2025zcf}
\begin{align}
  p_{\rm ge} (t) \simeq
  \eta^2 t^2 \int dv \mathcal{F} (v) W((\omega -E_v)t).
  \label{Pge_multifreq}
\end{align}
Importantly, the mean DM velocity in the Solar system is $v_{\rm DM}\sim 10^{-3}$ and the velocity dispersion of the DM is also $\sim v_{\rm DM}$, while the velocity components with $v\gg v_{\rm DM}$ are highly suppressed. Hence, the integral in Eq.~\eqref{Pge_multifreq} is dominated by the range $v\lesssim O(v_{\rm DM})$.

In order to clarify how the excitation probability depends on the exposure time $t$, it is convenient to define the DM coherence time as
\begin{align}
  \tau_{\rm DM} \equiv \frac{2\pi}{m v_{\rm DM}^2}.
  \label{tau_DM}
\end{align}
Numerically, the coherence time is
\begin{align}
  \tau_{\rm DM} \simeq 4\, {\rm msec}\times
  \left( \frac{m}{1\, \mu{\rm eV}} \right)^{-1}
  \left( \frac{v_{\rm DM}}{10^{-3}} \right)^{-2}.
\end{align}
When $t\ll \tau_{\rm DM}$, and in the resonance limit (i.e., $\omega\rightarrow m$), the argument of the function $W$ in Eq.~\eqref{Pge_multifreq} becomes much smaller than unity for the relevant region $v\sim O(v_{\rm DM})$. In this limit, it is reasonable to approximate $W\simeq 1$, so that
\begin{align}
  \left. p_{\rm ge}
  (t\ll \tau_{\rm DM})
  \right|_{\omega\simeq m}
  \simeq \eta^2 t^2,
  \label{Pge(small-t)}
\end{align}
which reproduces the result for a monochromatic external field.

\begin{figure}
  \centering
  \includegraphics[width=0.5\textwidth]{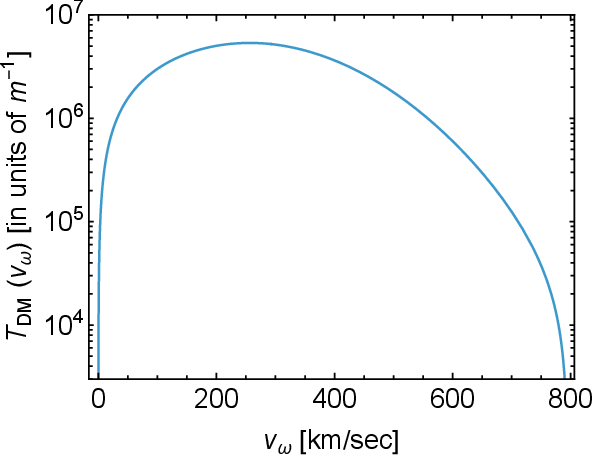}
  \caption{The coherence-time function $\mathcal{T}_{\rm DM}$ (in units of $m^{-1}$, the inverse of the DM mass) as a function of $v_\omega$, adopting the standard halo model (with $v_{\rm esc} = 544\ {\rm km/sec}$ \cite{Smith:2006ym}, $v_0 = 238\ {\rm km/sec}$, and $v_\odot = 251\ {\rm km/sec}$ \cite{Baxter:2021pqo}).}
  \label{fig:tauDM}
\end{figure}

On the other hand, the regime $t\gg \tau_{\rm DM}$ exhibits more intricate behavior. To discuss this case, it is convenient to parameterize the qubit frequency in the resonance region by introducing $v_\omega$ (with $v_\omega\lesssim O(v_{\rm DM})$), defined by
\begin{align}
  \omega = m + \frac{1}{2} m v_\omega^2.
\end{align}
For $t\gg 2\pi/m v_{\rm DM}^2$, the function $W((\omega -E_v)t)\simeq W(m v_\omega(v-v_\omega)t)$ is sharply peaked around $v= v_\omega$ (see Fig.\ \ref{fig:fnW}). Thus, the transition probability given in Eq.\ \eqref{Pge_multifreq} can be approximated as
\begin{align}
  p_{\rm ge} (t\gg \tau_{\rm DM})
  \simeq
  \eta^2 t \frac{2\pi \mathcal{F} (v_\omega)}{m v_\omega}.
  \label{Pge(large-t)}
\end{align}
Here, the factor $2\pi$ originates from the integral of $W$ (see Eq.\ \eqref{IntW=2pi}). This naturally motivates the introduction of the following quantity:
\begin{align}
  \mathcal{T}_{\rm DM} (v_\omega) \equiv
  \frac{2\pi \mathcal{F} (v_\omega)}{m v_\omega},
  \label{T_DM(v)}
\end{align}
with which $p_{\rm ge} (t\gg \tau_{\rm DM}) \simeq \eta^2\mathcal{T}_{\rm DM} t$. This quantity can be interpreted as the coherence time of the DM oscillation relevant to a qubit whose energy gap is $\omega = m + \frac{1}{2} m v_\omega^2$. In Fig.\ \ref{fig:tauDM}, we show $\mathcal{T}_{\rm DM}$ as a function of $v_\omega$, calculated based on the standard halo model. Typically, $\mathcal{F} (v_{\rm DM}) \sim O(v_{\rm DM}^{-1})$, and thus $p_{\rm ge} (t\gg \tau_{\rm DM}) \sim O(\eta^2 \tau_{\rm DM} t)$. Furthermore, $\mathcal{T}_{\rm DM} (v_{\rm DM})\sim\tau_{\rm DM}$, and Eqs.\ \eqref{Pge(small-t)} and \eqref{Pge(large-t)} are smoothly connected at $t\sim\tau_{\rm DM}$. 

\begin{figure}
  \centering
  \includegraphics[width=0.5\textwidth]{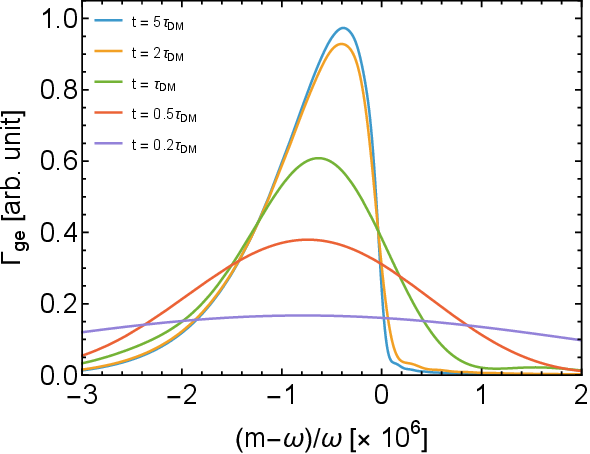}
  \caption{The effective transition rate $\Gamma_{\rm ge} (t)$ for several values of $t$ as a function of the normalized detuning $(m-\omega)/\omega$. The standard halo model with $v_{\rm esc} = 544\ {\rm km/sec}$ \cite{Smith:2006ym}, $v_0 = 238\ {\rm km/sec}$, and $v_\odot = 251\ {\rm km/sec}$ \cite{Baxter:2021pqo} is adopted.}
  \label{fig:evtrateDM}
\end{figure}

The qualitative behavior of the excitation probability changes around $t \sim \tau_{\rm DM}$. The quadratic growth of $p_{\rm ge}(t)$ with time, as shown in Eq.\ \eqref{Pge(small-t)}, ceases to hold once $t \sim \tau_{\rm DM}$. For $t \gtrsim \tau_{\rm DM}$, $p_{\rm ge}(t)$ instead exhibits a linear dependence on time. This leads to a significant suppression of the excitation probability in the large-$t$ regime compared to the scenario with a perfectly monochromatic background oscillation.

To make this suppression explicit, we introduce the effective transition rate, defined as the transition probability per unit time:
\begin{align}
  \Gamma_{\rm ge} (t) \equiv \frac{p_{\rm ge} (t)}{t},
\end{align}
which is relevant for $p_{\rm ge}\ll 1$. In Fig.\ \ref{fig:evtrateDM}, we plot $\Gamma_{\rm ge}$ as a function of the normalized detuning, $(m-\omega)/\omega$, for several values of the exposure time $t$. For a fixed DM mass, the transition rate increases monotonically until $t\sim O(\tau_{\rm DM})$, after which $\Gamma_{\rm ge}$ approaches a constant value, in agreement with the scaling behaviors discussed in Eqs.\ \eqref{Pge(small-t)} and \eqref{Pge(large-t)}. Furthermore, one finds that for fixed $\omega$, the sensitivity width in the DM mass becomes narrower as the exposure time $t$ is increased; this narrowing saturates at $t\sim\tau_{\rm DM}$. This detailed understanding of the excitation rate is crucial for optimizing the scan strategy in DM searches.

Here we have focused on the excitation process of the qubit. We emphasize, however, that the analysis above regarding the loss of DM coherence applies more broadly and is not specific to qubit excitations. In particular, the considerations are equally relevant to other types of haloscope DM detection experiments.  In those cases, too, the loss of coherence over a finite time interval affects the signal response, and the estimates for the transition probability must be appropriately modified to take into account the DM coherence properties. The impact of DM decoherence is a general feature that extends beyond the specific scenario of qubit excitations.

In practice, the field amplitude at the location of the experimental apparatus can be represented by the following single-frequency profile:
\begin{align}
  \sigma (t) \simeq \bar{\sigma} \cos (m t - \varphi),
  \label{sigma(t)}
\end{align}
where $m$ is the mass of $\sigma$ and $\varphi$ is a random phase. It should be emphasized, however, that the oscillation phase $\varphi$ must be regarded as random and uncorrelated between time intervals separated by the coherence time $\tau_{\rm DM}$. In other words, within each coherence interval, the DM oscillation maintains phase coherence, but the phase is effectively re-randomized from one interval to the next (see also Appendix~\ref{app:dm}). This treatment captures the essential features of the DM field and, up to an $O(1)$ numerical factor, reproduces Eqs.\ \eqref{Pge(small-t)} and \eqref{Pge(large-t)}.

In the approximation adopted above, we have assumed that various velocity modes interfere destructively, resulting in an effective cancellation of modes with $\vec{v}\neq\vec{v}'$. One should note that such a cancellation holds only after taking the stochastic and time averages; the same caveat applies to all the results derived under this approximation. In particular, the local amplitude of the wave-like DM varies by $O(1)$ relative to its mean value \cite{Centers:2019dyn, Lisanti:2021vij} (see also Appendix \ref{subsec:decoherence}), and the transition rate is expected to fluctuate correspondingly. The transition probability $p_{\rm ge}$ given in Eq.\ \eqref{pge_stochastic} (and the associated transition rate $\Gamma_{\rm ge}$) should therefore be understood as stochastic- and time-averaged quantities.

\section{Quantum System Interacting with Environments}
\label{sec:environments}
\setcounter{equation}{0}
\setcounter{figure}{0}
\setcounter{footnote}{1}

In actual experimental situations, qubits are not fully isolated but inevitably interact with their ``environments.'' Two important examples of such interactions are the measurement process and decoherence (i.e., noise effects). In general, interaction with the environment causes a non-unitary evolution of the qubit system. Such processes are universally described by a so-called completely positive, trace-preserving (CPTP) map acting on the density matrix. In this section, we introduce such maps and, with a particular focus on Markovian noise, derive the Lindblad equation, which plays an important role in describing the time evolution of open qubit systems. 

\subsection{Operator-Sum Representation}

Any physical process acting on a density matrix, including not only unitary time evolution, but also non-selective measurements and noise, can be regarded as a CPTP map on the density matrix. Such maps are often referred to as quantum channels. A quantum channel is a linear operation that transforms one density matrix to another while preserving the trace and positivity of the density matrix. 

The action of a CPTP map can be represented in the so-called operator-sum (or Kraus) representation \cite{KrausBook1983}:
\begin{align}
  \rho \rightarrow \rho' = \mathcal{E}[\rho] \equiv \sum_a K_a \rho K_a^\dagger,
  \label{operatorsum}
\end{align}
where $\rho$ and $\rho'$ denote the density matrices before and after the process, and $\{K_a\}$ are called Kraus operators (or measurement operators). These operators encode all the effects of the given physical process, including interactions with an environment, decoherence, or measurement.

For $\mathcal{E}$ to be a physical quantum channel, the Kraus operators must satisfy the completeness relation:
\begin{align}
  \sum_a K_a^\dagger K_a = \mathds{I}.
  \label{KK=1}
\end{align}
where $\mathds{I}$ denotes the identity operator. This condition guarantees that the total probability is preserved (i.e., the trace of $\rho$ remains one after the process).

When noise originates from an interaction with an environment, the operator-sum representation can be derived by considering the joint evolution of the system and the environment and then tracing out the environmental degrees of freedom. Let $\{\ket{e_a}\}$ be an orthonormal basis of the environment, and assume that the environment is initially prepared in the pure state $\rho_{\rm env}=\ketbra{e_0}{e_0}$. Then the initial state of the combined system is $\rho\otimes \ketbra{e_0}{e_0}$. The total time evolution is governed by a unitary operator $S=e^{-iH_{\rm tot}t}$, where $H_{\rm tot}$ is the Hamiltonian of the joint system. Accordingly, the total density matrix evolves as $\rho_{\rm tot}(t)=S\rho_{\rm tot}(0)S^\dagger$. The reduced density matrix of interest is obtained by tracing out the environment, yielding
\begin{align}
  \mathcal{E}[\rho] = \sum_a \mel{e_a}{\big[ S ( \rho \otimes \ketbra{e_0}{e_0} ) S^\dagger \big]}{e_a} = \sum_a K_a \rho K_a^\dagger,
\end{align}
where the Kraus operators are given explicitly by $K_a=\mel{e_a}{S}{e_0}$.

If, in addition, the environment is measured at the end of the process (i.e., a selective measurement is performed) and the outcome corresponding to the state $\ket{e_a}$ is obtained (for example, via a projective measurement), then the system state is updated as
\begin{align}
  \rho \rightarrow
  \frac{K_a \rho K_a^\dagger}{\mbox{tr}(K_a \rho K_a^\dagger)}.
\end{align}
The denominator provides the normalization required to ensure unit trace. Note also that the probability of obtaining the outcome associated with $\ket{e_a}$ is given by $\mbox{tr}(K_a \rho K_a^\dagger)$.

It is important to note that the choice of Kraus operators for a given physical process $\mathcal{E}$ is not unique. Specifically, suppose $\{K_a\}$ is one set of Kraus operators representing $\mathcal{E}$. For any unitary matrix $u$ acting on the index $a$, we can define a new set
\begin{align}
  K'_a \equiv \sum_b u_{ab} K_b,
  \label{KrausUnitaryTr}
\end{align}
where $u_{ab}$ denotes the $(a, b)$ entry of the unitary matrix $u$. The new set $\{K'_a\}$ also satisfies the completeness relation and yields the same map:
\begin{align}
  \sum_a K_a \rho K_a^\dagger = \sum_a K'_a \rho K'_a{}^\dagger.
\end{align}
This demonstrates that different sets of Kraus operators, related by unitary mixing, represent the same physical quantum channel. Consequently, Kraus representations are defined only up to such unitary transformations. This unitary freedom is often useful, for example, when choosing a Kraus representation that makes the physical interpretation and analytical calculations more transparent.

\subsection{Lindblad Equation}

We now consider the case where the evolution of the density matrix is local in time, i.e., the state at time $t+\Delta t$ depends only on the state at time $t$ (i.e., Markovian dynamics). In such a case, we can derive a differential equation, known as the Lindblad master equation (also known as the Gorini-Kossakowski-Sudarshan-Lindblad, or GKSL equation \cite{Lindblad:1975ef, Gorini:1975nb}), which governs the evolution of the density matrix under the influence of decoherence and noise. Hereafter, we derive the Lindblad equation. (The derivation of the Lindblad equation here follows the argument given in Ref.\ \cite{PreskillQCNotes}.)

For Markovian dynamics, the operator-sum representation in Eq.~\eqref{operatorsum} can be written as
\begin{align}
  \rho(t+\Delta t)
  = \sum_a K_a(t,\Delta t)\,\rho(t)\,K_a^\dagger(t,\Delta t).
  \label{rho(t+dt)}
\end{align}
Assuming that the evolution is smooth for small $\Delta t$, we expand each Kraus operator as a series in powers of $\sqrt{\Delta t}$:
\begin{align}
  K_a(t,\Delta t)
  = K_a^{(0)}(t) + K_a^{(1/2)}(t)\sqrt{\Delta t} + K_a^{(1)}(t)\Delta t + \cdots .
  \label{kraus_expand}
\end{align}
As the environmental influence must vanish in the limit $\Delta t\to 0$, the channel should approach the identity map. In other words, at zeroth order the Kraus operators must be proportional to the identity operator:
\begin{align}
  K_a^{(0)}(t) \propto \mathds{I}.
\end{align}
By using the unitary freedom in the operator-sum representation, we adopt the following Kraus operator basis:
\begin{align}
  K_0 (t, \Delta t) &= \mathds{I} + K_0^{(1)} (t) \Delta t + \cdots, \\
  K_{a \geq 1} (t, \Delta t) &= L_a (t) \sqrt{\Delta t} + \cdots.
\end{align}
Here, $K_0^{(1/2)} = 0$ is enforced to satisfy normalization and smooth time evolution, and the leading $\sqrt{\Delta t}$ terms for $a \geq 1$ are denoted by $L_a (t)$, called the Lindblad operators.

The trace-preserving condition for the map, given by Eq.~\eqref{KK=1}, imposes further constraints. It requires that the Hermitian part of $K_0^{(1)}$ is determined by the set of $L_a$. Denoting the anti-Hermitian part of $K_0^{(1)}$ as $-iH$ (with $H$ a Hermitian operator), we have
\begin{align}
  K_0^{(1)} = -i H - \frac{1}{2} \sum_a L_a^\dagger L_a.
\end{align}

Substituting these expansions into Eq.~\eqref{rho(t+dt)}, we obtain the Lindblad master equation:
\begin{align}
  \partial_t \rho = \mathcal{L} [\rho] \equiv
  -i [H, \rho] + \sum_a \mathcal{D}_{L_a} [\rho],
  \label{LindbladEq}
\end{align}
where the dissipator $\mathcal{D}_{L}$ acting on $\rho$ is defined as
\begin{align}
  \mathcal{D}_L [\rho] \equiv L \rho L^\dagger -
  \frac{1}{2} \left( L^\dagger L \rho + \rho L^\dagger L \right).
\end{align}
The complete time-evolution generator $\mathcal{L}$ is called the Lindblad superoperator (or Lindbladian). The Hermitian operator $H$ generates the unitary part of the dynamics (and hence $H$ corresponds to the system Hamiltonian), while the set of Lindblad operators $\{L_a\}$ characterizes the irreversible (dissipative and decohering) effects induced by the environment.

As we will discuss later, a wide class of Markovian noise processes can be modeled by Eq.~\eqref{LindbladEq} with an appropriate choice of Lindblad operators $L_a$.

\section{Qubit Evolution with Noise}
\label{sec:qubitwithnoise}
\setcounter{equation}{0}
\setcounter{figure}{0}
\setcounter{footnote}{1}

In this section, we analyze a simple qubit system subject to noise to understand the basic dynamical properties of open qubit systems. We use the Lindblad equation introduced in the previous section, assuming that the noise is Markovian.

\subsection{Simple Model with Amplitude Damping and Dephasing}

An important application of the Lindblad equation in DM-detection experiments (and, more generally, in qubit-based quantum sensing) is to describe the time evolution of a qubit in the presence of environmental noise. In this subsection, we introduce a minimal framework that incorporates both amplitude damping and dephasing.

We consider a qubit whose dynamics are governed only by the free Hamiltonian
\begin{align}
  H = \omega \ketbra{1}{1}.
\end{align}
We then study the Lindblad equation
\begin{align}
  \partial_t \rho = &\, -i [H, \rho]
  + \gamma_1 \mathcal{D}_a [\rho]
  + \frac{1}{2} \gamma_2 \mathcal{D}_Z [\rho],
\end{align}
where the dissipators are constructed from $a=\ketbra{0}{1}$ and $Z$. Here, $\gamma_1$ and $\gamma_2$ are positive constants. As discussed below, $\mathcal{D}_a$ describes amplitude damping, with the corresponding energy relaxation time $T_1$ determined by $\gamma_1$. Moreover, both $\mathcal{D}_a$ and $\mathcal{D}_Z$ contribute to dephasing, i.e., to the decay of the off-diagonal coherence of the density matrix.

We parameterize the qubit density matrix as in Eq.\ \eqref{rho_IJ}:
\begin{align}
  \rho \equiv
  \sum_{I,J=0,1}
  \rho_{IJ} (t) \ketbra{I}{J}.
\end{align}
The evolution equations for the coefficient functions are then given by
\begin{align}
  \partial_t \rho_{00} = &\, \gamma_1 \rho_{11},
  \\
  \partial_t \rho_{11} = &\, - \gamma_1 \rho_{11},
  \\
  \partial_t \rho_{01} = &\,
  \left( i\omega - \frac{1}{2} \gamma_1 - \gamma_2 \right) \rho_{01}.
\end{align}
Note that $\rho_{10}=\rho_{01}^*$. These equations can be solved, yielding
\begin{align}
  \rho_{00} (t) = &\, 1 - \rho_{11} (0) e^{-\gamma_1 t},
  \\
  \rho_{11} (t) = &\, \rho_{11} (0) e^{-\gamma_1 t},
  \\
  \rho_{01} (t) = &\, \rho_{01} (0)
  e^{\left( i\omega - \frac{1}{2} \gamma_1 - \gamma_2 \right)t},
\end{align}
where we have used the relation $\rho_{00} (0)+\rho_{11} (0)=1$.

From these damping behaviors, we find that $T_1$ and $T_2$ are related to $\gamma_1$ and $\gamma_2$ as
\begin{align}
  T_1 = &\, \frac{1}{\gamma_1},
  \\
  T_2 = &\, \left( \frac{1}{2} \gamma_1 + \gamma_2 \right)^{-1}.
\end{align}
Thus, within this simple model, one obtains
\begin{align}
  T_2 \leq 2 T_1.
\end{align}

\subsection{Excitation of a Qubit with Noise and Decoherence}
\label{subsec:qubitwithnoise}

We now consider the excitation dynamics of a qubit while systematically incorporating amplitude damping, dephasing, and spontaneous excitation (or dark count). This framework is particularly useful for gaining intuition about DM searches that rely on the direct excitation of qubits.

We consider a single qubit coupled to an external AC field, which may, for example, originate from interactions with DM. The excitation dynamics are governed by the Hamiltonian
\begin{align}
  H = \omega \ketbra{1}{1}
  - 2 \eta \cos(m t-\varphi)
  \left( \ketbra{0}{1} + \ketbra{1}{0} \right).
\end{align}
In DM search experiments, the drive strength $\eta$ is expected to be small, and the primary experimental objective is to establish a nonzero value of $\eta$.

To model the influence of noise and decoherence, we use the Lindblad equation
\begin{align}
  \partial_t \rho = &\, -i [H, \rho]
  + \gamma_0 \mathcal{D}_{a^\dagger} [\rho]
  + \gamma_1 \mathcal{D}_a [\rho]
  + \frac{1}{2} \gamma_2 \mathcal{D}_Z [\rho].
\end{align}
Relative to the model in the previous subsection, we have added the term $\mathcal{D}_{a^\dagger}$, which accounts for spontaneous excitation processes (for example, those induced by thermal noise).

The evolution equations for the components $\rho_{IJ}$, derived from the Lindblad equation above, are
\begin{align}
  \partial_t \rho_{00} = &\,
  \gamma_1 \rho_{11} - \gamma_0 \rho_{00}
  + 2 i \eta \left( \rho_{10} - \rho_{01} \right)
  \cos (m t - \varphi),
  \label{dot(rho00)}
  \\
  \partial_t \rho_{11} = &\,
  - \gamma_1 \rho_{11} + \gamma_0 \rho_{00}
  + 2 i \eta \left( - \rho_{10} + \rho_{01} \right)
  \cos (m t - \varphi),
  \label{dot(rho11)}
  \\
  \partial_t \rho_{01} = &\,
  \left( i\omega - \bar{\gamma} - \gamma_2 \right) \rho_{01}
  + 2 i \eta \left( \rho_{11} - \rho_{00} \right)
  \cos (m t - \varphi),
  \label{dot(rho01)}
\end{align}
where we have defined
\begin{align}
  \bar{\gamma} \equiv \frac{1}{2} \left( \gamma_1 + \gamma_0 \right).
\end{align}

In DM detection experiments, the near-resonant case $\omega\sim m$ is of particular importance because, as discussed in Section~\ref{sec:evolution}, the signal can be resonantly enhanced. In what follows, we focus on the regime $|\omega-m|\ll \omega+m$, where the qubit dynamics can be efficiently analyzed using the RWA.

To implement the RWA, it is convenient to define
\begin{align}
  \rho'_{01} \equiv \rho_{01} e^{-i(\omega t-\varphi)},~~~
  \rho'_{10} \equiv \rho_{10} e^{i(\omega t-\varphi)}.
\end{align}
By neglecting rapidly oscillating terms (i.e., terms oscillating with frequency $O(\omega)$ or $O(m)$), the equations reduce to
\begin{align}
  \partial_t \rho_{00} = &\,
  \gamma_1 \rho_{11} - \gamma_0 \rho_{00}
  + i \eta \left( e^{-i \Delta_\omega t} \rho'_{10} - e^{i \Delta_\omega t} \rho'_{01} \right),
  \\
  \partial_t \rho_{11} = &\,
  - \gamma_1 \rho_{11} + \gamma_0 \rho_{00}
  + i \eta \left( - e^{-i \Delta_\omega t} \rho'_{10} + e^{i \Delta_\omega t} \rho'_{01} \right),
  \\
  \partial_t \rho'_{01} = &\,
  -\left( \bar{\gamma} + \gamma_2 \right) \rho'_{01}
  + i \eta e^{-i \Delta_\omega t}
  \left( \rho_{11} - \rho_{00} \right),
\end{align}
where $\Delta_\omega$ denotes the detuning,
\begin{align}
  \Delta_\omega \equiv \omega - m.
\end{align}
From the damping behaviors of $\rho_{11}$ and $\rho'_{01}$ implied by the equations above, the coherence properties of the qubit are characterized by
\begin{align}
  T_1 = &\, \frac{1}{\gamma_1},
  \label{T1}
  \\
  T_2 = &\, \frac{1}{\gamma_2+\bar{\gamma}}.
  \label{T2}
\end{align}
In addition, for later convenience, we define the time scale of the dark count as
\begin{align}
  T_0 \equiv \frac{1}{\gamma_0}.
  \label{T0}
\end{align}

Next, we study the evolution relevant to DM excitation experiments, focusing on the case where the qubit is initially prepared in its ground state. The key observable is the excitation probability from $\ket{0}$ to $\ket{1}$, which is often used as a DM signal. The corresponding initial condition is
\begin{align}
  \rho_{00} (0) = 1,~~~
  \rho_{11} (0) = \rho_{01} (0) = 0.
\end{align}
Then, the transition probability from the ground state to the excited state is given by
\begin{align}
  p_{\rm ge} (t) = \rho_{11} (t).
\end{align}

Assuming $\eta$ is sufficiently small to justify a perturbative treatment, we expand
\begin{align}
  \rho_{IJ} = \sum_{p=0}^\infty \rho_{IJ}^{(p)},
\end{align}
where $\rho_{IJ}^{(p)}\sim O(\eta^p)$. A straightforward calculation gives
\begin{align}
  \rho_{00}^{(0)} = &\,
  \frac{1}{2\bar{\gamma}} \left( \gamma_1 + \gamma_0 e^{-2\bar{\gamma}t} \right),
  \\
  \rho_{11}^{(0)} = &\,
  \frac{\gamma_0}{2\bar{\gamma}} \left( 1 - e^{-2\bar{\gamma}t} \right),
  \\
  \rho_{01}^{(1)} = &\, i
  \frac{\eta}{2\bar{\gamma}} \left[
    (\gamma_0-\gamma_1) I_{\bar{\gamma}+\gamma_2-i\Delta_\omega} (t)
    - 2\gamma_0 I_{-\bar{\gamma}+\gamma_2-i\Delta_\omega} (t)
    \right] e^{(i\omega -\bar{\gamma}-\gamma_2)t-i\varphi},
  \label{p01(1)}
\end{align}
and
\begin{align}
  \rho_{11}^{(2)} = &\,
  \frac{\eta^2}{2\bar{\gamma}}
  \left[
    \frac{\gamma_1-\gamma_0}{\bar{\gamma}+\gamma_2-i\Delta_\omega}
    \left\{ I_{2\bar{\gamma}}(t)
    - I_{\bar{\gamma}-\gamma_2+i\Delta_\omega} (t) \right\}
    -
    \frac{2\gamma_0}{\bar{\gamma}-\gamma_2+i\Delta_\omega}
    \left\{ t - I_{\bar{\gamma}-\gamma_2+i\Delta_\omega} (t) \right\}
    \right] e^{-2\bar{\gamma} t}
  \nonumber \\[2mm] &\,
  + \mbox{h.c.},
  \label{rho11(2)}
\end{align}
where
\begin{align}
  I_\Gamma (t) \equiv \frac{1}{\Gamma} (e^{\Gamma t} - 1).
\end{align}

The asymptotic behavior of the excitation probability is of particular interest. In the large-$t$ limit, one finds
\begin{align}
  p_{\rm ge} (t\rightarrow\infty) = \rho_{11} (t \gg \bar{\gamma}^{-1}) \simeq
  \frac{\gamma_0}{\gamma_1+\gamma_0}
  +
  \eta^2
  \frac{(\gamma_1-\gamma_0)(\gamma_2+\bar{\gamma})}
       {2\bar{\gamma}^2[(\gamma_2+\bar{\gamma})^2+(\omega-m)^2]}.
       \label{Pge(t=inf)}
\end{align}
The first term corresponds to the excitation probability in the absence of the external AC field. This contribution can be interpreted as the dark-count probability, which we define as
\begin{align}
  p_{\rm ge}^{\rm (dark\mathhyphen count)} \equiv
  \frac{\gamma_0}{\gamma_1+\gamma_0}.
  \label{pnoise}
\end{align}
The excitation probability $p_{\rm ge} (t\rightarrow\infty)$ is maximized under the resonance condition $\omega = m$ and is strongly suppressed for large detuning, $|\omega-m|\gg T_2^{-1}$. 

In typical experimental situations, we have $\gamma_0 \ll \gamma_1,\,\gamma_2$. In this case, the density matrix at $t\gg\bar{\gamma}^{-1}$ becomes
\begin{align}
  \rho_{00} (t \gg \bar{\gamma}^{-1}) \simeq &\, 1 - \frac{T_1}{T_0} - \frac{2 \eta^2 T_1 T_2}{1+(\omega-m)^2 T_2^2},
  \label{rho00(simple)}
  \\
  \rho_{11} (t \gg \bar{\gamma}^{-1}) \simeq &\, \frac{T_1}{T_0} + \frac{2 \eta^2 T_1 T_2}{1+(\omega-m)^2 T_2^2},
  \label{rho11(simple)}
  \\
  \rho_{01} (t \gg \bar{\gamma}^{-1}) \simeq &\, 
  \frac{-i \eta T_2}{1 - i (\omega-m) T_2} e^{i(m t - \varphi)},
  \label{rho01(simple)}
\end{align}
and $p_{\rm ge} (t\rightarrow\infty)$ simplifies to
\begin{align}
  \left. p_{\rm ge} (t\rightarrow\infty) \right|_{\gamma_0\ll\gamma_1,\, \gamma_2}
  \simeq
  p_{\rm ge}^{\rm (dark\mathhyphen count)} + \frac{2 \eta^2 T_1 T_2}{1+(\omega-m)^2 T_2^2}.
  \label{Pge(small-gamma0)}
\end{align}
In this regime, the peak height of the excitation probability (after subtracting the dark-count contribution) is $2 \eta^2 T_1 T_2$. Notably, the qubit remains sensitive to the AC field even away from exact resonance, with the excitation probability being appreciable over the range $m-T_2^{-1} \lesssim \omega \lesssim m+T_2^{-1}$.

\begin{figure}[t]
  \centering
  \includegraphics[width=0.5\textwidth]{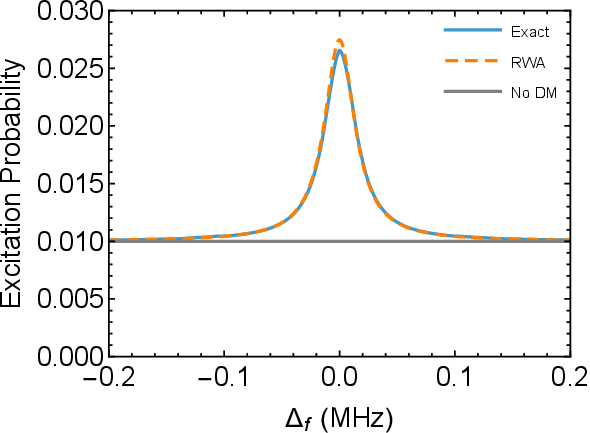}
  \caption{$p_{\rm ge} (t\rightarrow\infty)$ as a function of $\Delta_f \equiv \frac{1}{2\pi} \Delta_\omega$, with $T_1 = 100\ \mu{\rm sec}$, $T_2 = 10\ \mu{\rm sec}$, $p_{\rm ge}^{\rm (dark\mathhyphen count)} = 0.01$, and $\eta = 3\ \mu{\rm sec}^{-1}$. The blue solid and magenta dashed lines represent the exact result and the analytic result obtained from the RWA combined with perturbation theory, respectively. For comparison, the case without the DM contribution ($\eta=0$) is also shown by the gray solid line.}
  \label{fig:probge}
\end{figure}

Fig.\ \ref{fig:probge} shows the excitation probability in the large-$t$ limit as a function of the detuning, for $T_1 = 100\ \mu{\rm sec}$, $T_2 = 10\ \mu{\rm sec}$, $p_{\rm ge}^{\rm (dark\mathhyphen count)} = 0.01$, and $\eta=3\ \mu{\rm sec}^{-1}$. The dissipation rates $\gamma_0$, $\gamma_1$, and $\gamma_2$ are determined from Eqs.\ \eqref{T1}, \eqref{T2}, and \eqref{pnoise}. To obtain the theoretical prediction, we numerically solve the full evolution equations, Eqs.\ \eqref{dot(rho00)} -- \eqref{dot(rho01)}, and compare the results with the analytic approximation in Eq.\ \eqref{Pge(t=inf)}. As shown in the figure, Eq.\ \eqref{Pge(t=inf)} provides a good approximation to the transition probability; however, it should be stressed that this expression is valid only in the perturbative regime $\eta\ll\gamma_1,\,\gamma_2$.

\begin{figure}[t]
  \centering
  \includegraphics[width=0.5\textwidth]{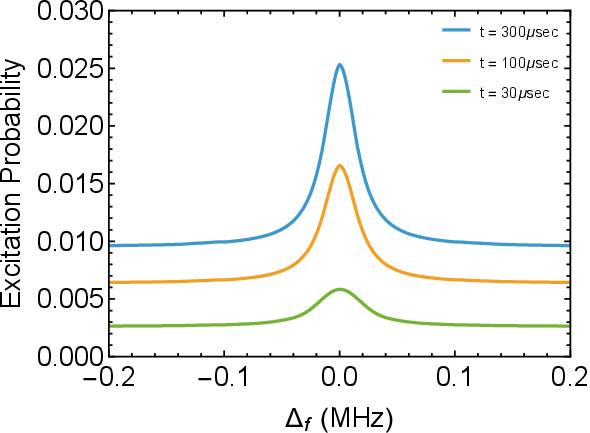}
  \caption{$p_{\rm ge} (t)$ for several values of $t$ as a function of $\Delta_f \equiv \frac{1}{2\pi} \Delta_\omega$, with $T_1 = 100\ \mu{\rm sec}$, $T_2 = 10\ \mu{\rm sec}$, $p_{\rm ge}^{\rm (dark\mathhyphen count)} = 0.01$, and $\eta = 3\ \mu{\rm sec}^{-1}$.}
  \label{fig:probgetdep}
\end{figure}

In Fig.\ \ref{fig:probgetdep}, we also plot $p_{\rm ge}(t)$ for several values of the exposure time $t$, using $T_1 = 100\ \mu{\rm sec}$, $T_2 = 10\ \mu{\rm sec}$, $p_{\rm ge}^{\rm (dark\mathhyphen count)} = 0.01$, and $\eta = 3\ \mu{\rm sec}^{-1}$. With this parameter set, we find that, for $t\gtrsim T_2$, the peak width of the excitation probability around $\omega=m$ is (almost) constant, whereas both the signal and dark-count probabilities increase with time, up to $t\sim T_1$.

It is also instructive to consider a more general situation in which the interaction term involves a superposition of different velocity modes, as discussed in Section~\ref{sec:evolution}. To illustrate this case, we take the interaction Hamiltonian in Eq.\ \eqref{Hamiltonian_1'} in the form
\begin{align}
  H = \omega \ketbra{1}{1}
  -2 \eta \left(\ketbra{0}{1}+\ketbra{1}{0}\right)
  \sum_{\vec{v}}
  f_{\vec{v}} \cos \left( E_v t - \varphi_{\vec{v}} \right),
\end{align}
where $E_v \simeq m + \frac{1}{2} m v^2$. Assuming that the phases $\varphi_{\vec{v}}$ are independent random phases, the transition probability from the ground state to the excited state can be expressed as the velocity-averaged counterpart of $\rho_{11}^{(2)}$ in Eq.\ \eqref{rho11(2)}, with the substitution $\Delta_\omega \rightarrow \omega - E_v$. It is convenient to reparameterize the qubit angular frequency as $\omega = m + \frac{1}{2} m v_\omega^2$, and we focus on the case $v_\omega \sim O(v_{\rm DM})$.

In the large-$t$ limit, the expression simplifies; in particular, for $\gamma_0 \ll \gamma_1$, we obtain
\begin{align}
  p_{\rm ge} (t\rightarrow\infty) \simeq
  p_{\rm ge}^{\rm (dark\mathhyphen count)}
  +
  \frac{8\eta^2 T_1}{m^2 T_2}
  \int dv \mathcal{F}(v)
  \frac{1}{(v^2-v_\omega^2)^2 + (4 / m^2 T_2^2)}.
\end{align}
In the regime $T_2 \ll 1/m v_{\rm DM}^2$, this reproduces Eq.\ \eqref{Pge(small-gamma0)}. By contrast, for $T_2 \gg 1/m v_{\rm DM}^2$, the behavior becomes more intricate; in this case, the integral is dominated by velocities near $v\sim v_\omega$, yielding
\begin{align}
  p_{\rm ge} (t\rightarrow\infty) \simeq
  p_{\rm ge}^{\rm (dark\mathhyphen count)}
  +
  \eta^2 T_1 \frac{2 \pi \mathcal{F}(v_\omega)}{m v_\omega}.
\end{align}
This result highlights the necessity of taking into account the DM coherence; in this regime, the signal depends sensitively on the coherence properties of the DM (see Eq.~\eqref{T_DM(v)}).

Before closing this section, we comment on the choice of observable for DM detection. So far, we have proposed to measure the number of excited qubits as the DM signal. This corresponds to a projective measurement onto the eigenstates of the Pauli-$Z$ operator. In principle, one may instead choose a different observable, for example, the Pauli-$Y$ (or Pauli-$X$) operator. The validity of using the Pauli-$Z$ measurement for DM detection will be discussed in Section~\ref{subsec:zvsy}.

\section{Qubit in a Cavity}
\label{sec:cavity}
\setcounter{equation}{0}
\setcounter{figure}{0}
\setcounter{footnote}{1}

In studying the evolution of a qubit, it is necessary to consider the effect of the conductive materials surrounding the qubit, i.e., the cavity, because qubits (in particular, superconducting qubits) are usually placed inside a cavity. The cavity imposes specific boundary conditions on the EM field inside it. Consequently, in order to understand the EM field acting on the qubit installed inside the cavity, the EM field must be evaluated by properly taking into account these boundary conditions, which may significantly affect the strength of the field. In this section, the properties of the EM field inside the cavity are discussed in connection with its interaction with a qubit placed therein. More detailed discussion can be found in, e.g., Ref.\ \cite{Blais:2020wjs}.

\subsection{EM Field in a Cavity}

Before discussing the interaction between a qubit and the EM field inside a cavity, we first clarify the behavior of the EM field itself within the cavity. Specifically, we consider the excitation of the EM field by an oscillating source. Here, we work in the Coulomb gauge:
\begin{align}
  \vec{\nabla}  \vec{A} = 0.
\end{align}

We consider the case in which the EM field inside a cavity is described by the Lagrangian
\begin{align}
  L = \int d^3 x
  \left[ -\frac{1}{4} F_{\mu\nu} F^{\mu\nu} + S(t)\, \vec{n}_S  \vec{A} \right],
  \label{L_cavity}
\end{align}
where
\begin{align}
  S(t) \equiv \bar{S} \sin(m t - \varphi),
\end{align}
with $\bar{S}$ a constant amplitude, $F_{\mu\nu} \equiv \partial_\mu A_\nu - \partial_\nu A_\mu$ the electromagnetic field-strength tensor, and $\vec{n}_S$ a fixed unit vector. The second term, proportional to $S(t)$, acts as an effective source term for the electromagnetic field within the cavity; such a source may arise from a wave-like DM background (or from an externally applied microwave drive field).\footnote
{The source term in Eq.\ \eqref{L_cavity} should be understood as an effective description after fixing the Coulomb gauge (and imposing the cavity boundary conditions), capturing the electromagnetic response relevant for our purpose.}

As a concrete example, for axion DM this source term is induced in the presence of an external magnetic field. Let us assume a uniform and static external field $\vec{B}^{\rm (ext)}=B^{\rm (ext)}\vec{n}_B$, where $B^{\rm (ext)}$ is a constant magnitude and $\vec{n}_B$ is a unit vector specifying its direction. Then, using Eq.\ \eqref{L_agammagamma}, the parameter $\bar{S}$ is evaluated as
\begin{align}
  \bar{S}^{\rm (axion)} = g_{a\gamma\gamma} m \bar{a} B^{\rm (ext)}
  = g_{a\gamma\gamma} B^{\rm (ext)} \sqrt{2\rho_{\rm DM}},
  \label{Sbar(axion)}
\end{align}
where $\bar{a}$ denotes the axion field amplitude while $\rho_{\rm DM}$ is the DM energy density. In this case, $\vec{n}_S$ coincides with $\vec{n}_B$; therefore, once the direction of the external magnetic field is specified, $\vec{n}_S$ is fixed and known.

Another representative example is dark photon DM. In the dark photon case, the effective source term arises directly from kinetic mixing with the ordinary photon. From the Lagrangian in Eq.\ \eqref{L_darkphoton}, the $\bar{S}$ parameter for the dark photon is given by
\begin{align}
  \bar{S}^{\rm (dark\mathhyphen{photon})} = \epsilon m^2 \bar{X} 
  = \epsilon m \sqrt{2\rho_{\rm DM}},
  \label{Sbar(darkphoton)}
\end{align}
where $\bar{X}$ is the dark photon field amplitude. In this case, $\vec{n}_S$ is aligned with the polarization direction of the dark photon field; accordingly, it is not fixed in time but varies stochastically, becoming randomized on the time scale set by the DM coherence time $\tau_{\rm DM}$.

In the Coulomb gauge, since the source couples only to spatial components of the vector potential, we can ignore the scalar component $A^0$. The electric and magnetic fields are expressed as
\begin{align}
  \vec{E} = &\, -\partial_t \vec{A},
  \\
  \vec{B} = &\, \vec{\nabla} \times \vec{A}.
\end{align}
From the Lagrangian given in Eq.\ \eqref{L_cavity}, the equation of motion for the EM field is derived as
\begin{align}
  \Box \vec{A} - S(t)\, \vec{n}_S = 0.
\end{align}

In the vacuum (with infinite volume), the above equation can be easily solved. In particular, we can find a solution without spatial dependence as
\begin{align}
  \vec{A}^{\rm (vac)} = - \frac{\bar{S}}{m^2} \, \vec{n}_S \sin(m t - \varphi).
  \label{A(vac)}
\end{align}
Then the electric field is evaluated as
\begin{align}
  \vec{E}^{\rm (vac)} = \frac{\bar{S}}{m} \, \vec{n}_S \cos(m t - \varphi).
  \label{E(vac)}
\end{align}
In discussing the effect of DM, it can be interpreted as the electric field induced by DM in free space. 

Now, we move on to discuss the EM field in the cavity. In this context, the spatial integration in Eq.\ \eqref{L_cavity} should be understood as being performed over the interior of the cavity. Inside the cavity, we expand the vector potential by using the mode functions $\vec{\tilde{A}}_n(\vec{x})$ as
\begin{align}
  \vec{A}(t, \vec{x}) = \sum_n q_n(t) \, \vec{\tilde{A}}_n(\vec{x}).
  \label{expansionA}
\end{align}
We normalize the mode functions according to
\begin{align}
  \int d^3 x\, \vec{\tilde{A}}_m(\vec{x})  \vec{\tilde{A}}_n(\vec{x}) = \delta_{m,n}.
\end{align}
The mode functions satisfy
\begin{align}
  \nabla^2 \vec{\tilde{A}}_n = -\omega_n^2 \vec{\tilde{A}}_n, \qquad
  \vec{\nabla}  \vec{\tilde{A}}_n = 0,
\end{align}
with $\omega_n$ the cavity eigenfrequencies.

Assuming that the cavity walls have sufficiently high conductivity, the boundary conditions require that the parallel component of the electric field and the perpendicular component of the magnetic field both vanish at the wall. This leads to the following boundary conditions at the surface of the cavity wall:
\begin{align}
  \left. \vec{\tilde{A}}_n \right|_\parallel = 0, ~~~
  \left. (\vec{\nabla} \times \vec{\tilde{A}}_n) \right|_\perp = 0,
\end{align}
for the chosen gauge, where the subscripts ``$\parallel$'' and ``$\perp$'' denote components tangential and normal to the cavity wall, respectively. 

Substituting the expansion \eqref{expansionA} into the Lagrangian, we find
\begin{align}
  L = \sum_n \left[
    \frac{1}{2} \left( \dot{q}_n^2 - \omega_n^2 q_n^2 \right)
    + S(t)\, I_n\, q_n
    \right],
\end{align}
where the ``dot'' denotes the derivative with respect to time, and
\begin{align}
  I_n \equiv \int d^3 x\, \vec{n}_S  \vec{\tilde{A}}_n(\vec{x}).
\end{align}

The Hamiltonian can be derived by identifying $q_n$ as canonical variables. Then, the canonical momentum is given by
\begin{align}
  p_n \equiv \frac{\partial L}{\partial \dot{q}_n} = \dot{q}_n,
\end{align}
and the Hamiltonian reads
\begin{align}
  H = \sum_n p_n \dot{q}_n - L
    = \sum_n \left[
      \frac{1}{2} (p_n^2 + \omega_n^2 q_n^2)
      - S(t)\, I_n\, q_n
    \right].
\end{align}

To study the EM field in the cavity, one may adopt either a classical approach (i.e., solving the classical equation of motion, as in the derivation of Eq.\ \eqref{A(vac)}) or a quantum approach. Here, we take the quantum approach, which allows us to use the Lindblad equation to take into account the effect of the dissipation of the EM field. We note here that, in the non-resonant case where $m$ is sufficiently far from any of the cavity frequencies and the effect of dissipation is unimportant, the classical and quantum calculations yield the same EM-field profile inside the cavity. 

To quantize the system, we promote $q_n$ and $p_n$ to operators and impose the canonical commutation relations:
\begin{align}
  [q_m, q_n] = [p_m, p_n] = 0, \qquad
  [q_m, p_n] = i \delta_{m,n}.
\end{align}
For convenience, we introduce annihilation and creation operators,
\begin{align}
  c_n \equiv \frac{1}{\sqrt{2\omega_n}} ( \omega_n q_n + i p_n ),~~~
  c_n^\dagger \equiv \frac{1}{\sqrt{2\omega_n}} ( \omega_n q_n - i p_n ),
\end{align}
which satisfy
\begin{align}
  [c_n, c_m^\dagger] = \delta_{m,n}.
\end{align}

With these, the field operator can be written as
\begin{align}
  \vec{A} = \sum_n \frac{1}{\sqrt{2\omega_n}} ( c_n + c_n^\dagger )
  \vec{\tilde{A}}_n,
\end{align}
and the Hamiltonian becomes
\begin{align}
  H = \sum_n H_n,
\end{align}
where
\begin{align}
  H_n = \omega_n c_n^\dagger c_n - 2\xi_n (c_n + c_n^\dagger) \sin(m t - \varphi),
\end{align}
with
\begin{align}
  \xi_n \equiv \frac{\bar{S} I_n}{2 \sqrt{2\omega_n}}.
\end{align}

With the Hamiltonian given above, we can analyze the cavity EM field. For this purpose, we introduce $\rho$, the density matrix for the field, assumed to obey the following Lindblad equation
\begin{align}
  \partial_t \rho = -i [H, \rho] + \sum_n \gamma_n \mathcal{D}_{c_n} [\rho],
\end{align}
where $\gamma_n$ are non-negative constants. As we see in the following, the term proportional to $\mathcal{D}_{c_n} [\rho]$ describes the photon loss in mode $n$. 

Because there are no mode-mode interactions, the density matrix factorizes as
\begin{align}
  \rho = \bigotimes_n \rho_n,
\end{align}
and each $\rho_n$ evolves according to
\begin{align}
  \partial_t \rho_n = -i [H_n, \rho_n] + \gamma_n \mathcal{D}_{c_n} [\rho_n].
  \label{dotrho_n(cavity)}
\end{align}

Now, let us derive the density matrix for the case with the source term. Suppose that the state remains in a coherent state at all times, with a time-dependent amplitude $\alpha_n(t)$:\footnote
{
Formally, the density matrix $\rho_n$ can be written as a superposition of the operators $\ketbra{\alpha}{\alpha}$ (see Appendix \ref{app:coherent}). From the discussion below, it follows that, if the initial density matrix is given by
\begin{align*}
  \rho_n (0) = \int d^2 \alpha P_n (\alpha) \ketbra{\alpha}{\alpha},
\end{align*}
then the time-evolved density matrix is
\begin{align*}
  \rho_n (t) = \int d^2 \alpha P_n (\alpha) \ketbra{\alpha_n(t)}{\alpha_n(t)},
\end{align*}
where $\alpha_n(t)$ is given by Eq.\ \eqref{alphant} with the replacement $\alpha_n(0)\rightarrow\alpha$. In obtaining this result, we used the fact that an initial coherent state remains a coherent state under the dynamics of the present model. Therefore, if $\gamma_n>0$, the density matrix approaches $\ketbra{\bar{\alpha}_n(t)}{\bar{\alpha}_n(t)}$ as $t\rightarrow\infty$, irrespective of the initial condition.
}
\begin{align}
  \rho_n(t) = \ketbra{\alpha_n(t)}{\alpha_n(t)},
\end{align}
where $\ket{\alpha}$ (with $\alpha$ being a complex number) is the normalized coherent state
\begin{align}
  \ket{\alpha} \equiv e^{-|\alpha|^2/2} e^{\alpha c_n^\dagger}\ket{0}_\gamma,
\end{align}
with $\ket{0}_\gamma$ being the cavity vacuum. Properties of the coherent state are summarized in Appendix \ref{app:coherent}.

After a straightforward calculation, using
\begin{align}
  c_n^\dagger \ket{\alpha} =
  \left( \partial_\alpha + \frac{1}{2} \alpha^* \right) \ket{\alpha},
\end{align}
the evolution equation for $\alpha_n(t)$ is found to be
\begin{align}
  \partial_t \alpha_n = \left(-i\omega_n - \frac{1}{2} \gamma_n \right) \alpha_n + 2 i \xi_n \sin(mt - \varphi).
\end{align}
The general solution is
\begin{align}
  \alpha_n(t) = \bar{\alpha}_n(t) + \left[ \alpha_n(0) - \bar{\alpha}_n(0) \right]
  \exp\left[ \left(-i\omega_n - \frac{1}{2} \gamma_n \right) t \right],
  \label{alphant}
\end{align}
where
\begin{align}
  \bar{\alpha}_n(t) = 2\xi_n \frac{(\omega_n - \frac{i}{2} \gamma_n)\sin(mt - \varphi) + i m \cos(mt - \varphi)}{(\omega_n^2 - m^2 - \frac{1}{4}\gamma_n^2) - i \omega_n \gamma_n}.
  \label{alphabar}
\end{align}

With the density matrix given, we can study various cavity quantities. In particular, in the following, we investigate the total energy inside the cavity, which is evaluated as
\begin{align}
  E = \mbox{tr} (\rho H).
\end{align}

First, let us consider the situation in which there is no external source present. In this case, the energy contained in mode $n$ is given by $E_n(t) = \omega_n|\alpha_n(t)|^2$. This quantity evolves in time according to $E_n(t) = e^{-\gamma_n t}E_n(0)$, illustrating an exponential decay as a result of dissipation. Here, $\gamma_n$ can be interpreted as the energy-loss rate; in other words, the coefficient of the dissipative term in Eq.\ \eqref{dotrho_n(cavity)} is normalized such that it corresponds directly to the rate of energy loss. Furthermore, $\gamma_n$ is related to the quality factor of the cavity. The quality factor $Q_n$ is defined as the ratio of the oscillation frequency to the energy-loss rate and can be expressed as
\begin{align}
  Q_n = \frac{\omega_n}{\gamma_n}.
\end{align}

Next, we study the asymptotic behavior of the density matrix. For $\gamma_n\neq 0$, the state of each mode relaxes to a coherent state, i.e.,
\begin{align}
  \rho_n \xrightarrow{t\to\infty}
  \ketbra{\bar{\alpha}_n(t)}{\bar{\alpha}_n(t)}.
  \label{rhotinf}
\end{align}
Accordingly, in the late-time regime it is natural to consider a factorized density matrix of the form
\begin{align}
  \rho = \bigotimes_n \ketbra{\bar{\alpha}_n(t)}{\bar{\alpha}_n(t)}.
\end{align}
For such a state, the total energy becomes
\begin{align}
  E(t)
  =
  \sum_n \left[
    \omega_n \bigl|\bar{\alpha}_n(t)\bigr|^2
    - 2 \xi_n \Bigl\{ \bar{\alpha}_n(t) + \bar{\alpha}_n^*(t) \Bigr\}
    \sin (mt-\varphi)
    \right].
\end{align}

Let us now consider the so-called non-resonant case, characterized by $\gamma_n \ll |\omega_n - m|$. In this regime, the effect of dissipation is negligible compared to the detuning, and thus dissipation can be ignored in the leading-order analysis. In this case, the expectation value of the field operator is given by
\begin{align}
  \ev{\vec{A}(t,\vec{x})} = \mbox{tr} \left[ \rho \vec{A}(t,\vec{x}) \right]
  = \sum_n \frac{\bar{S} I_n}{\omega_n^2-m^2} \vec{\tilde{A}}_n \sin (mt-\varphi).
  \label{<A(t,x)>}
\end{align}
It is worth noting that an equivalent expression for the vector potential is obtained via a direct analysis of the classical equations of motion for the system (without the dissipation). To estimate the energy in the cavity, we define the effective volume for mode $n$ as
\begin{align}
  \mathcal{V}_n \equiv I_n^2 =
  \left[ \int d^3 x\, \vec{n}_S \vec{\tilde{A}}_n(\vec{x}) \right]^2,
\end{align}
which quantifies the spatial overlap of the mode with the external source. Consequently, the total energy can be expressed as
\begin{align}
  E^{\rm (non\mathhyphen{res})} =
  \frac{1}{2}
  \sum_n
  \left( \frac{\bar{S}}{\omega_n^2-m^2} \right)^2
    \left[ m^2 + (m^2 - \omega_n^2) \sin^2 (mt-\varphi) \right]
    \mathcal{V}_n,
\end{align}
which, again, agrees with the classical result.

A particularly important limiting case is the resonance regime, in which the source frequency $m$ nearly matches one of the cavity frequencies $\omega_n$. Specifically, we assume $|\omega_n - m| \sim O(\gamma_n)$, while for all other modes $n'$ ($n' \neq n$), $|\omega_{n'}-m| \gg \gamma_{n'}$. In this situation, the total energy $E$ is dominated by the contribution from mode $n$. We find
\begin{align}
  E^{\rm (res)} =
  \frac{\bar{S}^2}{2\omega_n^2} Q_n^2 \mathcal{V}_n\, L (m, \omega_n, Q_n),
\end{align}
where the function
\begin{align}
  L (m, \omega, Q) \equiv
  \left[ 1 + 4 Q^2 \left( \frac{\omega - m}{\omega} \right)^2 \right]^{-1},
\end{align}
characterizes the resonance enhancement as a function of detuning and the quality factor.

In experimental setups, the energy stored in the cavity is often monitored using an antenna coupled to the cavity. In such a configuration, photon loss arises both from intrinsic dissipation in the cavity and from energy leakage into the antenna (external coupling). Accordingly, the dissipation coefficient can be decomposed as
\begin{align}
  \gamma_n = \gamma_n^{\rm (in)} + \gamma_n^{\rm (ext)},
\end{align}
where $\gamma_n^{\rm (in)}$ and $\gamma_n^{\rm (ext)}$ denote the internal dissipation rate and the loss rate due to leakage to the antenna, respectively. Introducing the antenna coupling coefficient
\begin{align}
  \beta_n \equiv \frac{\gamma_n^{\rm (ext)}}{\gamma_n^{\rm (in)}},
\end{align}
the power extracted through the antenna for the resonant case is given by
\begin{align}
  P = \gamma^{\rm (ext)} E^{\rm (res)} =
  \frac{\bar{S}^2}{2\omega_n} 
  \frac{\beta_n}{1+\beta_n} Q_n
  \mathcal{V}_n\, L (m, \omega_n, Q_n).
\end{align}
The quality factor appearing above,
\begin{align}
  Q_n = \frac{\omega_n}{\gamma_n^{\rm (in)}+\gamma_n^{\rm (ext)}},
\end{align}
is often called the ``loaded quality factor.'' By contrast, the quality factor in the absence of the antenna coupling (i.e., $\gamma_n^{\rm (ext)}=0$) is referred to as the ``unloaded quality factor,'' $Q_n^{\rm (unloaded)}\equiv \omega_n/\gamma_n^{\rm (in)}$.

\subsection{Cavity Coefficient}

We now consider a qubit placed inside a cavity. As discussed above, the EM field inside a cavity can be substantially modified by the cavity walls (or, more generally, by the conducting enclosure surrounding the qubit). In particular, in DM detection experiments that probe DM-induced EM fields using qubits, the EM field that actually couples to the qubit is generally modified by cavity (or packaging) effects, and can therefore be significantly enhanced or suppressed relative to the DM-induced EM field in vacuum.

Focusing on the qubit interaction with the electric field, we parameterize the cavity effect by introducing the cavity (or packaging) coefficient
\begin{align}
  \kappa \equiv \frac{|\vec{n}_{\rm q}\vec{E}(\vec{x}_{\rm qubit})|}{|\vec{n}_{\rm q}\vec{E}^{\rm (vac)}|},
\end{align}
where $\vec{E}(\vec{x}_{\rm qubit})$ is the electric field at the qubit position $\vec{x}_{\rm qubit}$, $\vec{E}^{\rm (vac)}$ is the DM-induced electric field in vacuum (see Eq.\ \eqref{E(vac)}), and $\vec{n}_{\rm q}$ is the unit vector along the direction of maximal qubit sensitivity to external electric fields (see Eq.\ \eqref{dipolevec}). Compared to the case in which the qubit is placed in vacuum, the qubit excitation rate in the cavity is therefore enhanced or suppressed by a factor of $\kappa^2$.

In the non-resonant case, using Eq.\ \eqref{<A(t,x)>}, the electric field in the cavity is evaluated as
\begin{align}
  \vec{E} (t,\vec{x}) =
  - \sum_n \frac{m \bar{S} I_n}{\omega_n^2-m^2} \vec{\tilde{A}}_n (\vec{x}) \cos (mt-\varphi).
\end{align}
Comparing with Eq.\ \eqref{E(vac)}, we obtain the following expression for the cavity coefficient:
\begin{align}
  \kappa = \frac{1}{|\vec{n}_{\rm q} \vec{n}_S|}
  \left| \sum_n \frac{m^2}{\omega_n^2-m^2} I_n \vec{n}_{\rm q} \vec{\tilde{A}}_n (\vec{x}) \right|.
\end{align}
This expression shows that $\kappa$ is enhanced when the drive frequency $m$ approaches one of the cavity eigenfrequencies. Furthermore, we find that $\kappa\rightarrow 0$ as $\vec{x}$ approaches the cavity wall when $\vec{n}_{\rm q}$ is parallel to the wall.

In the expression above, the cavity coefficient is written as a sum over contributions from individual modes. Alternatively, one may compute the cavity coefficient by directly solving the classical equation of motion. Since we are interested in the vector potential whose time dependence is coherent with the source, i.e., $\vec{A} (t,\vec{x})\propto\sin (mt-\varphi)$, we parameterize the vector potential as
\begin{align}
  \vec{A} (t,\vec{x}) = \frac{\bar{S}}{m^2} \left[ \vec{\alpha} (\vec{x}) - \vec{n}_S \right] \sin (mt-\varphi).
\end{align}
The vector field $\vec{\alpha} (\vec{x})$ obeys
\begin{align}
  \left( \nabla^2 + m^2 \right) \vec{\alpha} (\vec{x}) = 0,
  \label{BoxAlpha=0}
\end{align}
and satisfies the following boundary conditions at the surface of the cavity wall:
\begin{align}
  \left. \left( \vec{\alpha} - \vec{n}_S \right) \right|_\parallel = 0, ~~~
  \left. (\vec{\nabla} \times \vec{\alpha}) \right|_\perp = 0.
  \label{BC_alpha}
\end{align}
If the vector field $\vec{\alpha}$ is obtained by solving Eq.\ \eqref{BoxAlpha=0} subject to the boundary conditions given in Eq.\ \eqref{BC_alpha}, we can directly calculate the cavity coefficient.

Using $\vec{\alpha}$, the electric field that couples to the qubit is expressed as
\begin{align}
  \vec{E} (t,\vec{x}) = - \partial_t \vec{A} (t,\vec{x})
  = \frac{1}{m} \bar{S} \left[ - \vec{\alpha} (\vec{x}) + \vec{n}_S \right] \cos (mt-\varphi).
\end{align}
Then, the cavity coefficient can be derived as
\begin{align}
  \kappa = \left| \frac{\vec{n}_{\rm q} \vec{\alpha}(\vec{x}_{\rm qubit})}{\vec{n}_{\rm q} \vec{n}_S} -1 \right|.
\end{align}
From this expression, it is clear that the cavity coefficient is purely geometric; it depends on the shape and size of the cavity, the directions of the qubit $\vec{n}_{\rm q}$ and the source $\vec{n}_S$, and the location of the qubit, but it does not depend on the magnitude of the source $\bar{S}$.

\subsection{Jaynes-Cummings Model}

When a qubit is placed inside a cavity, the qubit-radiation interaction hybridizes the bare excitations, and the energy eigenstates of the combined qubit-cavity system become the so-called dressed states. Here we describe such a situation using the Jaynes-Cummings (JC) model \cite{Jaynes:1963zz}.

In many situations, a single cavity mode is dominant. In that case, the system is well approximated by a qubit coupled to a single radiation mode of fixed frequency. The Hamiltonian is
\begin{align}
  H
  = \omega\, a^\dagger a + \omega_\gamma\, c^\dagger c
  + g (a+a^\dagger)(c+c^\dagger),
\end{align}
where $a=\ketbra{0}{1}$ and $a^\dagger=\ketbra{1}{0}$ are the qubit lowering and raising operators (see Eqs.\ \eqref{op_a} and \eqref{op_adagger}), respectively, and $c$ and $c^\dagger$ are the annihilation and creation operators of the relevant cavity mode, obeying
\begin{align}
  [c,c^\dagger]=1.
\end{align}
Here, $\omega_\gamma$ is the cavity-mode frequency and $g$ is the coupling strength. States are represented as
\begin{align}
  \ket{I} \otimes \ket{n_\gamma}_\gamma,
\end{align}
where $\ket{I}$ (with $I=0,1$) denotes the qubit state, and $\ket{n_\gamma}_\gamma$ (where $n_\gamma$ is a non-negative integer) is a Fock state of the cavity mode defined as
\begin{align}
  \ket{n_\gamma}_\gamma
  = \frac{1}{\sqrt{n_\gamma!}} (c^\dagger)^{n_\gamma} \ket{0}_\gamma.
\end{align}

When the qubit and cavity frequencies are near resonance, it is common to neglect the $ac$ and $a^\dagger c^\dagger$ terms in the Hamiltonian, a simplification known as the RWA. In the interaction picture with respect to the free Hamiltonian $H_0=\omega\, a^\dagger a + \omega_\gamma\, c^\dagger c$, the $ac$ and $a^\dagger c^\dagger$ terms oscillate much more rapidly than the $a^\dagger c$ and $ac^\dagger$ terms. Consequently, their effects are effectively averaged out over time and become negligible. Under the RWA, we obtain the JC Hamiltonian:
\begin{align}
  H_{\rm JC}
  \equiv \omega\, a^\dagger a + \omega_\gamma\, c^\dagger c
  + g(a c^\dagger + a^\dagger c).
\end{align}

A key observation is that the total excitation number,
\begin{align}
  N_{\rm exc} \equiv a^\dagger a + c^\dagger c,
\end{align}
is conserved; indeed, one can verify that $[H_{\rm JC},N_{\rm exc}]=0$. The Hilbert space can therefore be decomposed into sectors of fixed $N_{\rm exc}$. The eigenvalue of $N_{\rm exc}$ for $\ket{I} \otimes \ket{n_\gamma}_\gamma$ is $I + n_\gamma$, and the ground state is $\ket{0} \otimes \ket{0}_\gamma$.

For the subspace with fixed total excitation number $n_{\rm exc}\geq 1$, the Hamiltonian reduces to a $2\times 2$ matrix. In the basis $(\ket{0}\otimes\ket{n_{\rm exc}}_\gamma,\ \ket{1}\otimes\ket{n_{\rm exc}-1}_\gamma)$, the Hamiltonian restricted to this subspace is
\begin{align}
  H_{\rm JC}^{(n_{\rm exc})}
  =
  \begin{pmatrix}
    n_{\rm exc}\omega_\gamma & \sqrt{n_{\rm exc}}\,g \\
    \sqrt{n_{\rm exc}}\,g & \omega+(n_{\rm exc}-1)\omega_\gamma
  \end{pmatrix}.
\end{align}
Its eigenvalues are
\begin{align}
  E_\pm^{(n_{\rm exc})}
  = \frac{1}{2}\left[
    \omega + (2n_{\rm exc}-1)\omega_\gamma
    \pm \sqrt{(\omega-\omega_\gamma)^2 + 4n_{\rm exc} g^2}
  \right],
\end{align}
and the corresponding (unnormalized) eigenstates can be written as
\begin{align}
  \ket{E_\pm^{(n_{\rm exc})}}
  \propto
  \left[
    \omega-\omega_\gamma \mp \sqrt{(\omega-\omega_\gamma)^2 + 4n_{\rm exc} g^2}
  \right]\ket{0}\otimes\ket{n_{\rm exc}}_\gamma
  -2\sqrt{n_{\rm exc}} g \ket{1}\otimes\ket{n_{\rm exc}-1}_\gamma.
\end{align}

In many applications, the coupling is weak, so that the qubit-cavity interaction can be treated perturbatively. In this regime, the energy eigenstates are well approximated by product states $\ket{I}\otimes\ket{n_\gamma}_\gamma$. We denote the energy eigenstate dominated by $\ket{I}\otimes\ket{n_\gamma}_\gamma$ as $\ket{I,n_\gamma}$. For $\sqrt{n_\gamma}\,g \ll |\omega-\omega_\gamma|$, the eigenvalues of $H_{\rm JC}$ and the corresponding eigenstates are approximated as follows:
\begin{itemize}
\item $\ket{0,n_\gamma}$:
Denoting its energy by $E_{0,n_\gamma}$, we obtain
\begin{align}
  E_{0,n_\gamma} \simeq n_\gamma(\omega_\gamma-\chi).
  \label{E_0n}
\end{align}
The corresponding eigenstate is
\begin{align}
  \ket{E_{0,n_\gamma}}
  \simeq \ket{0}\otimes\ket{n_\gamma}_\gamma
  - \frac{\sqrt{n_\gamma}\,g}{\omega-\omega_\gamma}\,
  \ket{1}\otimes\ket{n_\gamma-1}_\gamma.
\end{align}

\item $\ket{1,n_\gamma}$:
Denoting its energy by $E_{1,n_\gamma}$, we obtain
\begin{align}
  E_{1,n_\gamma} \simeq \omega_{\rm q} + n_\gamma(\omega_\gamma+\chi),
  \label{E_1n}
\end{align}
and the corresponding eigenstate is
\begin{align}
  \ket{E_{1,n_\gamma}}
  \simeq \ket{1}\otimes\ket{n_\gamma}_\gamma
  + \frac{\sqrt{n_\gamma+1}\,g}{\omega-\omega_\gamma}\,
  \ket{0}\otimes\ket{n_\gamma+1}_\gamma.
\end{align}
\end{itemize}
In the above expressions,
\begin{align}
  \omega_{\rm q} \equiv \omega + \chi,
\end{align}
with
\begin{align}
  \chi \equiv \frac{g^2}{\omega - \omega_\gamma}.
\end{align}

\begin{figure}[t]
\begin{center}
  \includegraphics[width=0.75\textwidth]{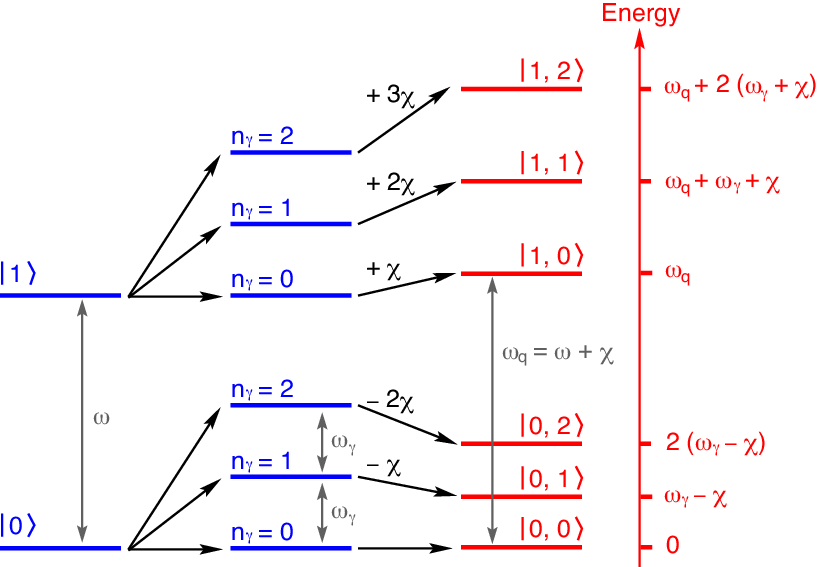}
\end{center}
\caption{Energy levels of the JC Hamiltonian.}
\label{fig:JClevels}
\end{figure}

In the weak-coupling limit, $\ket{0,n_\gamma}$ and $\ket{1,n_\gamma}$ can be viewed as states having $n_\gamma$ cavity photons with the qubit predominantly in its ground and excited states, respectively. The energy levels of the JC Hamiltonian are schematically shown in Fig.\ \ref{fig:JClevels}.

In the weak-coupling limit, the system is approximately described by the effective Hamiltonian
\begin{align}
  H_{\rm JC}^{\rm (eff)}
  \equiv
  -\frac{1}{2}\omega_{\rm q}(Z-\mathds{I})
  + \omega_\gamma c^\dagger c
  - \chi\, Z\, c^\dagger c
  = 
\omega_{\rm q} \ketbra{1}{1} + \omega_\gamma c^\dagger c
  + \chi \left( \ketbra{1}{1} - \ketbra{0}{0} \right) c^\dagger c.
  \label{HJC_eff}
\end{align}
Note that the JC Hamiltonian and the effective Hamiltonian are related by a unitary transformation:
\begin{align}
  H_{\rm JC}^{\rm (eff)} = U_{\rm JC} H_{\rm JC} U_{\rm JC}^\dagger + O(g^3),
\end{align}
with
\begin{align}
  U_{\rm JC} \equiv \exp\!\left[\frac{g}{\omega-\omega_\gamma}\,(a^\dagger c - a c^\dagger)\right].
\end{align}
Thus, neglecting effects of $O(g^3)$, eigenstates of $H_{\rm JC}$ and $H_{\rm JC}^{\rm (eff)}$ have a one-to-one correspondence (with identical eigenvalues); if $\ket{E}$ is the eigenstate of $H_{\rm JC}$ with the eigenvalue $E$ (i.e., $H_{\rm JC}\ket{E}=E\ket{E}$), the corresponding one for $H_{\rm JC}^{\rm (eff)}$ is given by $U_{\rm JC}\ket{E}$.

The effective Hamiltonian given in Eq.\ \eqref{HJC_eff} indicates that the level spacings differ from those in the absence of qubit-cavity coupling, as confirmed by Eqs.\ \eqref{E_0n} and \eqref{E_1n}. This coupling-induced shift of the energy levels is often referred to as the ``dispersive shift'' and its magnitude is characterized by the dispersive-shift parameter $\chi$. Notably, the qubit transition frequency becomes dependent on the cavity photon number, whereas the resonance frequency of the cavity mode depends on the qubit state. If the cavity photon number $n_\gamma$ is fixed, the qubit frequency is given by
\begin{align}
  \omega_{\rm q}^{(n_\gamma)} \simeq \omega_{\rm q} + 2 n_\gamma \chi.
  \label{omega_q(JC)}
\end{align}
If an external field is applied to the qubit, described by the interaction Hamiltonian
\begin{align}
  H_1 = -2 \eta \, (\ketbra{0}{1}+\ketbra{1}{0}) \cos (m t-\varphi),
\end{align}
then the qubit undergoes full Rabi oscillations between $\ket{0,n_\gamma}$ and $\ket{1,n_\gamma}$ when $m=\omega_{\rm q}^{(n_\gamma)}$, not $m=\omega$. This provides a basis for using a qubit as a photon counter (see the following discussion).

Conversely, if the qubit state is fixed, one obtains a ladder of states with different cavity photon numbers,
\begin{align*}
  \ket{I,0}, ~~\ket{I,1},~~ \ket{I,2}, ~~\ket{I,3},~~ \ket{I,4}, \cdots,
  ~~~(\mbox{with}\ I=0, 1),
\end{align*}
whose level spacing is $\omega_\gamma-\chi$ for $I=0$ and $\omega_\gamma+\chi$ for $I=1$. Thus, the cavity-mode frequency after the dispersive shift is denoted as
\begin{align}
  \omega_\gamma^{(\pm)} = \omega_\gamma \pm \chi = 
  \omega_\gamma \pm \frac{g^2}{\omega - \omega_\gamma},
  \label{omega_gamma(JC)}
\end{align}
where the positive and negative signs are for $I=1$ and $0$, respectively. This implies that the cavity-mode frequency depends on the qubit state. 

Various aspects of the energy-level shifts predicted by the JC model have already been exploited in ongoing DM search experiments. A particularly important application is the so-called dispersive readout \cite{Blais:2004hcj, Wallraff:2005bng}. By taking advantage of the fact that the cavity-mode frequencies depend on the qubit state, one can infer the qubit state from measurements of the resulting dispersive shift. In dispersive readout, the cavity is driven by a microwave tone, and the amplitude and phase of the reflected and/or transmitted signal are measured. These measured quantities depend on the cavity resonance frequency and, therefore, indirectly on the qubit state. Under an external drive (at frequency $\omega_{\rm d}$, which is experimentally controllable), the intra-cavity field can be described by a coherent state $\ket{\bar{\alpha}(t)}$ characterized by (see Eq.\ \eqref{rhotinf})
\begin{align}
  \bar{\alpha}(t) \simeq
  - \frac{\xi}
       {i (\omega'_\gamma - \omega_{\rm d}) + \frac{1}{2} \gamma}
       e^{-i(\omega_{\rm d} t - \varphi)},
\end{align}
where we consider the case in which $\omega_{\rm d}\sim \omega'_\gamma$. Here, $\xi$ and $\gamma$ (with $\gamma \ll \omega'_\gamma$ assumed) denote the drive strength and the dissipation rate, respectively, and $\omega'_\gamma$ is the frequency of the selected cavity mode (note that $\omega'_\gamma = \omega_\gamma$ in the absence of coupling to the qubit). When a qubit is placed inside the cavity, $\omega'_\gamma$ becomes dependent on the qubit state (see Eq.\ \eqref{omega_gamma(JC)}). As a consequence, both the amplitude and phase of the intra-cavity field, in particular, the prefactor multiplying $e^{-i(\omega_{\rm d} t - \varphi)}$, inherit a dependence on the qubit state. The complex field amplitude (including its phase) can be extracted using standard techniques like $I/Q$ measurement (see Appendix~\ref{subsec:cavityreadout}), thereby enabling the dispersive shift to be used for readout of the qubit state in the cavity.

It has been also demonstrated that a qubit can be used to infer the photon number in a cavity. In the dark-photon DM search reported in Ref.\ \cite{Dixit:2020ymh}, a coupled cavity-qubit system is prepared, and repeated quantum nondemolition measurements of cavity photons are performed by monitoring the shift of the qubit transition frequency, exploiting the fact that the qubit frequency depends on the photon number in the cavity. This procedure enabled the noise to be reduced to a level below the standard quantum limit.

\begin{figure}[t]
  \centering
  \includegraphics[width=0.6\textwidth]{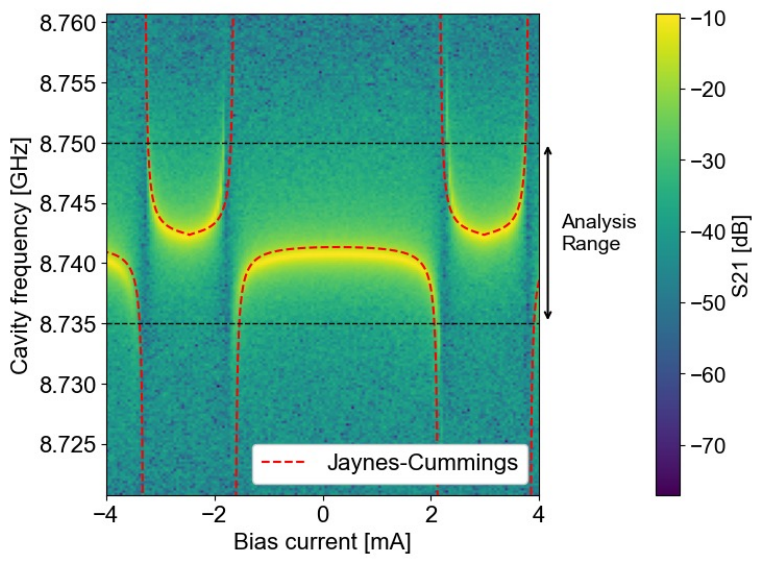}
  \caption{Relationship between the bias current applied to the coil (horizontal axis) and the cavity frequency (vertical axis), obtained in the DarQ-Lamb experiment \cite{Nakazono:2025tak}. The color scale indicates the cavity transmittance, with the yellow region representing the center of the cavity resonance. The red dashed line shows the prediction of the JC model. The analysis range adopted for the DarQ-Lamb experiment is also indicated.}
  \label{fig:KNakazono}
\end{figure}

Another application of the level shift is to use the qubit as a tuner of the cavity frequency, as reported by Refs.\ \cite{Zhao:2025thg, Nakazono:2025tak}. For example, for dark photon DM detection, Ref.\ \cite{Nakazono:2025tak} employed a cavity with a SQUID-based qubit located inside. By applying external magnetic field, the frequency parameter of the qubit $\omega$ was varied (see Eqs.\ \eqref{omega_qubit} and \eqref{EJ(eff)}). This, in turn, changes the cavity resonance frequency through the $\omega$ dependence of the dispersive-shift parameter $\chi$ (see Eq.\ \eqref{omega_gamma(JC)}). Compared to other methods of tuning the cavity frequency, this method has the advantage of avoiding the frictional heating and electromagnetic leakage associated with mechanical tuning. Fig.\ \ref{fig:KNakazono} shows the relation between the bias current flowing through the coil (which induces the external magnetic field) and the cavity frequency; one can see that the cavity frequency is successfully tuned and that the prediction of the JC model is well reproduced.\footnote
{Figure courtesy of K.~Nakazono (DarQ-Lamb collaboration).}

\section{Quantum Sensing: Basics}
\label{sec:sensing}
\setcounter{equation}{0}
\setcounter{figure}{0}
\setcounter{footnote}{1}

Up to this point, our arguments have focused on scenarios where DM detection is performed by treating each qubit individually and conducting a counting experiment to measure excited qubits. However, when employing multiple qubits, the quantum nature of these systems allows us, in principle, to entangle the qubits and pursue improved sensitivity through optimized measurement schemes.

To understand the maximal achievable sensitivity, it is essential to clarify how measurement processes are formally described. In particular, the description of measurements in terms of CPTP maps, or equivalently, in terms of so-called positive operator-valued measures (POVMs), plays a crucial role. These topics constitute the focus of this section. Readers can find a more complete and detailed overview of the topics in, e.g., Refs.\ \cite{PreskillQCNotes, Nielsen:2012yss}.

\subsection{POVM}

As we have mentioned in Section \ref{sec:environments}, a quantum measurement can be regarded as a CPTP map. If the measurement outcome is not recorded (or is traced out), the density matrix is mapped to the unconditional (or average) post-measurement state as
\begin{align}
  \rho \rightarrow \rho' = \sum_a K_a \rho K_a^\dagger.
\end{align}
Here, the index $a$ labels the measurement outcome. Furthermore, if the measurement outcome $a$ is obtained, the corresponding (post-selected) state is given by
\begin{align}
  \rho \rightarrow \rho' = \frac{K_a \rho K_a^\dagger}{\mbox{tr}(K_a \rho K_a^\dagger)},
  \label{rhodash:fixed-a}
\end{align}
and, using the Kraus operators, the probability for obtaining the outcome $a$ is given by $p_a = \mathrm{tr} (K_a \rho K_a^\dagger)$. 

Defining 
\begin{align}
  E_a\equiv K_a^\dagger K_a, 
\end{align}
the measurement is characterized by a set of POVM elements $\{E_a\}$; $E_a$ are positive semidefinite operators and satisfy the completeness relation
\begin{align}
  \sum_a E_a = \mathds{I}.
  \label{sumMa}
\end{align}
In terms of the POVM elements, the probability of obtaining outcome $a$ is
\begin{align}
  p_a = \mbox{tr}(\rho E_a).
\end{align}

Since each POVM element is a positive-semidefinite operator, one may construct Kraus operators for a given POVM $\{E_a\}$, for example, by choosing $K_a=\sqrt{E_a}$.\footnote
{For a given positive-semidefinite operator $A$, described as 
\begin{align*}
  A = \sum_i \lambda_i \ketbra{\psi_i}{\psi_i},
\end{align*}
with $\lambda_i\geq 0$ being the eigenvalues and $\ket{\psi_i}$ corresponding eigenstates of $A$, we define $\sqrt{A}$ as
\begin{align*}
  \sqrt{A} \equiv \sum_i \sqrt{\lambda_i} \ketbra{\psi_i}{\psi_i}.
\end{align*}
}
However, this is not a unique choice because sets of Kraus operators related by unitary transformations all give the same POVM element $E_a$; the Kraus operator $K'_a=V_a \sqrt{E_a}$ (with $V_a$ being an arbitrary unitary operator) gives $K^{\prime\dagger}_aK'_a=E_a$. This fact also implies that, even if a POVM is specified, the density matrix after the measurement has an ambiguity.

One often comes across the projection measurement, in which $\{E_a\}$ is given by a set of projection operators $\Pi_a$, satisfying 
\begin{align}
  \Pi_a \Pi_b = \delta_{a,b} \Pi_a.
\end{align}
It is often the case that $\Pi_a$ are chosen to be projection operators onto the eigenstates of a Hermitian operator, such as the Hamiltonian, spin operators, and so on. In the case of projection measurement, after the measurement, the state is mapped to one of the eigenstates of $\Pi_a$, which are mutually orthogonal. For general POVM measurements, on the contrary, such an orthogonality does not always hold.

Using Naimark's theorem, any POVM can be implemented as a projective measurement on an enlarged Hilbert space \cite{PreskillQCNotes, Nielsen:2012yss}. To see this, we extend the original Hilbert space by introducing an ancillary system:
\begin{align}
  \ket{\Psi} = \ket{\mbox{system}} ~~\rightarrow~~
  \ket{\mbox{system}} \otimes \ket{\mbox{ancilla}},
\end{align}
where $\ket{\mbox{system}}$ and $\ket{\mbox{ancilla}}$ denote the system state which we want to probe with the measurement and the newly-added ancilla state, respectively.  Accordingly, the total dimension of the Hilbert space is enlarged to $N_{\rm S}\times N_{\rm A}$, where $N_{\rm S}$ and $N_{\rm A}$ denote the dimensions of the system and the ancilla Hilbert spaces, respectively. The ancilla dimension should be at least as large as the number of POVM elements. For simplicity, we focus on the case where $N_{\rm A}$ equals the number of POVM elements; the discussion can be straightforwardly generalized to larger $N_{\rm A}$.

\begin{figure}[t]
  \centering
  \includegraphics[width=0.55\textwidth]{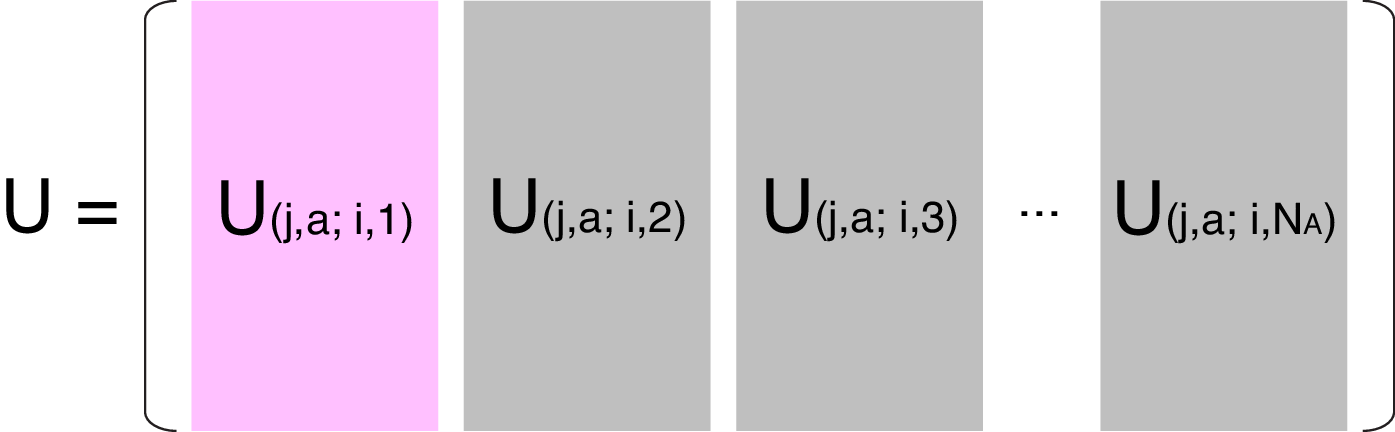}
  \caption{Schematic figure of the structure of the unitary matrix $U$, realizing the POVM measurement by a projection measurement. The elements $U_{j,a;i,1}$ are determined by Eq.\ \eqref{U_JaI1}, while others should be filled by column vectors which are orthonormal among themselves and orthogonal to the column vectors defined by $U_{j,a;i,1}$.}
  \label{fig:UMforPOVM}
\end{figure}

We can consider a unitary operation to the joint (system+ancilla) system such that
\begin{align}
  U \left( \ket{\psi_i} \otimes \ket{1} \right)
  = \sum_a \left( \sqrt{E_a} \ket{\psi_i} \right) \otimes \ket{a},
  \label{U(POVM)}
\end{align}
where $\{\ket{\psi_i}\}$ and $\{\ket{a}\}$ are orthogonal bases of the system and the ancilla, respectively. For any given set of POVM elements $\{E_a\}$, one can always construct a unitary matrix $U$ that satisfies Eq.\ \eqref{U(POVM)}. Indeed, Eq.\ \eqref{U(POVM)} fixes the $(j,a;i,1)$ matrix elements of $U$ as
\begin{align}
  U_{j,a;i,1} = \mel{\psi_j}{ \sqrt{E_a} }{\psi_i},
  \label{U_JaI1}
\end{align}
whereas the remaining elements $U_{j,a;i,b\neq 1}$ are not specified by Eq.\ \eqref{U(POVM)} (see Fig.\ \ref{fig:UMforPOVM}). The unitarity condition for $U$ is
\begin{align}
  \left[ U^\dagger U \right]_{i,a;j,b} = \delta_{i,j} \delta_{a,b}.
\end{align}
The above condition is satisfied for $a=b=1$ by using the POVM completeness relation given in Eq.\ \eqref{sumMa}:
\begin{align}
  \left[ U^\dagger U \right]_{i,1;j,1} =
  \sum_c \mel{\psi_i}{E_c}{\psi_j} = \delta_{i,j}.
\end{align}
Therefore, the matrix elements fixed by Eq.\ \eqref{U(POVM)} are consistent with unitarity; the entries $U_{j,a;i,1}$ specify $N_{\rm S}$ mutually orthonormal column vectors in a space of dimension $N_{\rm S}\times N_{\rm A}$. One can always complete these vectors to a full orthonormal basis by choosing an additional $N_{\rm S}\times (N_{\rm A}-1)$ column vectors that are orthonormal among themselves and orthogonal to the column vectors defined by $U_{j,a;i,1}$. By assigning these additional vectors to the previously undetermined entries $U_{j,a;i,b\neq 1}$, we obtain a unitary matrix $U$ satisfying Eq.\ \eqref{U(POVM)}. (This construction does not yield a unique $U$, but this non-uniqueness does not affect the present discussion.)

Now we can see that any POVM measurement is equivalent to the unitary evolution described by $U$, followed by the projection measurement of the ancilla state (with the measurement operator $\mathds{I}_{\rm S}\otimes \ketbra{a}{a}$, where $\mathds{I}_{\rm S}$ is the identity operator on the system of our interest). With such a projection measurement (with measurement outcome $a$), the density matrix of the system after the measurement is given by
\begin{align}
  \rho' = \frac{\sqrt{E_a} \rho \sqrt{E_a}}{\mbox{tr}(\rho E_a)},
\end{align}
which is equivalent to $\rho'$ given in Eq.\ \eqref{rhodash:fixed-a} for the case of $K_a=\sqrt{E_a}$.

\subsection{Cram\'er-Rao Bound}

When estimating an underlying parameter on which the probability distribution of the data and/or the density matrix depends, it is desirable to know a lower bound on the estimation error (or, more precisely, on the variance of an estimator). The so-called Cram\'er-Rao bound provides an absolute lower bound on the variance, and therefore plays an important role in parameter estimation.

Here we consider the case where the model contains a single unknown parameter, $\theta$. The distribution of the measurement outcome depends on $\theta$; we denote the probability of obtaining an outcome $a$ by $p_\theta (a)$, which satisfies
\begin{align}
  \sum_a p_\theta (a) = 1.
\end{align}
The parameter $\theta$ is estimated from the measurement outcome. We denote the estimator by $\hat{\theta}(a)$, which assigns an estimated value of $\theta$ to the observed data $a$. We assume that the estimator is unbiased, i.e.,
\begin{align}
  \ev*{\hat{\theta}}_\theta = \theta,
  \label{unbiasedestimator}
\end{align}
where $\ev{\cdots}_\theta$ denotes the expectation value with respect to $p_\theta(a)$:
\begin{align}
  \ev{f}_\theta \equiv \sum_a p_\theta (a) f(a).
\end{align}
Because we are interested in the Cram\'er-Rao bound around a particular value of $\theta$, the local unbiasedness is required, i.e., $\ev*{\hat{\theta}}_\theta = \theta$ and $\partial_\theta\ev*{\hat{\theta}}_\theta = 1$ should hold around a specific value of $\theta$ of our interest.

We begin with a derivation of the classical Cram\'er-Rao bound, for which we use the identity
\begin{align}
  \ev{ (\hat{\theta}-\theta) \partial_\theta \ln p_\theta}_{\theta}
  = 1,
\end{align}
together with the following inequality based on the Cauchy-Schwarz inequality,
\begin{align}
  \ev{ (\hat{\theta}-\theta) \partial_\theta \ln p_\theta}_\theta
  \leq \sqrt{\mbox{Var}[(\hat{\theta}-\theta)] \mbox{Var}[\partial_\theta \ln p_\theta]}
  = \sqrt{\mbox{Var}[\hat{\theta}] \mbox{Var}[\partial_\theta \ln p_\theta]},
  \label{ineqforCCRB}
\end{align}
where ``Var'' denotes the variance. In order to derive the above inequality, we have used the assumption that the estimator is unbiased (i.e., Eq.\ \eqref{unbiasedestimator}), as well as the relation $\ev{\partial_\theta \ln p_\theta}_\theta=0$. (Here and hereafter, we assume regularity conditions so that the differentiation $\partial_\theta$ and the summation (or integration) over the measurement result can be exchanged.) Combining the above relations, we obtain the classical Cram\'er-Rao bound \cite{Cramer1946}:
\begin{align}
  \mbox{Var}[\hat{\theta}] \geq \frac{1}{F_{\rm C}},
\end{align}
where
\begin{align}
  F_{\rm C} \equiv \mbox{Var}[\partial_\theta \ln p_\theta]
  = \sum_a \frac{(\partial_\theta p_\theta)^2}{p_\theta}.
  \label{defCFI}
\end{align}
We refer to $F_{\rm C}$ as the classical Fisher information (CFI). Under the regularity conditions, $\sum_a \partial_\theta^2 p_\theta=0$ follows, so that the CFI can also be written as
\begin{align}
  F_{\rm C} = - \ev{\partial_\theta^2 \ln p_\theta}_\theta.
\end{align}
The discussion up to this point applies both to measurements on classical systems and to measurements on quantum systems once a measurement scheme (hence a classical outcome distribution) is specified.

Hereafter, we focus on measurements on a quantum system, where the density matrix depends on the parameter $\theta$. Considering the measurement with the POVM $\{E_a\}$, the outcome distribution is given by
\begin{align}
  p_\theta (a) = \mbox{tr} (\rho_\theta E_a),
\end{align}
where $\rho_\theta$ is the $\theta$-dependent density matrix. We write the spectral decomposition of $\rho_\theta$ as
\begin{align}
  \rho_\theta = \sum_i \lambda_i (\theta) \ketbra{\psi_i (\theta)}{\psi_i (\theta)},
  \label{decomp_rhotheta}
\end{align}
where $\lambda_i$ and $\ket{\psi_i}$ are the $i$-th eigenvalue and the corresponding normalized eigenstate of $\rho_\theta$, respectively. In the following, we assume that the null space of $\rho_\theta$ is (locally) preserved under variations of $\theta$. This imposes the constraint
\begin{align}
  \mel{\psi_i}{\partial_\theta\rho_\theta}{\psi_j} = 0, ~~~\mbox{if $\lambda_i=\lambda_j=0$}.
  \label{nullofrho}
\end{align}
This constraint implies $\partial_\theta\lambda_i=0$ whenever $\lambda_i=0$, and thus it prevents eigenvalues from becoming negative under variations of $\theta$.

For further discussion, we introduce the symmetric logarithmic derivative (SLD) $L_\theta$ through
\begin{align}
  \partial_\theta \rho_\theta = \frac{1}{2}
  \left( L_\theta \rho_\theta + \rho_\theta L_\theta \right).
\end{align}
The SLD is a Hermitian operator and, under the constraint \eqref{nullofrho}, it is well-defined on the support of $\rho_\theta$. Using the spectral decomposition in Eq.~\eqref{decomp_rhotheta}, the SLD can be expressed as
\begin{align}
  L_\theta = \sum_{\lambda_i+\lambda_j>0} \frac{2}{\lambda_i+\lambda_j}
  \mel{\psi_i}{\partial_\theta\rho_\theta}{\psi_j}
  \ketbra{\psi_i}{\psi_j}.
\end{align}
In terms of the SLD, the CFI can be written as
\begin{align}
  F_{\rm C} = \sum_a \frac{1}{\mbox{tr}(\rho_\theta E_a)}
  \left| \mbox{Re} [ \mbox{tr}(\rho_\theta L_\theta E_a) ] \right|^2.
  \label{CFI}
\end{align}
In order to derive a bound on the CFI, we use the following two inequalities
\begin{align}
  \left| \mbox{Re} [ \mbox{tr}(\rho_\theta L_\theta E_a) ] \right|^2\leq
  \left| \mbox{tr}(\rho_\theta L_\theta E_a) \right|^2,
  \label{ineqforQCRB1}
\end{align}
and 
\begin{align}
  \left| \mbox{tr}(\rho_\theta L_\theta E_a) \right|^2 \leq
  \mbox{tr}(\rho_\theta L_\theta E_a L_\theta)
  \mbox{tr}(\rho_\theta E_a ).
  \label{ineqforQCRB2}
\end{align}
Notice that the inequality \eqref{ineqforQCRB2} can be obtained from the operator Cauchy-Schwarz inequality,
\begin{align}
  \left| \mbox{tr}(A^\dagger B) \right|^2 \leq
  \mbox{tr}(A^\dagger A) \mbox{tr}(B^\dagger B),
\end{align}
which holds for arbitrary operators $A$ and $B$, by substituting $A=\sqrt{E_a}L_\theta\sqrt{\rho_\theta}$ and $B=\sqrt{E_a}\sqrt{\rho_\theta}$; the inequality is saturated when $A\propto B$. Then, we can find
\begin{align}
  F_{\rm C} \leq
  \sum_a \frac{1}{\mbox{tr}(\rho_\theta E_a)} \left|\mbox{tr}(\rho_\theta L_\theta E_a)\right|^2
  \leq
  \sum_a \mbox{tr}(\rho_\theta L_\theta E_a L_\theta)
  = \mbox{tr}(\rho_\theta L_\theta^2),  
\end{align}
where, in the last equality, the POVM completeness relation has been used (see Eq.\ \eqref{sumMa}).

We now define the quantum Fisher information (QFI) $F_{\rm Q}$ by
\begin{align}
  F_{\rm Q} \equiv \mbox{tr}(\rho_\theta L_\theta^2),
  \label{defQFI}
\end{align}
which bounds the CFI from above:
\begin{align}
  F_{\rm C} \leq F_{\rm Q}.
\end{align}
Then, we obtain the quantum Cram\'er-Rao bound (QCRB) \cite{Braunstein:1994zz} (see also \cite{Paris:2008zgg}):
\begin{align}
  \mbox{Var}[\hat{\theta}] \geq \frac{1}{F_{\rm C}} \geq \frac{1}{F_{\rm Q}}.
\end{align}

For the case of a pure state, for which we denote
\begin{align}
  \rho_\theta^{\rm (pure)} = \ketbra{\Psi_\theta}{\Psi_\theta},
\end{align}
the SLD is found to be
\begin{align}
  L_\theta^{\rm (pure)} = 2 \sum_{\braket{\Psi_\theta}{\psi}=0}
  \left( \braket{\psi}{\partial_\theta\Psi_\theta}
  \ketbra{\psi}{\Psi_\theta} + \mathrm{h.c.}
  \right),
\end{align}
where the summation is taken over the states $\ket{\psi}$ that are perpendicular to $\ket{\Psi_\theta}$. The QFI then simplifies to
\begin{align}
  F_{\rm Q}^{\rm (pure)} = 4
  \left( \braket{\partial_\theta\Psi_\theta}{\partial_\theta\Psi_\theta}
  - \left| \braket{\Psi_\theta}{\partial_\theta\Psi_\theta} \right|^2
  \right).
  \label{FQpure}
\end{align}
Furthermore, we are often interested in the case in which the state $\ket{\Psi_\theta}$ takes the following form:
\begin{align}
  \ket{\Psi_\theta} = e^{-i\theta G} \ket{\Psi_0},
\end{align}
where $G$ is a Hermitian operator, while $\ket{\Psi_0}$ is a $\theta$-independent state. In such a case, the QFI is given by
\begin{align}
  F_{\rm Q}^{(G)} = 4 \mbox{Var}_{\Psi_\theta} [G] \equiv
  4 \left( \mel{\Psi_\theta}{G^2}{\Psi_\theta} - \mel{\Psi_\theta}{G}{\Psi_\theta}^2 \right) = 
  4 \left( \mel{\Psi_0}{G^2}{\Psi_0} - \mel{\Psi_0}{G}{\Psi_0}^2 \right).  
\end{align}
Note that, from the last expression, it is evident that $F_{\rm Q}^{(G)}$ is $\theta$-independent.

For a single qubit, whose state is represented by a $\theta$-dependent Bloch vector $\vec{\rho}_\theta$, one can derive a convenient closed-form expression for the QFI. In the case of a mixed state, i.e., $|\vec{\rho}_\theta|<1$, the SLD is given by
\begin{align}
  L_\theta^{\rm (qubit)} = - \frac{\vec{\rho}_\theta \partial_\theta \vec{\rho}_\theta}{1 - |\vec{\rho}_\theta|^2} \mathds{I} +
  \left( \partial_\theta \vec{\rho}_\theta +
  \frac{\vec{\rho}_\theta \partial_\theta \vec{\rho}_\theta}{1 - |\vec{\rho}_\theta|^2} \vec{\rho}_\theta \right) \vec{\sigma},
\end{align}
where $\vec{\sigma}=(X,Y,Z)$ denotes the vector of Pauli operators. The corresponding QFI is then expressed as
\begin{align}
  F_{\rm Q}^{\rm (qubit)} =
  |\partial_\theta \vec{\rho}_\theta|^2 +
  \frac{(\vec{\rho}_\theta \partial_\theta \vec{\rho}_\theta)^2}{1 - |\vec{\rho}_\theta|^2}.
\end{align}

It is important to note that the CFI depends on the choice of POVM elements; this is explicit in Eq.~\eqref{CFI}, which contains $E_a$. Thus, the CFI provides a bound on the variance for a given measurement scheme. By contrast, the QFI is independent of the choice of measurement operators and is determined based only on the $\theta$-dependence of the density matrix $\rho_\theta$. Indeed, the QFI is the supremum of the CFI over all POVMs:
\begin{align}
  F_{\rm Q} = \sup_{\{E_a\}} F_{\rm C}.
\end{align}
The QFI bounds the CFI from above over all measurement methods.

An optimal measurement that achieves $F_{\rm C}=F_{\rm Q}$ is given by the projective measurement onto the eigenbasis of the SLD:
\begin{align}
  E_a^{\rm (optimal)} = \ketbra{\ell_a}{\ell_a},
  \label{E_a(optimal)}
\end{align}
where $\ket{\ell_a}$ are eigenstates of the SLD:
\begin{align}
  L_\theta \ket{\ell_a} = \ell_a \ket{\ell_a}.
\end{align}
With Eq.~\eqref{E_a(optimal)}, one can verify that the inequalities \eqref{ineqforQCRB1} and \eqref{ineqforQCRB2} are both saturated, as required for optimality. We note here that the optimal measurement method generally depends on $\theta$; in practice one uses adaptive schemes or operates near a known working point $\theta_0$.

It is also instructive to consider the situation in which we have $N$ copies of an identical quantum system. If there is no entanglement among the systems and, moreover, if there are no interactions between different systems, the total density matrix can be written as the tensor product of the density matrices of the $N$ subsystems:
\begin{align}
  \rho_\theta^{\rm (prod)} = \bigotimes_{n=1}^N \rho_{n,\theta},
\end{align}
where $\rho_{n,\theta}$ is the density matrix acting on the $n$-th subsystem. In this case, the SLD for the total system is given by
\begin{align}
  L_\theta^{\rm (prod)} =
  L_{1,\theta} \otimes \mathds{I}_2 \otimes \cdots \otimes \mathds{I}_N +
  \mathds{I}_1 \otimes L_{2,\theta} \otimes \cdots \otimes \mathds{I}_N +
  \cdots +
   \mathds{I}_1 \otimes \mathds{I}_2 \otimes \cdots \otimes L_{N,\theta},
\end{align}
where $\mathds{I}_n$ denotes the identity operator on the $n$-th subsystem, and $L_{n,\theta}$ is the SLD associated with $\rho_{n,\theta}$. The QFI for the total system is then expressed as
\begin{align}
  F_{\rm Q}^{\rm (prod)} =
  \sum_{n=1}^N \mbox{tr} \left( \rho_{n,\theta} L_{n,\theta}^2 \right) +
  \sum_{n\neq n'} \mbox{tr} \left( \rho_{n,\theta} L_{n,\theta} \right)
  \mbox{tr} \left( \rho_{n',\theta} L_{n',\theta} \right).
\end{align}
Using the relation $\mbox{tr} \left( \rho_\theta L_\theta \right) = \mbox{tr} \left( \partial_\theta \rho_\theta \right) = 0$, we obtain
\begin{align}
  F_{\rm Q}^{\rm (prod)} = \sum_{n=1}^N F_{n,{\rm Q}},
\end{align}
where $F_{n,{\rm Q}}$ denotes the QFI of the $n$-th subsystem. In particular, if the density matrices of all the subsystems are identical, the QFI of the total system is $N$ times larger than that of the subsystem and hence is proportional to $N$. Thus, due to the QCRB, the lower bound on the uncertainty in the determination of $\theta$, i.e., $\sqrt{\mbox{Var}[\hat{\theta}]}$, scales as $N^{-1/2}$ in the case of an (unentangled) product state.

The aforementioned argument can also be extended to the case where the measurement is repeated multiple times. If all measurements are performed independently, this situation is equivalent to having a number of unentangled, identical quantum systems, each subjected to an independent measurement. Therefore, for multiple independent repetitions, the total QFI becomes $N_{\rm rep}$ times larger than that for a single measurement, where $N_{\rm rep}$ denotes the number of measurement repetitions.

\subsection{Parameter Estimation and Hypothesis Testing}

We consider the task of determining an unknown parameter $\theta$ from measurement data of an observable $\mathcal{A}$. By performing a measurement of $\mathcal{A}$, one obtains information about its expectation value and variance in the state $\rho$; in particular,
\begin{align}
  \ev{\mathcal{A}}_\theta = \mbox{tr} (\rho_\theta \mathcal{A}).
\end{align}
After $N$ independent measurements, an estimator for $\ev{\mathcal{A}}_\theta$ is given by the sample mean
\begin{align}
  \hat{\mathcal{A}} = \frac{1}{N} \sum_{n=1}^N \mathcal{A}_n,
  \label{A-mean}
\end{align}
where $\mathcal{A}_n$ denotes the outcome of the $n$-th measurement. By the central limit theorem, for large $N$ the distribution of $\hat{\mathcal{A}}$ approaches a normal distribution with mean $\ev{\mathcal{A}}_\theta$ and variance $\mathrm{Var}[\mathcal{A}]/N$. Therefore, the standard deviation of $\hat{\mathcal{A}}$ is
\begin{align}
  \delta \ev{\mathcal{A}}_\theta \equiv
  \sqrt{\frac{\mbox{Var}[\mathcal{A}]}{N}}.
\end{align}
Using the linear (error-propagation) approximation, the uncertainty of the corresponding parameter estimator $\hat{\theta}=\hat{\theta}(\hat{\mathcal{A}})$ can be written as
\begin{align}
  \delta\theta
  = \frac{\delta\langle \mathcal{A} \rangle}{\left|d\langle \mathcal{A} \rangle / d\theta\right|}
  = \frac{\sqrt{\mbox{Var}[\mathcal{A}]}}{\sqrt{N}\,\left|d\langle \mathcal{A} \rangle / d\theta\right|},
  \label{error-propagation}
\end{align}
where $\delta\theta$ should be interpreted as the root-mean-square (RMS) estimation error within this linear-response approximation.

The discussion so far concerns point estimation, i.e., how small the RMS error $\delta\theta$ can be made. A complementary and equally important perspective is hypothesis testing: given a specific value $\theta_0$, can we reject it based on the measurement outcomes? Specifically, suppose we wish to test
\begin{align}
  H_0: \theta = \theta_0 \qquad \text{vs.} \qquad H_1: \theta \neq \theta_0,
\end{align}
where $H_0$ is the null hypothesis and $H_1$ is the alternative hypothesis. 

After $N$ measurements of the observable $\mathcal{A}$, the sample mean $\hat{\mathcal{A}}$ in Eq.\ \eqref{A-mean} satisfies the asymptotic normal approximation
\begin{align}
  \hat{\mathcal{A}} \xrightarrow{N\to\infty}
  \mathcal{N}\!\left( \ev{\mathcal{A}}_\theta, \frac{\mbox{Var}[\mathcal{A}]}{N}\right).
\end{align}
Thus, under $H_0$, the standardized test statistic is
\begin{align}
  z = \frac{\hat{\mathcal{A}} - \ev{\mathcal{A}}_{\theta_0}}
  {\sqrt{\mbox{Var}[\mathcal{A}]_{\theta_0}/N}}
  \;\xrightarrow{H_0} \mathcal{N}(0,1),
\end{align}
where $\mbox{Var}[\mathcal{A}]_{\theta_0}$ is the variance estimated at $\theta_0$ in $H_0$. The corresponding two-sided $p$-value is
\begin{align}
    p = 2 \left[1 - \Phi(|z|) \right],
\end{align}
where $\Phi$ denotes the cumulative distribution function of the standard normal distribution. The null hypothesis $H_0$ is rejected at significance level $\alpha$ when $p<\alpha$, equivalently when $|z|>z_{\alpha/2}$, where $z_{\alpha/2}$ denotes the $(1-\alpha/2)$ quantile of the standard normal distribution (e.g., $z_{0.025}\simeq 1.96$ for $\alpha=0.05$).

Using the error-propagation relation in Eq.\ \eqref{error-propagation}, one may write, to leading order,
\begin{align}
  \hat{\mathcal{A}} - \langle\mathcal{A}\rangle_{\theta_0}\simeq
  (\hat{\theta}-\theta_0)
|\partial_\theta\langle\mathcal{A}\rangle|.
\end{align}
This approximation is valid when $\delta\theta\ll|\partial_\theta\ev{\mathcal{A}}_{\theta}/\partial_\theta^2\ev{\mathcal{A}}_{\theta}|$, i.e., when the response is approximately linear over the scale of the estimation error. The test statistic can then be rewritten as
\begin{align}
  z \simeq
  \frac{\hat{\theta} - \theta_0}{\delta\theta}.
\end{align}

The $N$ independent measurements considered above can also be viewed as a separable measurement performed on a total system consisting of $N$ identical subsystems. In this sense, $N$ represents the number of subsystems (or copies) available for determining $\theta$. Then, according to the QCRB,
\begin{align}
  \delta\theta \geq \frac{1}{\sqrt{F_{\rm Q}(\theta)}},
\end{align}
where $F_{\rm Q}$ is the QFI of the total system (with $N$ subsystems). As discussed above, when the subsystems are unentangled and the global state is separable, the QFI is additive and takes the form $F_{\rm Q}=N F_{\rm Q}^{\rm (single)}$, where $F_{\rm Q}^{\rm (single)}$ is the QFI for a single subsystem. In this separable setting, the achievable precision therefore improves with $N$ as $\propto N^{-1/2}$. 

The scaling implied by $F_{\rm Q}\propto N$ is referred to as the standard quantum limit (SQL):
\begin{align}
  \delta\theta^{\mathrm{(SQL)}} \propto \frac{1}{\sqrt{N}}.
\end{align}
By contrast, if there is entanglement among the subsystems, the QFI can in principle scale as $F_{\rm Q}\propto N^2$. This enhanced scaling is known as the Heisenberg limit (HL):
\begin{align}
  \delta\theta^{\mathrm{(HL)}} \propto \frac{1}{N}.
\end{align}
If achievable, the HL scaling yields a substantially stronger improvement in precision with increasing $N$ than the SQL scaling. We note, however, that in many practical situations the HL cannot be attained due to noise and decoherence, finite experimental duration, or resource constraints such as limits on the number of available queries.

\subsection{\textit{Z} Measurement vs.\ \textit{Y} Measurement for DM Detection}
\label{subsec:zvsy}

It is instructive to use the arguments developed so far to infer the optimized measurement strategy for performing the DM search experiment with a qubit. Our proposal for extracting the DM signal has been to count the number of excited qubits after exposure to the evolution driven by the DM. This corresponds to performing measurement of $Z$; in other words, the projection operators onto the eigenstates of the Pauli-$Z$ operator are chosen as the measurement operators. One may also consider other types of measurements; in particular, the measurement of $Y$ (or $X$), which may alter the sensitivity. Here, we examine which measurement is relevant in the DM detection experiment. 

Here, we consider the simple noise model introduced in Section~\ref{subsec:qubitwithnoise}, which includes spontaneous excitation (giving rise to dark counts), amplitude damping, and dephasing. The corresponding time scales are denoted by $T_0$, $T_1$, and $T_2$, respectively. In the weak-signal limit, we employ the density matrix given in Eqs.\ \eqref{rho00(simple)} -- \eqref{rho01(simple)}. In particular, we focus on the regime where the time scales associated with dark counts and amplitude damping are much shorter than that of the Rabi oscillation, i.e., $T_0, T_1 \ll \eta^{-1}$. We further assume $T_1 \ll T_0$, so that the dark-count probability remains small. Then, in the resonance limit $\omega \rightarrow m$, which yields the best sensitivity for DM detection, the density matrix is approximated as
\begin{align}
  \rho \simeq &\, \left( 1 - \frac{T_1}{T_0} - 2 \eta^2 T_1 T_2 \right) \ketbra{0}{0} +
  \left( \frac{T_1}{T_0} + 2 \eta^2 T_1 T_2 \right) \ketbra{1}{1}
  \nonumber \\ &\,
  -i \eta T_2 e^{i(\omega t - \varphi)} \ketbra{0}{1}
  +i \eta T_2 e^{-i(\omega t - \varphi)} \ketbra{1}{0}.
\end{align}

Before performing an analysis using the CFI, let us provide an intuitive argument based on the counting of the expected numbers of signal $S$ and background $B$. For the case of the experiment counting the number of excited qubits, the expected numbers of signal and background are given by the $\eta$-dependent and $\eta$-independent parts of the $\ketbra{1}{1}$ term in the density matrix, respectively. Thus, for the case of the Pauli-$Z$ measurement, the significance parameter after $\nu$ repetitions of the measurement cycle, given by the ratio $S/\sqrt{B}$, is estimated as
\begin{align}
  \left[ \frac{S}{\sqrt{B}} \right]^{(Z)} \sim 
  \sqrt{\nu}\,  \eta^2 \sqrt{T_0 T_1} T_2,
\end{align}
where an overall factor of $O(1)$, which is irrelevant to the following argument, has been neglected. Here we have used the expectation that $S$ and $B$ both linearly grow with $\nu$.

For the case of the Pauli-$Y$ measurement, the expectation value is $\ev{Y} \simeq 2 \eta T_2 \cos (\omega t - \varphi)$ while the variance is $\ev{Y^2}-\ev{Y}^2\simeq 1$ for a single measurement. Thus, after $\nu$ repetitions of the measurement cycle, the expected significance is
\begin{align}
  \left[ \frac{S}{\sqrt{B}} \right]^{(Y)} \sim 
  \sqrt{\nu}\, \eta T_2 \cos (\omega t - \varphi).
\end{align}
If the phase $\varphi$ were known and time-independent, one could optimize the exposure time $t$ to achieve $[S/\sqrt{B}]^{(Y)} \sim \sqrt{\nu}\, \eta T_2$. In such a case, the Pauli-$Y$ measurement would yield better sensitivity than the Pauli-$Z$ measurement.

The advantage of the Pauli-$Y$ measurement can be discussed more precisely using the CFI. If the phase $\varphi$ were known and time-independent, the CFIs with respect to $\eta$ for the Pauli-$Z$ and Pauli-$Y$ measurements would be
\begin{align}
  F_{\rm C}^{(Z)} \simeq 16 \nu \eta^2 T_0 T_1 T_2^2,
\end{align}
and
\begin{align}
  F_{\rm C}^{(Y)} \simeq 4 \nu T_2^2 \cos^2(\omega t - \varphi),
  \label{F_C^Y}
\end{align}
respectively. Again, the Pauli-$Y$ measurement yields better sensitivity than the Pauli-$Z$ measurement for $T_0, T_1\ll \eta^{-1}$. Note that the QFI is $F_{\rm Q}\simeq 4 \nu T_2^2$, and the Pauli-$Y$ measurement saturates the QCRB if we can optimize the exposure time $t$.

The advantage of the Pauli-$Y$ measurement, however, is unlikely to be realized in a realistic DM detection experiment. This is because the CFI for the Pauli-$Y$ measurement given in Eq.~\eqref{F_C^Y} is valid only when the entire measurement, i.e., all $\nu$ independent measurement cycles, is completed within the DM coherence time. If the total measurement duration exceeds $\tau_{\rm DM}$, as is likely to occur in an actual experimental setup, the unknown phase $\varphi$ becomes randomized over the course of the measurement, and the density matrix obtained after each exposure varies accordingly from one cycle to the next. In particular, under this phase randomization, the expectation value of the Pauli-$Y$ measurement outcome averages to zero. By contrast, the expectation value of the Pauli-$Z$ measurement outcome retains a definite sign on top of the noise contribution, and therefore remains informative even in this regime. Consequently, an analysis based on the Pauli-$Y$ measurement, relying on an observable that is linear in $\eta$, cannot be relied upon for DM detection, whereas the Pauli-$Z$ measurement remains a viable option.

\section{Challenges to Heisenberg Limit}
\label{sec:HL}
\setcounter{equation}{0}
\setcounter{figure}{0}
\setcounter{footnote}{1}

As the number $N$ of copies of the quantum system increases, it becomes desirable to employ a measurement protocol that attains the HL to maximize the sensitivity to the unknown parameter $\theta$. However, as mentioned earlier, the HL is often unachievable in practice. In this section, we examine the basic properties of the QFI to clarify the challenges in achieving the HL.

\subsection{Bound on the QFI}

Let us consider the case where we have $N$ copies of a quantum system; the initial density matrix of the total system is taken to be $\rho^{\rm (init)}$, which does not depend on $\theta$. For simplicity, we assume that the initial state is pure:
\begin{align}
  \rho^{\rm (init)} = \ketbra{\Psi^{\rm (init)}}{\Psi^{\rm (init)}}. 
  \label{rhoinit=pure}
\end{align}
The generalization to the case with a mixed initial state is straightforward, as we can always purify $\rho^{\rm (init)}$.

Hereafter, we focus on the scenario where there is no interaction between the quantum systems, and thus each quantum system evolves independently. Under this assumption, the density matrix after the evolution can be written as
\begin{align}
  \rho = \sum_{\vec{a}} K_{\vec{a}} \rho^{\rm (init)} K_{\vec{a}}^\dagger,
\end{align}
where the Kraus operators for the total system are given by
\begin{align}
  K_{\vec{a}} = \bigotimes_{n=1}^N K_{n,a_n}.
\end{align}
Here, $K_{n,a}$ is the Kraus operator acting on the $n$-th subsystem, and $\vec{a} \equiv (a_1, a_2, \cdots, a_N)$.

A useful upper bound on the QFI can be derived by means of the so-called channel extension method \cite{Escher:2011eff, Demkowicz-Dobrzanski:2012dnq, Kolodynski:2013lka}. The first step of this method is to extend the Hilbert space as follows:
\begin{align}
  \mathcal{H}_{\rm S} ~~\rightarrow~~
  \mathcal{H}_{\rm SE} \equiv \mathcal{H}_{\rm S} \otimes \mathcal{H}_{\rm E},
\end{align}
where $\mathcal{H}_{\rm S}$ is the Hilbert space of the original system and $\mathcal{H}_{\rm E}$ is that of an auxiliary environment. We then define
\begin{align}
  \tilde{\rho} \equiv \sum_{\vec{a},\vec{b}} K_{\vec{a}} \rho^{\rm (init)} K_{\vec{b}}^\dagger \otimes \ketbra{e_{\vec{a}}}{e_{\vec{b}}},
\end{align}
where $\ket{e_{\vec{a}}}$ are orthonormal states in the environment. The density matrix of the system can be recovered by tracing out the environment:
\begin{align}
  \rho = \mathrm{tr}_{\rm E} \left( \tilde{\rho} \right),
\end{align}
where $\mathrm{tr}_{\rm E}$ denotes the partial trace over $\mathcal{H}_{\rm E}$. Upon substituting Eq.~\eqref{rhoinit=pure}, one finds that $\tilde{\rho}$ describes a pure state:
\begin{align}
  \tilde{\rho} = \ketbra{\tilde{\Psi}}{\tilde{\Psi}},
\end{align}
where
\begin{align}
  \ket{\tilde{\Psi}} = \sum_{\vec{a}}
  K_{\vec{a}} \ket{\Psi^{\rm (init)}} \otimes \ket{e_{\vec{a}}}.
\end{align}

Next, we compare the QFI of $\rho$ with that of $\tilde{\rho}$, hypothetically allowing measurements on the enlarged Hilbert space $\mathcal{H}_{\rm SE}$; the CFI and QFI associated with $\tilde{\rho}$ are denoted as $\tilde{F}_{\rm C}$ and $\tilde{F}_{\rm Q}$, respectively. For any POVM $\{E_a\}$ on $\mathcal{H}_{\rm S}$, one can always construct a corresponding POVM $\{\tilde{E}_a\}$ on $\mathcal{H}_{\rm SE}$ defined by
\begin{align}
  \tilde{E}_a \equiv E_a \otimes \mathds{I}_{\rm E},
\end{align}
where $\mathds{I}_{\rm E}$ denotes the identity operator on $\mathcal{H}_{\rm E}$. This construction ensures that any measurement outcome achievable on $\mathcal{H}_{\rm S}$ is also accessible on $\mathcal{H}_{\rm SE}$, and consequently
\begin{align}
  F_{\rm Q} = \sup_{\{E_a\}} F_{\rm C} \leq
  \sup_{\{\tilde{E}_a\}} \tilde{F}_{\rm C} = \tilde{F}_{\rm Q}.
  \label{FQmax}
\end{align}
Applying the expression of the QFI for pure states given in Eq.~\eqref{FQpure}, we obtain
\begin{align}
  F_{\rm Q} \leq
  \tilde{F}_{\rm Q} =
  4 \left(
  \braket{\partial_\theta \tilde{\Psi}}{\partial_\theta \tilde{\Psi}}
  - \left| \braket{\tilde{\Psi}}{\partial_\theta \tilde{\Psi}} \right|^2
  \right).
  \label{tilde-FQ}
\end{align}
This inequality establishes a useful upper bound on the QFI for the system under consideration, and is employed in the following discussion. For the calculation, the following equations are useful:
\begin{align}
  &\braket{\partial_\theta \tilde{\Psi}}{\partial_\theta \tilde{\Psi}}
  = 
  \sum_{n=1}^N
  \mel{\Psi^{\rm (init)}}
  {\left[
      \mathds{I}_1\otimes \cdots \otimes
      \sum_{a_n} ( \partial_\theta K_{n,a_n}^\dagger )
        ( \partial_\theta K_{n,a_n} ) \otimes \cdots
    \right]
  }
  {\Psi^{\rm (init)}}
  \nonumber \\ &\, \quad +
  \sum_{n\neq n'}
  \mel{\Psi^{\rm (init)}}
      {\left[
          \cdots \otimes
          \sum_{a_n} K_{n,a_n}^\dagger
          ( \partial_\theta K_{n,a_n} ) \otimes \cdots \otimes
          \sum_{a_{n'}} ( \partial_\theta K_{n',a_{n'}}^\dagger )
          K_{n',a_{n'}}
          \otimes \cdots
          \right]
      }
      {\Psi^{\rm (init)}},
      \label{<Psi'|Psi'>}
\end{align}
and
\begin{align}
  \braket{\tilde{\Psi}}{\partial_\theta \tilde{\Psi}}
  = &\,
  \sum_{n=1}^N
  \mel{\Psi^{\rm (init)}}
      {\left[
          \mathds{I}_1\otimes \cdots \otimes
          \sum_{a_n} K_{n,a_n}^\dagger
          ( \partial_\theta K_{n,a_n} ) \otimes \cdots
          \right]
      }
      {\Psi^{\rm (init)}}.
      \label{<Psi|Psi'>}
\end{align}
In the above equations, identity operators on each subsystem are, for the most part, denoted by ``$\cdots$'' for brevity. We emphasize that, to achieve the HL, $\sum_{a} K_{n,a}^\dagger( \partial_\theta K_{n,a} )$ must be non-vanishing; otherwise, the expression for $\tilde{F}_{\rm Q}$ contains only $O(N)$ terms, yielding at most the SQL irrespective of the initial state.

When applying the above argument, it is important to remember that the choice of the set of Kraus operators possesses a unitary ambiguity; the same physical process can be equivalently described by different sets of Kraus operators related through unitary transformations (see Section \ref{sec:environments}). Notably, $\tilde{F}_{\rm Q}$ generally depends on the choice of these Kraus operators. Thus, the most stringent upper bound on the QFI is obtained by optimizing over possible choices of the Kraus operators via such unitary transformations.

\subsection{Case without Entanglement}

Based on the discussion above, one can argue that the HL cannot be achieved without entanglement. In the present setup, each quantum system evolves independently, and thus there is no entanglement if the initial state is unentangled.

In the absence of entanglement, the initial state is a product state, and we may write
\begin{align}
  \ket{\Psi^{\rm (init)}} = \bigotimes_{n=1}^{N} \ket{\psi_n}.
\end{align}
Then, using Eqs.\ \eqref{<Psi'|Psi'>} and \eqref{<Psi|Psi'>}, we obtain
\begin{align}
  \tilde{F}_{\rm Q}^{\rm (prod)} = 4 \sum_{n=1}^N \left[ \sum_{a_n}
    \mel{\psi_n}{( \partial_\theta K_{n,a_n}^\dagger ) ( \partial_\theta K_{n,a_n} )}{\psi_n}
    -
    \left| \sum_{a_n} \mel{\psi_n}{K_{n,a_n}^\dagger ( \partial_\theta K_{n,a_n} )}{\psi_n} \right|^2
    \right].
\end{align}
Because the expression above is a sum of $N$ terms, it scales at most linearly with $N$. This linear scaling clearly indicates that the HL cannot be attained in the absence of entanglement.

\subsection{A Model with Simple Noise}
\label{subsec:simplemodel}

Cases involving entanglement are of particular interest because quantum coherence may enhance the signal sensitivity, as we will see in the following section. However, it is also well known that achieving the HL is challenging in the presence of noise. Here, we illustrate this difficulty by considering a specific class of noise. Since our primary focus is on using qubits for DM detection, we study a simple model of $N$ qubits that is directly relevant to the DM-detection scenario of interest. We adopt a model in which local Markovian noise acts independently on each qubit. We will show that, in the setting adopted here, the QFI cannot attain the HL.

We take the Hamiltonian to be
\begin{align}
  H = -\eta \sum_{n=1}^N X_n,
  \label{H0(N-qubits)}
\end{align}
where $X_n$ is the Pauli-$X$ operator acting on the $n$-th qubit, and $\eta$ is the unknown parameter to be estimated. This Hamiltonian can be viewed as an effective description of an $N$-qubit system coupled to oscillating DM in the resonance limit with $\varphi=0$. We further assume that the dominant noise source is described by local Markovian dynamics, with jump operators given by the Pauli-$X$, $Y$, and $Z$ operators, acting independently on each qubit. 

Denoting the density matrix of the total system by $\rho$, we adopt the following Lindblad master equation:
\begin{align}
  \partial_t \rho = -i [H, \rho]
  + \frac{1}{2} \gamma_X \sum_{n=1}^N \left( X_n \rho X_n - \rho \right)
  + \frac{1}{2} \gamma_Y \sum_{n=1}^N \left( Y_n \rho Y_n - \rho \right)
  + \frac{1}{2} \gamma_Z \sum_{n=1}^N \left( Z_n \rho Z_n - \rho \right),
  \label{mastereq_simple}
\end{align}
where $\gamma_X$, $\gamma_Y$, and $\gamma_Z$ are real and non-negative constants, interpreted as noise rates.

The purpose of the following argument is to show that, in the present model, the HL cannot be achieved regardless of the choice of the initial state, even when entanglement is allowed \cite{Huelga:1997mw, Giovannetti:2011chh, Escher:2011eff, Demkowicz-Dobrzanski:2012dnq, Kolodynski:2013lka}. To see this, we first derive a set of Kraus operators which describes the evolution of the system. Since each qubit evolves independently in the present model, the Kraus operators for the total system can be written as tensor products of Kraus operators describing the evolution of each qubit.

The Kraus operators for the evolution of each qubit can be obtained by solving the following Lindblad equation:
\begin{align}
  \partial_t \rho_n = i \eta [X_n, \rho_n]
  + \frac{1}{2} \gamma_X \left( X_n \rho_n X_n - \rho_n \right)
  + \frac{1}{2} \gamma_Y \left( Y_n \rho_n Y_n - \rho_n \right)
  + \frac{1}{2} \gamma_Z \left( Z_n \rho_n Z_n - \rho_n \right),
\end{align}
where $\rho_n$ is the (auxiliary) density matrix used to derive the Kraus operators. Writing $\rho_n$ as in Eq.\ \eqref{bloch_rep_rho}, the evolution equations for the components of the Bloch vector are given by
\begin{align}
  \partial_t \rho_X = &\, -(\gamma_Y + \gamma_Z) \rho_X,
  \label{dot-rhox}
  \\
  \partial_t \rho_Y = &\, 2 \eta \rho_Z - (\gamma_X + \gamma_Z) \rho_Y,
  \label{dot-rhoy}
  \\
  \partial_t \rho_Z = &\, -2 \eta \rho_Y - (\gamma_X + \gamma_Y) \rho_Z.
  \label{dot-rhoz}
\end{align}
The solution is given by
\begin{align}
  \rho_X (t) = &\, e^{-\Gamma_X t} \rho_X (0),
  \\
  \rho_Y (t) = &\, e^{-\Gamma_{YZ} t}
  \left[
    \left\{ \cos 2\eta' t + \frac{(\gamma_Y-\gamma_Z) \sin 2\eta' t}{4\eta'} \right\} \rho_Y (0) + \frac{\eta\sin 2\eta' t}{\eta'} \rho_Z (0) \right],
  \\
  \rho_Z (t) = &\, e^{-\Gamma_{YZ} t}
  \left[
    - \frac{\eta\sin 2\eta' t}{\eta'} \rho_Y (0)
    + 
    \left\{ \cos 2\eta' t - \frac{(\gamma_Y-\gamma_Z) \sin 2\eta' t}{4\eta'} \right\} \rho_Z (0) \right],
\end{align}
with
\begin{align}
  \eta' \equiv \sqrt{\eta^2 - \frac{1}{16} (\gamma_Y-\gamma_Z)^2},
\end{align}
and
\begin{align}
  \Gamma_X \equiv &\, \gamma_Y + \gamma_Z,
  \\
  \Gamma_{YZ} \equiv &\, \gamma_X + \frac{1}{2} \left( \gamma_Y + \gamma_Z \right).
\end{align}

A complete treatment of the present model is rather involved and beyond the scope of this article. Nevertheless, by focusing on the weak-noise regime, $\gamma_{X,Y,Z} \ll \eta$, we can obtain useful insight into how noise affects the achievable sensitivity. In particular, we will see that even very weak noise is sufficient to prevent one from attaining the HL. In the limit $\gamma_{X,Y,Z} \ll \eta$, for which $\eta'\simeq \eta$, we obtain
\begin{align}
  \rho_Y (t) \simeq &\, e^{-\Gamma_{YZ} t}
  \left[ \cos 2\eta t \rho_Y (0) + \sin 2 \eta t \rho_Z (0) \right],
  \label{rhoY_gen}
  \\
  \rho_Z (t) \simeq &\, e^{-\Gamma_{YZ} t}
  \left[ -\sin 2\eta t \rho_Y (0) + \cos 2 \eta t \rho_Z (0) \right].
  \label{rhoZ_gen}
\end{align}
We note that when $\gamma_Y=\gamma_Z$ the system can be solved without assuming $\gamma_{X,Y,Z} \ll \eta$; in that case, Eqs.\ \eqref{rhoY_gen} and \eqref{rhoZ_gen} are exact solutions of the Lindblad equation. Therefore, in such cases, the arguments below hold without taking the weak-noise limit. These include a model with only bit-flip noise represented by the Pauli-$X$ operator (i.e., $\gamma_Y=\gamma_Z=0$) and a model with only depolarization noise (i.e., $\gamma_X=\gamma_Y=\gamma_Z$).\footnote
{Another case of interest is the model discussed in Section \ref{subsec:qubitwithnoise}, which describes a model with amplitude damping and spontaneous excitation, characterized by the operators $a=\ketbra{0}{1}$ and $a^\dagger=\ketbra{1}{0}$, respectively, together with phase damping. In this case, the Lindblad equation is given by
\begin{align*}
  \partial_t \rho_n = &\,i \eta [X_n, \rho_n]
  + \gamma_0
  \left[ a_n^\dagger \rho_n a_n - \frac{1}{2} (a_n a_n^\dagger \rho_n + \rho_n a_n a_n^\dagger) \right]
  + \gamma_1
  \left[ a_n \rho_n a_n^\dagger - \frac{1}{2} (a_n^\dagger a_n \rho_n + \rho_n a_n^\dagger a_n) \right]
  \\ &\,
  + \frac{1}{2} \gamma_2 \left( Z_n \rho_n Z_n - \rho_n \right),
\end{align*}
with $\gamma_1$ and $\gamma_0$ denoting the de-excitation and excitation rates, respectively. The corresponding evolution equations for the Bloch-vector components are
\begin{align*}
  \partial_t \rho_X = -(\bar{\gamma} + \gamma_2) \rho_X,~~~
  \partial_t \rho_Y = 2 \eta \rho_Z - (\bar{\gamma}+\gamma_2) \rho_Y,~~~
  \partial_t \rho_Z = -2 \eta \rho_Y - 2\bar{\gamma} \rho_Z +(\gamma_1-\gamma_0),
\end{align*}
where $\bar{\gamma}\equiv\frac{1}{2}(\gamma_1+\gamma_0)$. Noting that the above differential equations admit the steady-state solution
\begin{align*}
  \rho_X^{\rm (s.s.)} = 0,~~~
  \rho_Y^{\rm (s.s.)} = \frac{\eta}{2\eta^2+\bar{\gamma}(\bar{\gamma}+\gamma_2)} (\gamma_1-\gamma_0),~~~
  \rho_Z^{\rm (s.s.)} = \frac{\bar{\gamma}+\gamma_2}{4\eta^2+2\bar{\gamma}(\bar{\gamma}+\gamma_2)} (\gamma_1-\gamma_0),
\end{align*}
one can write the general solution as
\begin{align*}
  (\rho_X, \rho_Y, \rho_Z) =
  (\rho'_X, \rho'_Y, \rho'_Z) +
  (\rho_X^{\rm (s.s.)}, \rho_Y^{\rm (s.s.)}, \rho_Z^{\rm (s.s.)}),
\end{align*}
where $(\rho'_X, \rho'_Y, \rho'_Z)$ satisfies the corresponding homogeneous equation. In the weak-noise regime, the steady-state contribution is suppressed, and we may therefore approximate the Bloch vector by $(\rho'_X, \rho'_Y, \rho'_Z)$. In such a case, the system may be well approximated by the one described by Eqs.\ \eqref{dot-rhox} -- \eqref{dot-rhoz} (with $\gamma_X=\gamma_Y=\bar{\gamma}$ and $\gamma_Z=\gamma_2$), and the same argument (as given in the main text) applies as well.
}

From these expressions, one can construct Kraus operators for the evolution of the $n$-th qubit as
\begin{align}
  K_{n,0} \equiv &\, \sqrt{f_0} e^{i\eta t X_n},
  \label{K_n0}
  \\
  K_{n,1} \equiv &\, \sqrt{f_1} e^{i\eta t X_n} X_n,
  \label{K_n1}
  \\
  K_{n,2} \equiv &\, \sqrt{f_2} e^{i\eta t X_n} Y_n,
  \\
  K_{n,3} \equiv &\, \sqrt{f_3} e^{i\eta t X_n} Z_n,
\end{align}
where
\begin{align}
  f_0 = &\, \frac{1+e^{-\Gamma_Xt}}{4} + \frac{e^{-\Gamma_{YZ} t}}{2},
  \\
  f_1 = &\, \frac{1+e^{-\Gamma_Xt}}{4} - \frac{e^{-\Gamma_{YZ} t}}{2},
  \\
  f_2 = &\, \frac{1-e^{-\Gamma_Xt}}{4},
  \\
  f_3 = &\, \frac{1-e^{-\Gamma_Xt}}{4}.
\end{align}
The single-qubit density matrix then evolves as
\begin{align}
  \rho_n (t) = \sum_{a=0}^3 K_{n,a} \rho_n (0) K_{n,a}^\dagger.
\end{align}

To derive an upper bound on the QFI, we employ the following unitary transformation of Kraus operators:
\begin{align}
  \begin{pmatrix}
    K_{n,0}^\prime \\ K_{n,1}^\prime \\ K_{n,2}^\prime \\ K_{n,3}^\prime
  \end{pmatrix}
  =
  \begin{pmatrix}
    \cos \mu \eta t & i \sin \mu \eta t & 0 & 0\\
    i \sin \mu \eta t & \cos \mu \eta t & 0 & 0\\
    0 & 0 & \cos\eta t & \sin\eta t \\
    0 & 0 & -\sin\eta t & \cos\eta t
  \end{pmatrix}
  \begin{pmatrix}
    K_{n,0} \\ K_{n,1} \\ K_{n,2} \\ K_{n,3}
  \end{pmatrix},
\end{align}
with
\begin{align}
  \mu = -\frac{f_0+f_1}{2 \sqrt{f_0 f_1}}
  = -
  \frac{1+e^{-\Gamma_Xt}}
  {\sqrt{(1+e^{-\Gamma_Xt})^2-4e^{-2\Gamma_{YZ}t}}}.
\end{align}
With this transformation, one can verify that $\sum_{a=0}^3 K_{n,a}^{\prime\, \dagger} ( \partial_\eta K_{n,a}^\prime )=0$, which in turn implies that the HL is not achievable in the present case. 

For this choice of the Kraus operators, we find
\begin{align}
  \sum_{a=0}^3 ( \partial_\eta K_{n,a}^{\prime \dagger}) ( \partial_\eta K_{n,a}^\prime ) = \frac{2 (1+e^{-\Gamma_X t}) e^{-2\Gamma_{YZ} t}}
  {(1+e^{-\Gamma_Xt})^2-4e^{-2\Gamma_{YZ}t}} t^2 \mathds{I}.
\end{align}
Thus, for a system of $N$ qubits, the bound in Eq.\ \eqref{tilde-FQ} becomes
\begin{align}
  F_{\rm Q} \leq
  \frac{8 (1+e^{-\Gamma_X t}) e^{-2\Gamma_{YZ} t}}
  {(1+e^{-\Gamma_Xt})^2-4e^{-2\Gamma_{YZ}t}}
  N t^2,
  \label{FQbound_general}
\end{align}
which holds irrespective of the choice of the initial state. From the above expression, the maximal achievable QFI (after optimizing over $t$) in the present model is found to be of $O(N\gamma^{-2})$, with $\gamma$ being a typical size of the noise rates. 

So far, we have investigated simple cases with a local Markovian noise model. Bounds on the QFI for major noise models can be found in Ref.\ \cite{Kolodynski:2013lka}. In general, in the presence of noise and/or decoherence, achieving the HL is difficult even when entangled states are available. In contrast, in the absence of noise, the HL can be attained with an appropriate choice of the initial state \cite{Bollinger:1996ces, Giovannetti:2004cas, Giovannetti:2005bdr}.

We also note that the QFI scaling $O(N\gamma^{-2})$ is consistent with the optimized CFI obtainable via individual measurements on each qubit; this optimum is achieved when the exposure time is chosen as $t\sim \gamma^{-1}$. Thus, in the simple scenarios considered here, individual qubit measurements, without exploiting entanglement, already achieve the same scaling as the QFI bound for estimating $\eta$.

Importantly, however, this does not imply that entanglement is never advantageous for DM detection. In particular, realistic DM detection requires careful treatment of the decoherence of the DM oscillations, an effect that has not yet been incorporated in the present discussion. Quantum error correction is another possible avenue to mitigate noise. A more detailed investigation will be presented in the next section.

\section{Entangled States for Dark Matter Detection}
\label{sec:ghz}
\setcounter{equation}{0}
\setcounter{figure}{0}
\setcounter{footnote}{1}

So far, we have not discussed in detail the use of entangled states for DM detection. Entangled states can, however, potentially enhance the signal and/or reduce the noise, and therefore the advantages of employing entanglement are of significant interest. In particular, in a future regime where high-quality qubits and high-fidelity gate operations are available, the use of entangled states, together with suitable quantum operations on the qubit system used for DM detection, may provide a substantial improvement in the sensitivity of DM detection experiments.

In this section, we explore these possibilities. Specifically, we investigate whether the HL can be achieved in an $N$-qubit system. To be concrete, we focus on the so-called Greenberger-Horne-Zeilinger (GHZ) state \cite{Greenberger:1989tfe, Greenberger:1990uox}, adopting a model described by the following Hamiltonian:
\begin{align}
  H = -\eta \sum_{n=1}^N X_n,
  \label{HX_Nqubits}
\end{align}
where $\eta$ is treated as the unknown parameter to be estimated. As mentioned previously, this Hamiltonian can be regarded as an effective interaction-picture description of a qubit coupled to DM in the resonance limit, under the assumption that the phase $\varphi$ is fixed (see Eq.~\eqref{H_I(resonance)}). When $\varphi$ is fixed, the effective Hamiltonian can be brought into the form of Eq.~\eqref{HX_Nqubits} by an appropriate redefinition (i.e., a rotation) of the Pauli-$X$ and $Y$ operators. The effects of randomization of $\varphi$ will be discussed later. In DM-motivated applications, $\eta$ is expected to be extremely small, and our goal is to determine whether measurements can resolve a non-zero value of $\eta$.

For many of the relevant protocols, various quantum (i.e., unitary) operations on the qubit system are required, as we discuss below. In the current analysis, we simply assume that any unitary operation is possible on the quantum system of interest, and we neglect the associated operational costs.

\subsection{GHZ State: Case without Noise}

We first consider the idealized limit in which noise and decoherence are absent. In this limit, the quantum state evolves unitarily according to the Schr\"odinger equation with the Hamiltonian in Eq.~\eqref{HX_Nqubits}. Since there are no qubit-qubit interactions, the system evolution factorizes as
\begin{align}
  \ket{\Psi (t)} = U_t^{\otimes N} \ket{\Psi (0)},
\end{align}
where $U_t$ is the single-qubit time-evolution operator,
\begin{align}
  U_t = e^{i\eta t X} =
  \begin{pmatrix}
    \cos\eta t & i \sin \eta t \\
    i \sin \eta t & \cos \eta t
  \end{pmatrix}.
\end{align}
It is useful to note that the eigenstates of $U_t$ are
\begin{align}
  \ket{\pm} \equiv \frac{1}{\sqrt{2}}
  \left( \ket{0} \pm \ket{1} \right),
\end{align}
which satisfy
\begin{align}
  U_t \ket{\pm} = e^{\pm i\delta} \ket{\pm},
\end{align}
with
\begin{align}
  \delta = \eta t.
\end{align}
For the single-qubit case ($N=1$), the state can be written as
\begin{align}
  \ket{\psi (t)} = \psi_0 (t) \ket{0} + \psi_1 (t) \ket{1},
\end{align}
and the unitary evolution is explicitly
\begin{align}
  \begin{pmatrix}
    \psi_0 (t) \\ \psi_1 (t)
  \end{pmatrix}
  =
  \begin{pmatrix}
    \cos \delta & i \sin \delta \\
    i \sin \delta & \cos \delta
  \end{pmatrix}
  \begin{pmatrix}
    \psi_0 (0) \\ \psi_1 (0)
  \end{pmatrix}.
\end{align}
Therefore, if the initial state is $\ket{\psi (0)}=\ket{0}$, then the excitation probability at time $t$ is approximately $\delta^2$ (assuming $\delta\ll 1$).

A simple strategy to probe a non-zero $\eta$ is as follows: (i) prepare $N$ qubits, each initialized in $\ket{0}$, and hence the state is initially a product state $\ket{0}^{\otimes N}$, (ii) let them evolve unitarily, and (iii) perform independent projective measurements on each qubit and count the number of excitations. In this scheme, the expected number of signal events (i.e., excited qubits) is
\begin{align}
  S^{\rm (prod)} \simeq N \delta^2,
\end{align}
in the small-$\delta$ limit.

In the absence of noise, the GHZ state provides a substantial improvement in sensitivity. Let us assume that we prepare, at $t=0$, a GHZ state
\begin{align}
  \ket{\rm GHZ} \equiv \frac{1}{\sqrt{2}} \left( \ket{+}^{\otimes N} + \ket{-}^{\otimes N} \right).
\end{align}
Since $\ket{\pm}$ are eigenstates of $U_t$ with eigenvalues given by pure phases, the time evolution is
\begin{align}
  \ket{\rm GHZ} \xrightarrow{~U_t^{\otimes N}~} \frac{1}{\sqrt{2}} \left( e^{iN\delta} \ket{+}^{\otimes N} + e^{-iN\delta} \ket{-}^{\otimes N} \right),
\end{align}
showing that the phase accumulates coherently across all $N$ qubits. This coherent phase accumulation is the essential mechanism behind HL scaling.

\begin{figure}[t]
  \begin{center}
    \begin{quantikz}
      \lstick[1]{$\ket{0}$} \slice[style=black]{$t_{\rm i}$}
      & \gate{H} \slice[style=black]{$t_{\rm 0}$} & \ctrl{1} & \ctrl{2} & \ctrl{4} & \ctrl{5}
      \slice[style=black]{$t_{\rm 1}$} & \gate[1][1cm]{U_t} \slice[style=black]{$t_{\rm 2}$} &
      \ctrl{5} & \ctrl{4} & \ctrl{2} & \ctrl{1}  \slice[style=black]{$t_{\rm 3}$} &
       \gate{H} \slice[style=black]{$t_{\rm f}$} &\meter{}
      \\
      \lstick[5]{$\ket{0}^{\otimes N}$}
      & \gate{H} & \gate{Z} &  &  &
      & \gate[1][1cm]{U_t} &
      &  &  & \gate{Z} &  &
      \\
      & \gate{H} &  & \gate{Z} &  &
      & \gate[1][1cm]{U_t} &
      &  &\gate{Z}  &  &  &
      \\
      \wave&&&&&&&&&&&&&
      \\
      & \gate{H} &  &  & \gate{Z} &
      & \gate[1][1cm]{U_t} &
      & \gate{Z} &  &  &  &
      \\
      & \gate{H} &  &  &  & \gate{Z}
      & \gate[1][1cm]{U_t} &
      \gate{Z} &  &  &  &  &
    \end{quantikz}
  \end{center}
  \caption{Quantum circuit for achieving HL scaling. The gate labeled $H$ denotes the Hadamard gate, while the symbols ``$\bullet$'' and ``$Z$'' connected by a line represent the Controlled-$Z$ gate. The block $U_t$ denotes the unitary time evolution.}
  \label{fig:quantumcircuit_CZ}
\end{figure}
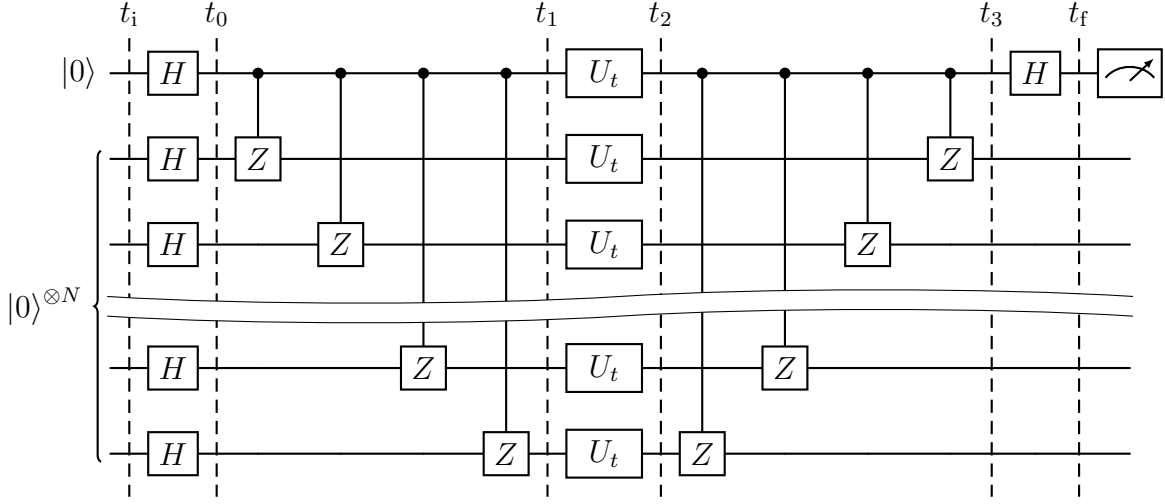

In practice, even when starting from a simple product state such as $\ket{\Psi(0)} = \ket{0}^{\otimes N}$, one can design a suitable quantum circuit to achieve HL sensitivity. To illustrate this, we introduce one ancilla qubit together with $N$ sensor qubits, and we assume that the ancilla does not undergo the time evolution. The relevant states are of the form
\begin{align*}
  \ket{\rm ancilla} \otimes \left[ \bigotimes_{n=1}^N \ket{\psi_n} \right].
\end{align*}
An example of a circuit of interest is shown in Fig.~\ref{fig:quantumcircuit_CZ}. It includes the Hadamard gate $H$ (not to be confused with the Hamiltonian),
\begin{align}
  \begin{pmatrix}
    \ket{0} \\ \ket{1}
  \end{pmatrix}
  \xrightarrow{~H~}
  \frac{1}{\sqrt{2}}
  \begin{pmatrix}
    1 & 1 \\ 1 & -1
  \end{pmatrix}
  \begin{pmatrix}
    \ket{0} \\ \ket{1}
  \end{pmatrix},
\end{align}
and the controlled-$Z$ (C$Z$) gate,
\begin{align}
  \mbox{C}Z \equiv \ketbra{0}{0} \otimes \mathds{I} + \ketbra{1}{1} \otimes Z,
\end{align}
where the first qubit is the control and the second is the target. In this example, starting from $\ket{\Psi (0)} = \ket{0}\otimes\ket{0}^{\otimes N}$, the state evolves through the circuit steps as
\begin{align}
  \ket{ \Psi (t_0) } = &\,
  \frac{1}{\sqrt{2}} \ket{0} \otimes \ket{+}^{\otimes N}
  + \frac{1}{\sqrt{2}} \ket{1} \otimes \ket{+}^{\otimes N},
  \\
  \ket{ \Psi (t_1) } = &\,
  \frac{1}{\sqrt{2}} \ket{0} \otimes \ket{+}^{\otimes N}
  + \frac{1}{\sqrt{2}} \ket{1} \otimes \ket{-}^{\otimes N},
  \\
  \ket{ \Psi (t_2) } = &\,
  \frac{1}{\sqrt{2}} e^{iN\delta} \ket{0} \otimes \ket{+}^{\otimes N}
  + \frac{1}{\sqrt{2}} e^{-iN\delta} \ket{1} \otimes \ket{-}^{\otimes N},
  \\
  \ket{ \Psi (t_f) } = &\,
  \left( \cos N\delta \ket{0} + i \sin N\delta \ket{1} \right) \otimes \ket{+}^{\otimes N},
\end{align}
where $\delta=\eta(t_2-t_1)$, and we assume that $t_1-t_{\rm i}$ and $t_{\rm f}-t_2$ are negligibly short. Note that $\ket{ \Psi (t_1) }$ is a maximally entangled GHZ state.

The probability to observe an excitation of the ancilla qubit is then
\begin{align}
  p_{0\rightarrow 1}= \sin^2 (N \delta) \simeq N^2 \delta^2,
\end{align}
where we assume $N\delta\ll 1$. Accordingly, when using the GHZ protocol, the expected signal is estimated as
\begin{align}
  S^{\rm (GHZ)} \simeq N^2 \delta^2.
\end{align}
In addition, in this case, the QFI is given by
\begin{align}
  F_{\rm Q}^{\rm (GHZ)} (t) = 4 N^2 t^2,
\end{align}
thereby achieving the HL.

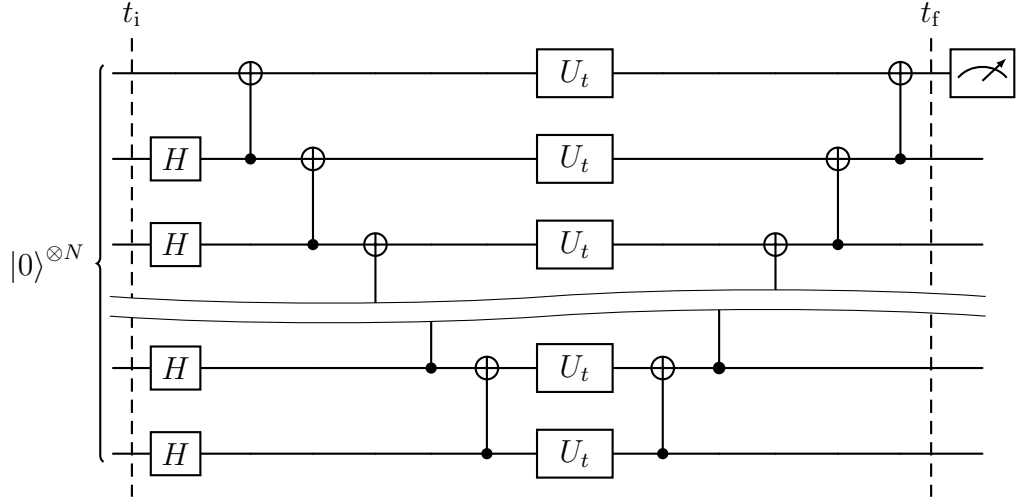
\begin{figure}[t]
  \begin{center}
    \begin{quantikz}
      \lstick[6]{$\ket{0}^{\otimes N}$} \slice[style=black]{$t_{\rm i}$} && \targ{}  &  & &&  & \gate[1][1cm]{U_t} & && &   &  \targ{ } \slice[style=black]{$t_{\rm f}$} &\meter{} 
      \\
      &\gate{H}& \ctrl{-1} & \targ{}& & &&\gate[1][1cm]{U_t} & & & & \targ{} & \ctrl{-1}&
      \\
      &\gate{H}& &\ctrl{-1} & \targ{}\vqw{1} & &&\gate[1][1cm]{U_t} & & &  \targ{}\vqw{1} & \ctrl{-1} &&
      \\
      \wave&&&&&&&&&&&&&
      \\
      &\gate{H}& &&& \ctrl{-1} & \targ{} & \gate[1][1cm]{U_t}&\targ{}& \ctrl{-1} \ & & & & 
      \\
      &\gate{H}& && && \ctrl{-1} & \gate[1][1cm]{U_t} &\ctrl{-1} & & & & &
    \end{quantikz}
  \end{center}
  \caption{Quantum circuit for HL. The gate with $H$ represents the Hadamard gate, while that with ``$\bullet$'' and ``$\oplus$'' connected by the line is the Controlled-NOT (CNOT) gate (where ``$\bullet$'' is the control qubit), which is defined as $\mbox{CNOT} \equiv \ketbra{0}{0} \otimes \mathds{I} + \ketbra{1}{1} \otimes X$. The $U_t$ represents the unitary time evolution.}
  \label{fig:quantumcircuit_CNOT}
\end{figure}

So far, we have focused on the circuit shown in Fig.\ \ref{fig:quantumcircuit_CZ}. However, a circuit that realizes HL scaling (in the noiseless case) is not unique. In particular, one can design a circuit based on one-dimensional nearest-neighbor interactions using $O(N)$ gates. An example is shown in Fig.\ \ref{fig:quantumcircuit_CNOT}; as shown in Ref.\ \cite{Chen:2023swh}, this circuit also achieves HL scaling.

\subsection{GHZ State: Case with Noise}

The discussion above applies to the noiseless case. In the presence of noise, by contrast, achieving HL scaling is generally difficult \cite{Chen:2023swh, Bodas:2025vff}. To understand the effect of noise on the GHZ state, it is instructive to study the model with bit-flip noise represented by the Pauli-$X$ noise; such a case can be analyzed by using the results given in Section~\ref{subsec:simplemodel} with $\gamma_Y=\gamma_Z=0$. Adopting the master equation given in Eq.~\eqref{mastereq_simple}, and using the Kraus operators given in Eqs.\ \eqref{K_n0} and \eqref{K_n1}, we study the evolution of the density matrix. With the initial condition
\begin{align}
  \rho^{\rm (GHZ)} (0) = \ketbra{\rm GHZ}{\rm GHZ},
\end{align}
the density matrix at time $t$ is given by
\begin{align}
  \rho^{\rm (GHZ)} (t) = &\,
  \frac{1}{2} \left(
    \ket{+}^{\otimes N}\bra{+}^{\otimes N} +
    \ket{-}^{\otimes N}\bra{-}^{\otimes N}
    \right.
    \nonumber \\ &\,
    \left.
    +
    e^{-N(\gamma_X-2i\eta)t}
    \ket{+}^{\otimes N}\bra{-}^{\otimes N} +
    e^{-N(\gamma_X+2i\eta)t}
    \ket{-}^{\otimes N}\bra{+}^{\otimes N}
    \right),
\end{align}
from which the QFI is evaluated as
\begin{align}
  F_{\rm Q}^{\rm (GHZ)} (t) = 4 N^2 t^2 e^{-2N\gamma_X t}.
\end{align}
One should note that the damping rate of the QFI for the case of GHZ initial state is enhanced by $N$, implying a strong suppression of the sensitivity for $t\gtrsim \frac{1}{N\gamma_X}$. By contrast, if the state is initially given by a non-entangled product state as
\begin{align}
  \rho^{\rm (prod)} (0) = \ket{0}^{\otimes N}\bra{0}^{\otimes N},
\end{align}
we obtain
\begin{align}
  \rho^{\rm (prod)} (t) = &\,
  \left[ \frac{1}{2} \left\{
    \mathbf{I} + e^{-\gamma_X t} \left( \cos 2\eta t Z + \sin 2 \eta t Y \right)
    \right\} \right]^{\otimes N},
\end{align}
and the corresponding QFI is
\begin{align}
  F_{\rm Q}^{\rm (prod)} (t) = 4 N t^2 e^{-2\gamma_X t}.
\end{align}
In this case, the damping rate of the QFI is of the order of that of a single qubit.

In realistic experimental situations, we are interested in the expected sensitivity for a given total experimental time $T$. Fixing the exposure time per measurement cycle to be $t$, the total number of measurement cycles is $\nu=\frac{T}{t}$. Since the QFI adds over independent repetitions, the total QFI is
\begin{align}
  \bar{F}_{\rm Q} (t) \equiv \nu F_{\rm Q} (t).
\end{align}
For the cases of direct-product and GHZ initial states, we obtain
\begin{align}
  \bar{F}_{\rm Q}^{\rm (prod)} (t) = 4 N T t e^{-2\gamma_X t},~~~
  \bar{F}_{\rm Q}^{\rm (GHZ)} (t) = 4 N^2 T t e^{-2N\gamma_X t}.
\end{align}
The optimal exposure time that maximizes $\bar{F}_{\rm Q}$ is
\begin{align}
  t_*^{\rm (prod)} = \frac{1}{2 \gamma_X},~~~
  t_*^{\rm (GHZ)} = \frac{1}{2N \gamma_X}.
\end{align}
Accordingly, the optimal value of the QFI is found to be
\begin{align}
  \bar{F}_{\rm Q}^{\rm (prod)} (t_*^{\rm (prod)}) = 
  \bar{F}_{\rm Q}^{\rm (GHZ)} (t_*^{\rm (GHZ)}) =
  \frac{2 N T}{e \gamma_X}.
\end{align}
This result shows that, if the exposure time $t$ can be freely optimized without external constraints, the GHZ state offers no advantage over the product state. In addition, the optimal QFI scales linearly with $N$, corresponding to SQL scaling rather than HL scaling.

The discussion so far, however, does not imply that entanglement is unimportant for DM detection in the presence of noise. Generically, without error correction, we expect that, in the presence of noise, $\bar{F}_{\rm Q}^{\rm (GHZ)}$ and $\bar{F}_{\rm Q}^{\rm (prod)}$ behave as
\begin{align}
  \bar{F}_{\rm Q}^{\rm (prod)} (t) \sim N T t e^{-\gamma t},~~~
  \bar{F}_{\rm Q}^{\rm (GHZ)} (t) \sim N^2 T t e^{-N\gamma t},
\end{align}
where factors of $O(1)$ are neglected. Here, we assume that the noise acts independently and identically on each qubit, with rate $\gamma$. (We denote by $\gamma$ a generic noise rate, which reduces to $\gamma_X$ in the bit-flip model considered above.) Thus, if the exposure time can be chosen arbitrarily, the optimal values of the QFIs in these two cases are of the same order.

For DM detection, however, we must take into account $\tau_{\rm DM}$, i.e., the coherence time of the DM field. Because the coherence of the DM oscillation is lost on time scales longer than $\sim\tau_{\rm DM}$, we cannot choose the exposure time $t$ to be longer than $\sim\tau_{\rm DM}$. Taking this constraint into account, the optimal exposure times become
\begin{align}
  t_*^{\rm (prod)} \sim \mbox{min} (\tau_{\rm Q}, \tau_{\rm DM}),~~~
  t_*^{\rm (GHZ)} \sim \mbox{min} (N^{-1}\tau_{\rm Q}, \tau_{\rm DM}),
\end{align}
with $\tau_{\rm Q}\sim\gamma^{-1}$ being the single-qubit coherence time. The expected sensitivity then depends on the relative sizes of $\tau_{\rm Q}$ and $\tau_{\rm DM}$:
\begin{itemize}
\item For $\tau_{\rm Q}\lesssim\tau_{\rm DM}$, the argument so far is unchanged.
\item For $\tau_{\rm DM}\lesssim\tau_{\rm Q}\lesssim N\tau_{\rm DM}$, we have $t_*^{\rm (prod)} \sim \tau_{\rm DM}$ and $t_*^{\rm (GHZ)} \sim N^{-1}\tau_{\rm Q}$, which gives
  \begin{align}
    \bar{F}_{\rm Q}^{\rm (prod)} (t_*^{\rm (prod)}) \sim N T \tau_{\rm DM},~~~
    \bar{F}_{\rm Q}^{\rm (GHZ)} (t_*^{\rm (GHZ)}) \sim N T \tau_{\rm Q}.
  \end{align}
\item For $\tau_{\rm Q}\gtrsim N\tau_{\rm DM}$, $t_*^{\rm (prod)}\sim t_*^{\rm (GHZ)} \sim\tau_{\rm DM}$, and hence
  \begin{align}
    \bar{F}_{\rm Q}^{\rm (prod)} (t_*^{\rm (prod)}) \sim N T \tau_{\rm DM},~~~
    \bar{F}_{\rm Q}^{\rm (GHZ)} (t_*^{\rm (GHZ)}) \sim N^2 T \tau_{\rm DM}.
  \end{align}
\end{itemize}
Thus, we expect that, when $\tau_{\rm Q}\gtrsim\tau_{\rm DM}$, an entangled state can enhance the maximal possible sensitivity. Even though the HL may be unlikely to be achieved, entanglement protocols can still be useful in DM detection to enhance the signal particularly with high-quality qubits.

\subsection{Other Entangled States}

So far, we have focused on the use of the GHZ state for DM detection. Of course, there are many other entangled states of qubits that could be exploited for this purpose. Examples include the W state \cite{Dur:2000zz} (or, more generally, Dicke states \cite{Dicke:1954zz}). Indeed, there is ongoing active discussion on employing various entangled qubit states for DM detection \cite{Chen:2025tgj, Bodas:2025vff, Fukuda:2025afi, He:2025ovo, Zheng:2025qgv, Bogorad:2026ggm} (see also Refs.\ \cite{HAYSTAC:2020kwv, Agrawal:2023umy, Freiman:2025tse} for quantum-enhanced states in cavity-based setups).

Another interesting possibility is the use of quantum error correction (QEC). In our discussion up to now, we have assumed a setup in which no quantum operations are applied to the system between state preparation and state readout. One may, however, incorporate QEC procedures during this interval, which could suppress a significant fraction of the noise \cite{Kessler:2014ozk, Zhou:2017kxr}.

Here, following Ref.\ \cite{Fukuda:2025afi}, we outline a recent proposal for using the W state in combination with a QEC protocol to enhance the sensitivity of wave-like DM detection. For an $N$-qubit system, the W state is defined as the equal superposition of all states with a single excitation:
\begin{align}
  \ket{{\rm W}} \equiv
  \frac{1}{\sqrt{N}} \sum_{n=1}^N X_n \ket{0}^{\otimes N}
  = \frac{1}{\sqrt{N}}
  \left(
  \ket{100 \cdots 0} + \ket{010 \cdots 0} + \ket{001 \cdots 0} + \cdots +
  \ket{000 \cdots 1} \right).
\end{align}
Here we adopt the bit-string notation for the computational basis, $\ket{I_1I_2I_3\cdots I_N}\equiv\ket{I_1}\otimes\ket{I_2}\otimes\ket{I_3}\otimes\cdots\otimes\ket{I_N}$. It is straightforward to verify that $\ket{{\rm W}}$ is fully symmetric under the exchange of any two qubits. 

More formally, we introduce the permutation operator $P_{nn'}$, defined by its action on single-qubit Pauli operators as
\begin{align}
  P_{nn'} \vec{\sigma}_{n''} P_{nn'}^\dagger =
  \left\{ \begin{array}{ll}
      \vec{\sigma}_{n'} & :~ n'' = n \\
      \vec{\sigma}_n & :~ n'' = n' \\
      \vec{\sigma}_{n''} & :~ n'' \neq n, n'
    \end{array} \right. ,
\end{align}
where $\vec{\sigma}_n=(X_n, Y_n, Z_n)$. Note that $P_{nn'}$ is unitary. Furthermore, $P_{nn'}^2=\mathds{I}$ and hence $P_{nn'}^{-1}=P_{nn'}$. We further assume that the reference state $\ket{0}^{\otimes N}$ is invariant under such a permutation; that is, for any $(n,n')$,
\begin{align}
  P_{nn'} \ket{0}^{\otimes N} = \ket{0}^{\otimes N}.
\end{align}
Then, by construction, the W state is likewise invariant under any $P_{nn'}$:
\begin{align}
  P_{nn'} \ket{{\rm W}} = \ket{{\rm W}}.
\end{align}
The reason why the W state (and, more generally, Dicke states) is of interest for DM detection is that, for $N$ identical qubit sensors, the Hamiltonian describing the DM-qubit interaction satisfies
\begin{align}
  P_{nn'}HP_{nn'}^\dagger=H.
\end{align}
The Hamiltonian given in Eq.\ \eqref{HX_Nqubits} is an example of this class of Hamiltonians. For such a Hamiltonian, the permutational symmetry of the state is preserved under the unitary time evolution induced by the DM field. Thus, if the system is initialized in a state respecting the permutation symmetry, e.g., $\ket{0}^{\otimes N}$, the state remains within the permutation-symmetric subspace, which constitutes only a very small fraction of the total Hilbert space \cite{Chen:2025tgj}. 

These symmetric states are, in turn, markedly different from states produced by typical noise processes. For instance, if bit-flip noise acts on the $n_\star$-th qubit of the state $\ket{0}^{\otimes N}$, one obtains
\begin{align}
  \ket{0}^{\otimes N} \rightarrow \ket{n_\star} \equiv X_{n_\star} \ket{0}^{\otimes N}.
\end{align}
Obviously, the state $\ket{n_\star}$ does not respect the permutation symmetry. Even though the number of excited qubits is the same, $\ket{\rm W}$ and $\ket{n_\star}$ are quantum mechanically very different; their overlap is
\begin{align}
  \braket{\rm W}{n_\star} = \frac{1}{\sqrt{N}}.
\end{align}
Thus, in the large-$N$ limit, these states become nearly orthogonal. Motivated by this observation, one can perform a QEC-like operation to mitigate the effect of the noise.

We now discuss in detail how the QEC-like protocol can enhance sensing sensitivity. To make this discussion concrete, let us focus on a model described by the Hamiltonian in Eq.\ \eqref{HX_Nqubits}, subject to Markovian noise parallel to the signal direction. In this scenario, the time evolution of the system density matrix is governed by the following Lindblad equation:
\begin{align}
  \partial_t \rho
  = \mathcal{L}[\rho] \equiv
  -i [H, \rho]
  + \frac{1}{2} \gamma \sum_{n=1}^N \left( X_n \rho X_n - \rho \right).
  \label{dotrho_woqec}
\end{align}
It should be noted that alternative noise mechanisms, such as phase-damping noise (described by the Pauli-$Z$ operator) or amplitude-damping noise (described by the qubit-lowering operator), may also be mapped onto bit-flip noise through an appropriate QEC-like protocol \cite{Fukuda:2025afi}. Therefore, the arguments developed below can be generalized to those situations as well.

In the present model, we can naturally utilize Pauli-$Z$ measurements on each qubit to extract information about the parameter $\eta$. When adopting such separable measurements, and for a fixed total experimental duration $T$, the CFI obtained from $\nu = T/t$ repetitions of the measurement cycle is given by
\begin{align}
  \nu F_{\rm C}^{\rm (sep)} \simeq 16 N \eta^2 t^3 T \frac{e^{-2\gamma t}}{1 - e^{-2\gamma t}} + O(\eta^4),
\end{align}
where we focus on the leading-order contribution in $\eta$. The CFI attains its maximum at $t \sim \gamma^{-1}$, and at this optimal exposure time, we find $\nu F_{\rm C}^{\rm (sep)} \sim O(N \eta^2 \gamma^{-3} T)$. In the following, we will demonstrate that it is possible to construct a measurement protocol, using QEC, that surpasses this bound.

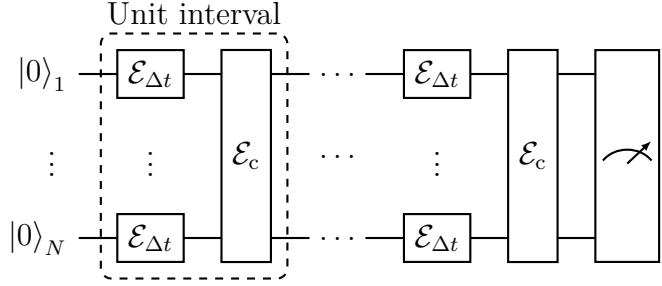
\begin{figure}[tbp]
  \centering
  \begin{quantikz}
    \lstick{$\ket{0}_1$}
    & \gate{\mathcal{E}_{\Delta t}} \gategroup[3,steps=2,style={dashed, rounded corners, inner sep=2pt}]{Unit interval} & \gate[3]{\mathcal{E}_\text{c}} & \ \ldots\ &  \gate{\mathcal{E}_{\Delta t}} & \gate[3]{\mathcal{E}_\text{c}}  & \meter[3]{}\\
    \lstick{\ \vdots\ } \setwiretype{n} & \ \vdots\   & & \ \ldots\  & \ \vdots\  & &\\
    \lstick{$\ket{0}_N$}&  \gate{\mathcal{E}_{\Delta t}} & & \ \ldots\ & \gate{\mathcal{E}_{\Delta t}} & &
  \end{quantikz}
  \caption{
    Schematic quantum circuit for our QEC-like error-mitigation protocol.
  }
  \label{fig:qec}
\end{figure}

Let us assume that the system is initialized in the state $\ket{0}^{\otimes N}$. We further suppose that the signal strength $\eta$ is sufficiently small such that, starting from $\ket{0}^{\otimes N}$, the probability of populating doubly excited states remains negligible. In the absence of noise, the dynamics then remains confined to the subspace spanned by $\{\ket{0}^{\otimes N}, \ket{\rm W}\}$ throughout the protocol. We refer to this subspace as the ``logical code space'' or ``logical qubit'' and define
\begin{align}
  \ket{0_{\rm L}} \equiv \ket{0}^{\otimes N},~~~
  \ket{1_{\rm L}} \equiv \ket{\rm W}.
\end{align}

Our sensing protocol consists of the repeated application of adaptive steps \cite{Sekatski:2017xdg, Demkowicz-Dobrzanski:2014gvc, Demkowicz-Dobrzanski:2017stg, Zhou:2017kxr}, as depicted schematically in Fig.\ \ref{fig:qec}. The central idea is to divide the total sensing duration $\tau$, which is set by the system's coherence time, into shorter intervals of duration $\Delta t$. Throughout each interval, the state undergoes evolution according to the Lindblad equation Eq.\ \eqref{dotrho_woqec}, with the corresponding quantum channel denoted as $\mathcal{E}_{\Delta t}$:
\begin{align}
  \mathcal{E}_{\Delta t}^{\otimes N}[\rho(t)] = \rho(t + \Delta t)
  \simeq \rho(t) +  \mathcal{L}[\rho(t)] \Delta t.
  \label{eq:channel_E_dt}
\end{align}
As noted above, this evolution can cause the state to exit the code space, primarily due to bit-flip noise. Specifically, noise events may drive the state to $X_n \ket{0_{\rm L}}$ or $X_n \ket{1_{\rm L}}$, which clearly lie outside the code space.

The QEC-like protocol aims to (i) detect such noise events via a suitable POVM and (ii) apply a recovery operation conditional on the measurement outcome when the state is found outside the logical code space. The POVM we use is specified by the following Kraus operators:
\begin{align}
  K_0 \equiv &\, \ketbra{0_{\rm L}}{0_{\rm L}} + \ketbra{1_{\rm L}}{1_{\rm L}},
  \\
  K_n \equiv &\, \sqrt{\frac{N-1}{N}}
  \left( \, \ket{n_{(1)}}_\perp \!\! \bra{n_{(1)}}_\perp +
  \ket{n_{(2)}}_\perp \!\! \bra{n_{(2)}}_\perp \, \right),
  \\
  K_{\rm S} \equiv &\, \ketbra{\rm S}{\rm S},
\end{align}
where $n=1,2,\cdots,N$, and
\begin{align}
  \ket{n_{(1)}}_\perp &\equiv
  \sqrt{\frac{N}{N-1}} X_n \ket{0_{\rm L}} - \frac{1}{\sqrt{N-1}} \ket{\rm W},
  \\
  \ket{n_{(2)}}_\perp &\equiv
   \sqrt{\frac{N}{(N-1)(N-2)}} X_n \sum_{n'\neq n} X_{n'} \ket{0_{\rm L}} -
   \sqrt{\frac{2}{N-2}}\ket{S},
   \\
  \ket{S} &\equiv
  \frac{1}{\sqrt{2N (N-1)}} \sum_{n \neq n'} X_n X_{n'} \ket{0_{\rm L}}.
\end{align}
It is straightforward to verify that these states are normalized and satisfy $\braket{\rm W}{n_{(1)}}_\perp = \braket{\rm S}{n_{(2)}}_\perp = 0$. The POVM elements are defined as $E_a \equiv K_a^\dagger K_a$ ($a=0,1,\cdots,N,\mathrm{S}$), and the remainder is given by
\begin{align}
  E_{\rm R}\equiv \mathds{I} - E_0 - \sum_{n=1}^N E_n - E_{\rm S}.
\end{align}
By construction, the POVM $\{E_0, E_n, E_{\rm S}, E_{\rm R}\}$ is complete, and one can check that all elements, including $E_{\rm R}$, are positive semidefinite.

Suppose now that the state is initially within the code space. In this case, it is found that
\begin{align}
  \mathrm{tr} \left[
    E_{\rm R} \left\{ \rho(t) + \mathcal{L}[\rho (t)] \Delta t \right\}
    \right] \sim  O(\Delta t^2),
\end{align}
indicating that the probability of obtaining outcome $E_{\rm R}$ is negligible to order $O(\Delta t)$. Thus, for the purposes of the analysis below, it suffices to consider the measurement outcomes $\{0,1,2,\cdots,N, {\rm S}\}$. Here, outcome $a=0$ indicates that the state remains in the code space, while $a=n$ corresponds to detection of a bit-flip error on the $n$-th qubit.

Following the POVM measurement, we apply a recovery operation determined by the measurement outcome. These recovery operations are defined as the following unitary operators:
\begin{align}
  R_0 &\equiv \mathds{I},
  \\
  R_n &\equiv
  \left( \ketbra{0_{\rm L}}{n_{(1)}}_\perp + \ketbra{1_{\rm L}}{n_{(2)}}_\perp + \mbox{h.c.} \right)
  + \mathds{I}_n,
  \\
  R_{\rm S} &\equiv \left( \ketbra{1_{\rm L}}{\rm S} + \mbox{h.c.} \right)
  + \mathds{I}_{\rm S},
\end{align}
where $\mathds{I}_n$ and $\mathds{I}_{\rm S}$ denote identity operators acting on the respective orthogonal subspaces, ensuring that $R_n$ and $R_{\rm S}$ are genuinely unitary. Using these, we define the quantum channel describing the state update for incorporating both the POVM measurement and the conditional recovery operation:
\begin{align}
  \widehat{\mathcal{E}}_{\Delta t} [\rho]
  \equiv
  \sum_{a=0,n,S} R_a K_a \left\{ \rho (t) + \mathcal{L}[\rho (t)] \Delta t \right\} K_a^\dagger R_a^\dagger.
\end{align}
Assuming that the density matrix evolves smoothly over time, we may take the continuous-time limit to obtain the following effective Lindblad equation:
\begin{align}
  \partial_t \rho = &\,
  \lim_{\Delta t\rightarrow 0}
  \frac{1}{\Delta t}
  \left( \widehat{\mathcal{E}}_{\Delta t}[\rho] - \rho \right)
  \nonumber \\ = &\,
  -i \sqrt{N} \eta [X_{\rm L}, \rho]
  + \frac{1}{2} \gamma \left( X_{\rm L} \rho X_{\rm L} - \rho \right)
  + \frac{1}{2} \frac{N-1}{N+\sqrt{N(N-2)}} \gamma
  \left( Z_{\rm L} \rho Z_{\rm L} - \rho \right),
  \label{Lindblad_L}
\end{align}
where
\begin{align}
  X_{\rm L} \equiv &\, \ketbra{0_{\rm L}}{1_{\rm L}} +
  \ketbra{1_{\rm L}}{0_{\rm L}},
  \\
  Z_{\rm L} \equiv &\, \ketbra{0_{\rm L}}{0_{\rm L}}
  - \ketbra{1_{\rm L}}{1_{\rm L}}.
\end{align}
Notice that the noise rates of this effective Lindblad equation are $O(\gamma)$, not enhanced by the number of qubits.

In the small-$\eta$ limit, if we perform a measurement of Pauli-$Z_{\rm L}$ on the logical qubit, the CFI is found to be
\begin{align}
  \nu F_{\rm C}^{({\rm QEC})} \simeq
  256 N^2 \eta^2 T
  \frac{1}{\gamma^4 t}
  \frac{e^{-2\gamma t}}{1-e^{-2\gamma t}}
  \left( 2 - 2e^{-\gamma t/2} - \gamma t \right)^2 + O(\eta^4),
\end{align}
where we have taken the large-$N$ limit. Importantly, the CFI scales as $N^2$, corresponding to the HL scaling. The optimal exposure time is given by $t \sim \gamma^{-1}$, yielding $\nu F_{\rm C}^{({\rm QEC})} \sim O(N^2 \eta^2 \gamma^{-3} T)$. Consequently, a significant enhancement in sensing sensitivity is achieved for large $N$.

We emphasize, however, that this argument breaks down for $N \gtrsim (\gamma/\eta)^2$. In this regime, as one can see from the Lindblad equation Eq.\ \eqref{Lindblad_L}, the system undergoes multiple Rabi oscillations within the code space for $t \sim \gamma^{-1}$, and the perturbative small-$\eta$ expansion ceases to be valid. (Moreover, transitions to higher-excitation states can no longer be neglected.) In such cases, the optimal strategy is to partition the $N$ sensor qubits into subsets of size $(\gamma/\eta)^2$, and to carry out the above QEC protocol within each subset for $t \sim \gamma^{-1}$. Under this protocol, the attainable CFI becomes $\sim O(N \gamma^{-1} T)$. As a result, the HL cannot be realized in the extreme large-$N$ limit. Nevertheless, even in this regime, the sensitivity remains substantially improved over the separable-measurement approach, especially for $\eta \ll \gamma$, which is typically the regime of greatest practical interest.

\section{Dark Matter Detection with Qubits}
\label{sec:DMsearch}
\setcounter{equation}{0}
\setcounter{figure}{0}
\setcounter{footnote}{1}

Finally, we discuss the potential of detecting DM through the direct excitation of superconducting qubits\,\cite{Chen:2022quj, Chen:2024aya}. Superconducting qubits constitute one of the most important realizations of qubits, and they are widely implemented in present-day quantum computers. (For an overview of the basic properties of superconducting qubits, see Appendix\,\ref{app:superconducting}.) In addition to their essential role in quantum computation, superconducting qubits are also recognized as promising quantum sensors for DM searches. In this section, we focus on the idea of utilizing superconducting qubits to sense electric fields induced by DM interactions.

In such a scenario, improving the signal rate generally requires employing qubits with stronger coupling to the electric field. However, increasing the coupling strength inevitably leads to several challenges: it tends to enhance noise rates, increase decoherence rates, and strengthen the coupling to the surrounding environment. These factors make the actual implementation of DM detection experiments more difficult. Therefore, identifying or engineering qubits that can offer strong signal coupling while mitigating noise and decoherence represents a key challenge for DM detection using this approach. It is worth noting that this trade-off is reminiscent of the dilemmas encountered in the development of (fault-tolerant) quantum computers, where considerable efforts are devoted to overcoming similar issues. Continued progress in the field of quantum sensing is expected, and advances made may be leveraged for DM detection experiments as well.\footnote
{S. Chen, private communication.}

In the following, we outline the procedure for employing superconducting qubits in DM detection. We especially focus on the specific cases of dark-photon and axion DM, summarizing experimental protocols for their detection and reviewing the expected sensitivities of these approaches.

\subsection{Detection Protocol}

The detection experiment proposed here aims to directly observe qubit excitations caused by the DM-induced electric field. The process is described by the following effective Hamiltonian:
\begin{align}
  H &= \omega \ketbra{1}{1} -2\eta\left(\ketbra{0}{1}+\ketbra{1}{0}\right)\cos(mt-\varphi),
\end{align}
where $\eta$ denotes the coupling constant between the DM field and the qubit (see Section \ref{sec:evolution}), and the drive frequency $m$ should be understood as the DM mass. Formulas necessary to evaluate $\eta$ for both dark photon and axion cases are provided in Appendices \ref{app:dm} and \ref{app:superconducting}. 

To achieve maximal sensitivity, it is important to tune the qubit frequency to coincide with the oscillation frequency of the DM-induced electric field, which is set by the DM mass. Under such resonant conditions, the excitation probability is significantly enhanced. For this reason, we consider utilizing frequency-tunable superconducting qubits; this is achievable, for example, by employing SQUID-based qubits in which the qubit frequency is tuned via an externally applied magnetic field.

We consider the case in which all the qubits are initialized to the ground state. The excitation probability of a single qubit in the resonant case can be estimated as (see Eq.~\eqref{Pge(small-t)})
\begin{align}
  p_{\rm ge} \simeq (\eta \tau)^2,
\end{align}
where $\tau$ denotes the relevant coherence time; in the experimental context, we take $\tau = \min(\tau_{\rm Q}, \tau_{\rm DM})$, where $\tau_{\rm Q}=Q/\omega$ (with $Q$ being the qubit quality factor) is the coherence time of the qubit. The exposure time for one cycle is thus set to approximately $\tau$, which also sets the scan step of the angular frequency $\omega$ to about $\tau^{-1}$.

To detect a small excitation probability, a repetitive counting experiment is performed according to the following sequence, using either a single or multiple SQUID-based qubits. For the multi-qubit setup, it is assumed that independent readout for each qubit is available.
\begin{itemize}
\item[1.] All qubits are prepared in the ground state at $t=0$ and allowed to evolve until $t = \tau$.
\item[2.] A single-shot state readout is performed for all qubits. Here, a typical readout duration of $O(100)~{\rm ns}$ is assumed.
\item[3.] For a fixed frequency $\omega$, steps 1 and 2 are repeated $n_{\rm rep}$ times. The total number of qubit measurements is given by $N_{\rm try} \equiv N n_{\rm rep}$, where $N$ is the number of qubits involved. The expected number of excitations is $N_{\rm sig} = p_{\rm ge}N_{\rm try}$, neglecting possible inter-qubit interference effects.
\item[4.] The above measurements are repeated at different qubit frequencies in order to scan over the possible DM mass range.
\end{itemize}

\begin{figure}[t]
  \centering
  \includegraphics[width=0.4\textwidth]{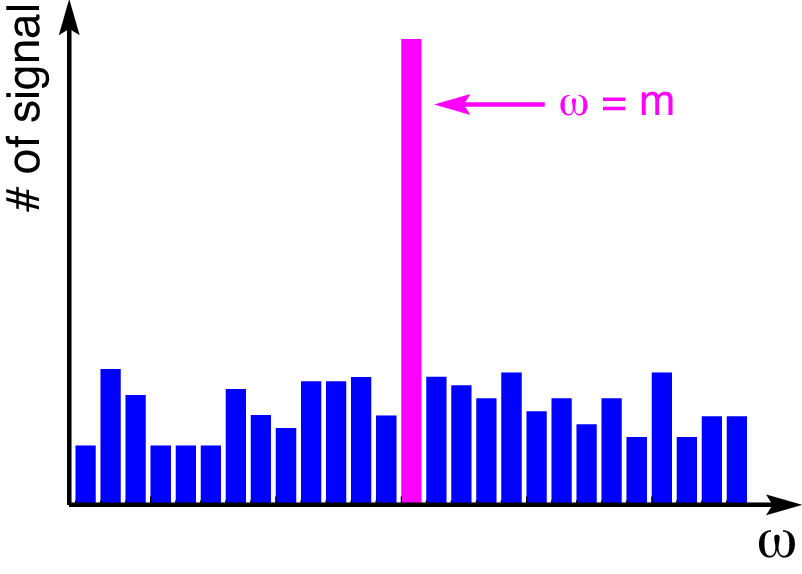}
  \caption{Schematic figure of the result of the frequency scan, assuming there exists a signal peak at $\omega\sim m$. The signal bin is indicated in magenta, while the noise bins are in blue.}
  \label{fig:scan_sideband}
\end{figure}

This search strategy is essentially a resonance search, where one looks for a sharp peak in the excitation rate as a function of qubit frequency. Such a peak signals the DM-induced electric field at a DM mass corresponding to that frequency (see Fig.~\ref{fig:scan_sideband}).

\subsection{Dark Photon DM}

The first target for the detection of wave-like DM in experiments is the dark photon, a massive vector boson that kinetically mixes with the ordinary photon. The interaction Lagrangian describing its kinetic mixing with the photon is given by
\begin{align}
  \mathcal{L}_{\rm int} = \frac{1}{2} \epsilon F_{\mu\nu} X^{\mu\nu},
\end{align}
where $X_{\mu\nu}\equiv\partial_\mu X_\nu-\partial_\nu X_\mu$ with $X_\mu$ denoting the vector potential associated with the dark photon field. The dark photon is a hypothetical field, and if it has a sub-eV (or smaller) mass accompanied by an appreciable oscillation amplitude in the present Universe, it constitutes an attractive and well-motivated candidate for wave-like DM. Through its kinetic mixing with the ordinary EM field, oscillations of the dark photon can induce an ordinary electric field. This effect can then be probed using superconducting qubits. The basic properties of the dark photon are summarized in Appendix \ref{subsec:darkphoton}.

The effective coupling strength $\eta$ can be evaluated using Eq.\ \eqref{EfieldDPH} and Eq.\ \eqref{eta_DM}. This allows us to calculate the transition probability for the qubit. In the resonance limit where $\omega\rightarrow m_X$, the transition probability from the ground state to the excited state is given numerically by
\begin{align}
  p_{\rm ge} \simeq &\, 0.12 \times
  \kappa^2 \cos^2 \Theta
  \left( \frac{\epsilon}{10^{-11}} \right)^2
  \left( \frac{f}{1\ {\rm GHz}} \right)
  \left( \frac{\tau}{100\ \mu{\rm sec}} \right)^2
  \left( \frac{C}{0.1\ {\rm pF}} \right)  
  \left( \frac{d}{100\ \mu{\rm m}} \right)^2
  \nonumber \\ &\, \times
  \left( \frac{\rho_{\rm DM}}{0.45\ {\rm GeV/cm}^3} \right),
\label{pge_DPH}
\end{align}
where $C$ is the capacitance of the qubit, $d$ represents the separation between the capacitance pads, and $\Theta$ denotes the angle between the direction of the induced electric field and the axis of the conductor.

Based on this probability, we can estimate the expected experimental sensitivity for detecting dark photon DM. Specifically, we calculate the signal rate using Eq.\ \eqref{pge_DPH}, with the parameters set to $C=0.1\ {\rm pF}$, $d=100\ \mu{\rm m}$, $\kappa=1$, $\tau_{\rm Q}=100\ \mu{\rm sec}$, and $\rho_{\rm DM}=0.45\ {\rm GeV}/{\rm cm}^3$. We also take the angular average $\cos^2\Theta\rightarrow\frac{1}{3}$. The number of qubits is assumed to be either $N=1$ or $100$. For the background, we assume a flat dark-count probability of $0.1\%$. 

The sensitivity of the search is determined by comparing the expected number of signal events, $N_{\rm sig}$, with the expected number of dark counts, $N_{\rm dark}$. For simplicity, we use $N_{\rm sig}/\sqrt{N_{\rm dark}}$ to estimate the sensitivity, corresponding to the significance in units of the Gaussian-equivalent standard deviation. The following criterion is adopted for DM detection in this study:
\begin{align}
  N_{\rm sig} > \max \left(3, 5 \sqrt{N_{\rm dark}}\right),
  \label{signalobs}
\end{align}
which corresponds to a $5\sigma$ criterion when dark counts are non-negligible, or requires at least three signal events in regimes with negligible dark count background.

The qubit frequency scan is conducted over the range $1\leq f\leq10~{\rm GHz}$, corresponding to a dark photon mass of $4-40~\mu{\rm eV}$. We assume $\tau_{\rm Q}=100\ \mu{\rm sec}$. With this setup, the number of scan points required is $\sim 6\times 10^6$. If we consider a one-year operational period, the available measurement time per scan point is about 5 seconds, assuming that an equal amount of time is spent at each frequency bin. The readout time is neglected, assuming that it is much shorter than the coherence time $\tau$.

\begin{figure}[t]
  \centering
  \includegraphics[width=0.55\textwidth]{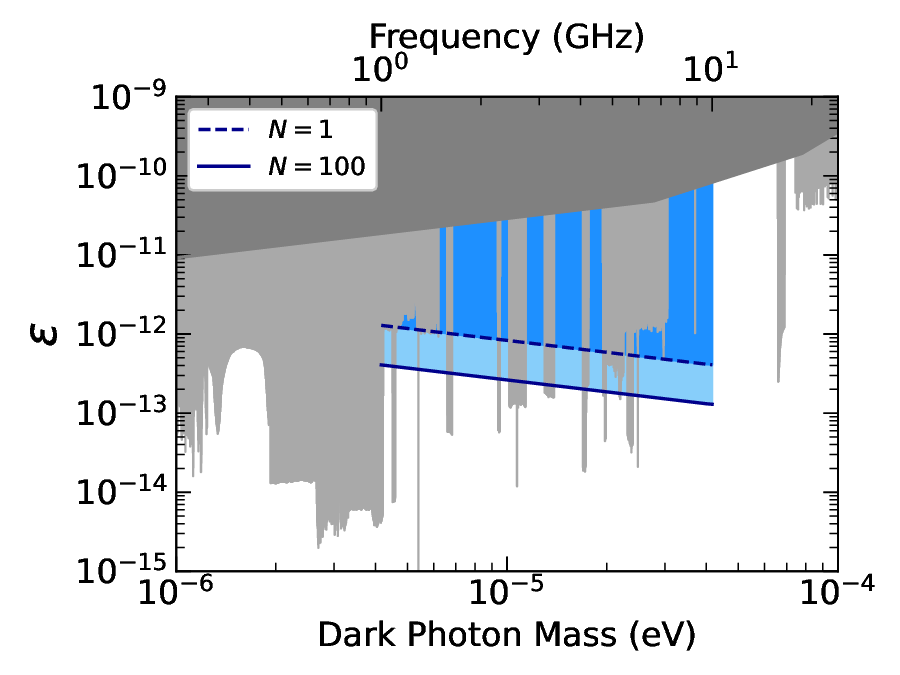}
  \caption{Expected sensitivity of the dark photon search experiment with superconducting qubits. The blue-shaded regions indicate the sensitivity with the 1-year scan over the frequency range from $1$ to $10\ {\rm GHz}$ for $N=1$ (dark blue) and $100$ (light blue). Parameters of $C=0.1\ {\rm pF}$, $d=100\ \mu{\rm m}$, $\kappa=1$, $\tau_{\rm Q}=100\ \mu{\rm sec}$, and $\rho_{\rm DM}=0.45\ {\rm GeV}/{\rm cm}^3$ are assumed. The gray-shaded region is excluded by the cosmological and astrophysical constraints \cite{McDermott:2019lch} (dark gray) and the existing hidden-photon search experiments (light gray) based on the summary in Ref.\ \cite{Caputo:2021eaa}. A flat dark-count probability of $0.1\ \%$ is assumed.}
  \label{fig:reach_dph_dc010}
\end{figure}

Fig.\ \ref{fig:reach_dph_dc010} shows the projected sensitivity of the proposed experiment. We can see that the proposed experiment can probe the parameter region which has not been explored yet. 

\subsection{Axion DM}

Another important target is the axion $a$. In particular, the QCD axion is a pseudo-Nambu-Goldstone boson associated with the Peccei-Quinn (PQ) mechanism \cite{Peccei:1977hh, Peccei:1977ur, Weinberg:1977ma, Wilczek:1977pj}, which was originally proposed as a solution to the strong CP problem. If coherent axion oscillations persist in the present Universe with an appropriate amplitude, the axion can constitute DM. In addition, axion-like particles (ALPs) are also important targets.  (Hereafter, we collectively refer to both the QCD axion and ALPs as ``axions.'') For a summary of the axion properties, see Appendix~\ref{subsec:axion}.

The axion couples to the EM fields through the interaction
\begin{align}
  \mathcal{L}_{a\gamma\gamma} = g_{a\gamma\gamma} a\, \vec{E}\vec{B},
\end{align}
so, in the presence of an external magnetic field, axion oscillations induce an oscillating electric field. This oscillating electric field can, in principle, be detected using sensitive quantum devices such as qubits. 

In this context, we explore the possibility of detecting axion DM using superconducting qubits\,\cite{Chen:2024aya}. To effectively convert axion oscillations into a detectable electric field, the experimental setup should apply a strong static magnetic field in the vicinity of the qubits. One of the main experimental challenges is that strong magnetic fields generally degrade the coherence of qubits, making conventional operation difficult. Nevertheless, it has been demonstrated that qubit coherence can be preserved under static magnetic fields as high as $O(1)$~T, provided that the magnetic field is aligned in-plane with respect to the thin films of the qubit\,\cite{Krause:2021llk}. Notably, such an arrangement coincides with the direction required to excite the qubit via axion-induced oscillations. In the following, we assume that sufficiently strong magnetic fields can be applied to the qubits and estimate the expected sensitivity, with the expectation that future technological advances will further improve qubit operation in strong magnetic environments.

The drive strength $\eta$ is evaluated using Eqs.\ \eqref{EfieldAxion} and \eqref{eta_DM}. On this basis, the transition probability is estimated as
\begin{align}
  p_{\rm ge} \simeq &\,
  0.026 \times \kappa^2
  \left( \frac{g_{a\gamma \gamma}}{10^{-10} \ \mathrm{GeV^{-1}}}\right)^2
  \left( \frac{f}{1 \ \mathrm{GHz}}\right)^{-1}
  \left( \frac{\tau}{100 \ \mathrm{\mu sec}} \right)^2
  \left( \frac{C}{0.1 \ {\rm pF}}\right)
  \left( \frac{d}{100 \ {\rm \mu m}}\right)^2
  \nonumber \\ &\,
  \times \left( \frac{B^{\rm (ext)}}{1 \ \mathrm{T}}\right)^2
  \left( \frac{\rho_{\rm DM}}{0.45 \ {\rm GeV/cm^3}}\right),
\end{align}
where $B^{\rm (ext)}$ is the external magnetic field. 

\begin{figure}
  \centering
  \includegraphics[width=0.55\textwidth]{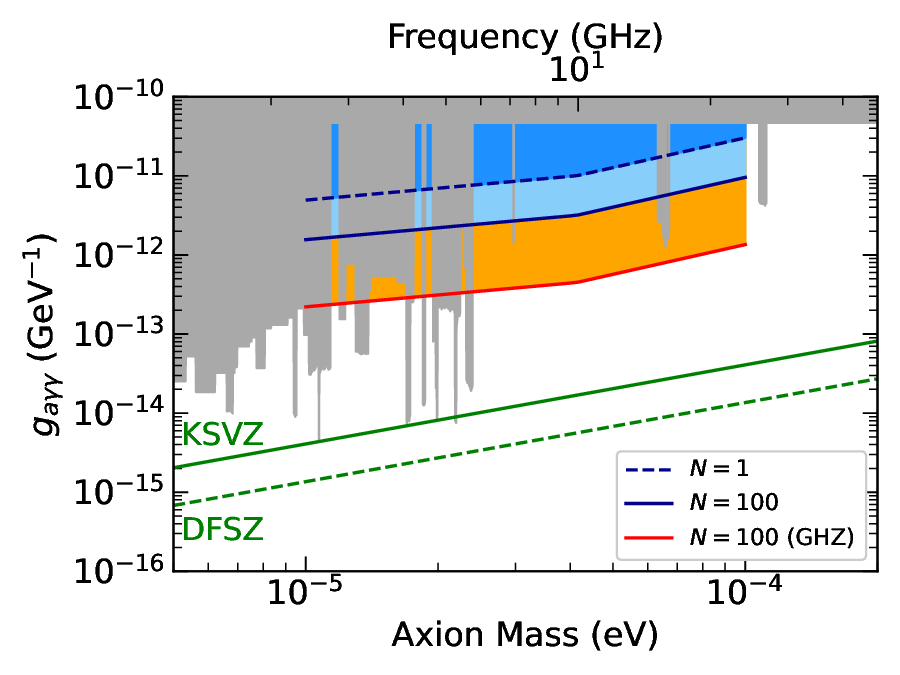}
  \caption{Expected sensitivity of the axion DM search experiment with superconducting qubits. The blue-shaded regions indicate the sensitivity with the 1-year scan over the DM mass range from $10$ to $100\ \mu{\rm eV}$ for $N=1$ (dark blue) and $100$ (light blue) with $\kappa=1$. If the enhancement from the GHZ state can be exploited, the orange region can be covered for $N=100$. Parameters of $C=0.1\ {\rm pF}$, $d=100\ \mu{\rm m}$, $\kappa=1$, $\tau_{\rm Q}=100\ \mu{\rm sec}$, $B^{\rm (ext)}=5\ {\rm T}$, and $\rho_{\rm DM}=0.45\ {\rm GeV}/{\rm cm}^3$ are assumed. The grey area shows the astrophysical constraint from the observation of the stellar population in the Globular Clusters \cite{Dolan:2022kul} and the constraints from Haloscope experiments \cite{AxionLimits}. The solid and dashed green lines indicate parameter regions suggested by the KSVZ model and DFSZ model of QCD axions, respectively.}
  \label{fig:reach_alp_dc010k001}
  \vspace{7mm}
  \includegraphics[width=0.55\textwidth]{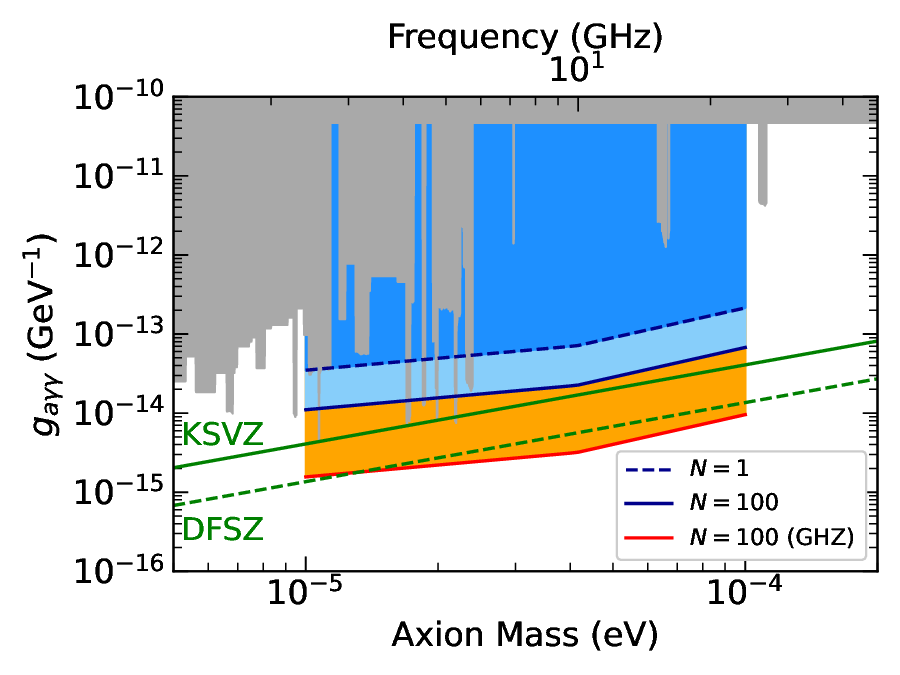}
  \caption{Same as Fig.\ \ref{fig:reach_alp_dc010k001}, except for $\kappa=100$.}
  \label{fig:reach_alp_dc010k100}
\end{figure}

The expected sensitivity is shown in Figs.\ \ref{fig:reach_alp_dc010k001} and \ref{fig:reach_alp_dc010k100} for $\kappa=1$ and $\kappa=100$, respectively, assuming $B^{\rm (ext)}=5$~T. We consider a setup in which the mass range $10<m_a<100\ \mu{\rm eV}$ (with $m_a$ denoting the axion mass) is scanned within one year. In both figures, we also present the projected sensitivity with quantum enhancement employing a GHZ state of $N=100$ qubits (see Section~\ref{sec:ghz}); achieving this performance requires a qubit quality factor of at least $\mathcal{O}(10^8)$. We find that the sensitivity exhibits a change in behavior around $m_a \simeq 40\ \mu{\rm eV}$. This feature reflects a transition in the relevant coherence time $\tau$; for $m_a \lesssim 40~\mu{\rm eV}$, $\tau$ is limited by the qubit coherence time $\tau_{\rm Q}$, whereas for $m_a \gtrsim 40\ \mu{\rm eV}$ it is set by the DM coherence time $\tau_{\rm DM}$.

The results shown in Figs.\ \ref{fig:reach_alp_dc010k001} and \ref{fig:reach_alp_dc010k100} indicate that, even with $N=1$ (i.e., a single qubit), this experimental approach can probe currently unexplored regions of parameter space. Moreover, by incorporating both quantum enhancement and cavity enhancement, the experiment may reach the parameter space suggested for the QCD axion.

\section{Conclusions and Discussion}
\label{sec:conclusions}
\setcounter{equation}{0}
\setcounter{figure}{0}
\setcounter{footnote}{1}

In this article, we have presented an introductory review of qubits and several related topics, with particular emphasis on their relevance to DM detection; we have provided a self-contained bridge connecting DM searches with qubits, circuit QED, and quantum-sensing formalisms.

We have focused on wave-like DM (particularly axion DM and dark-photon DM), whose local field can be treated as a classical oscillation with a finite coherence time. These DM candidates can induce an oscillating electric field around the qubits and can thereby excite the qubits. We have discussed and analyzed the evolution of qubits under the influence of such a DM-induced electric field, employing minimal and simple models that capture the essential features of qubit dynamics in qubit-based experiments.

In contrast to quantum sensing of a purely harmonic electromagnetic field, one of the difficulties in detecting the DM signal arises from the fact that the DM oscillation has a finite coherence time, a consequence of the DM field having a velocity distribution. We have discussed the effect of the decoherence of the DM oscillation, and have seen that, in order to optimize the detection protocol, particularly the choice of exposure time, it is important to take into account both the coherence time of the DM, $\tau_{\rm DM}$, and the various coherence times of the qubit (such as $T_1$, $T_2$, and $T_0$).

One interesting possibility for enhancing the sensitivity is to use entangled states (such as the GHZ state). When the exposure time can be chosen arbitrarily, it is known that protocols using entangled states cannot gain much advantage over a simple counting experiment using separable states, since a single qubit can already reach comparable precision simply by extending the exposure time. In the case of DM detection, however, the exposure time is effectively bounded by the DM coherence time $\tau_{\rm DM}$, beyond which the phase information of the DM-induced drive is lost; this removes the option of compensating for the lack of entanglement by indefinitely extending the exposure time. It is in this regime that protocols employing entangled states may offer a genuine advantage. In particular, if high-quality qubits with a coherence time significantly longer than the DM coherence time become available, an improvement in sensitivity may be achieved through the use of entangled states.

As mentioned above, we have employed minimal and simple models, which may not capture several effects relevant to realistic experimental situations. In particular, we have considered only Markovian noise acting independently on each qubit. In actual multi-qubit devices, however, noise can be correlated across qubits, e.g., cosmic-ray-induced quasiparticle bursts are known to cause simultaneous relaxation of many qubits on the same chip \cite{Vepsalainen:2020trd, Cardani:2020vvp, McEwen:2021wdg}, and such correlated noise is not captured by the present analysis. The advantage of entangled-state and QEC-like protocols relies on the assumption of independent noise. Correlated noise of this kind could therefore degrade the sensitivity gain expected from entanglement, and its impact deserves careful study in future work. More generally, non-Markovian noise may also become important in realistic settings; readers interested in non-Markovian noise are referred to, e.g., Refs.\ \cite{Breuer:2015zlm, deVega:2017gfm}.

We also note that, throughout this article, we have focused on the statistical sensitivity determined by qubit noise and DM coherence; we did not consider systematic effects, such as RF interference, cosmic-ray events, and temperature drifts, nor the practical cost and fidelity of the multi-qubit operations required for entangled-state protocols. A quantitative assessment of these issues will be important for translating the ideas discussed here into an actual DM search. Furthermore, we have neglected the effects arising from the motion of the Earth, which, in principle, can induce daily and annual modulations in the DM signal. A detailed study of these modulations may also be of interest, as they could provide valuable information for characterizing the DM signal.

The application of quantum technologies, including qubits and other quantum platforms, is expected to become increasingly important in DM detection and, more broadly, in the search for physics beyond the Standard Model. We anticipate that continued advances in quantum-sensing technologies and associated experimental techniques, if properly utilized, could substantially expand our sensitivity to phenomena that have not yet been observed. Such developments should open new frontiers in our understanding of the Universe and play a key role in probing various unexplored areas of new physics.

\section*{Acknowledgements}

The author is grateful to H.~Fukuda and S.~Thanaporn for stimulating discussions throughout our collaborations so far, as well as for their helpful suggestions and comments during the preparation of this article. The idea of writing this review grew out of conversations with them. The author also thanks S.~Chen, T.~Inada, K.~Nakazono, T.~Nitta, and K.~Watanabe for fruitful discussions. This work was supported by JSPS KAKENHI Grant No.~23K22486. The author acknowledges the use of UTokyo Azure (\url{https://utelecon.adm.u-tokyo.ac.jp/en/research_computing/utokyo_azure/}). Although this article was prepared with the assistance of AI tools ({\tt ChatGPT} and {\tt Claude}), the author takes full responsibility for its contents.

\appendix

\section{Wave-Like Dark Matter}
\label{app:dm}
\setcounter{equation}{0}
\setcounter{figure}{0}

In this Appendix, we first provide a brief summary of the generic features of 
wave-like DM, including a discussion of its decoherence during cosmological 
evolution. We then summarize the basic properties of two major candidates for 
wave-like DM, i.e., the axion (or ALPs) and the dark photon.

\subsection{General Properties}

DM is a hypothetical form of matter that accounts for about $26\ \%$ of the energy density of the present Universe \cite{Planck:2018vyg}. Although there is substantial evidence for the existence of DM from astrophysical and cosmological observations, its particle-physics properties remain largely unknown. The direct (and even indirect) detection of DM is therefore one of the most important goals in particle physics, astrophysics, and cosmology today. Many experimental and observational efforts have been devoted to its detection; however, no definitive discovery has been achieved so far.

Key properties of DM include:
\begin{itemize}
\item It interacts only very weakly with Standard Model particles.
\item It should be ``cold,'' i.e., its pressure should be sufficiently small. Equivalently, under cosmic expansion its energy density should scale as $a^{-3}$ (where $a$ denotes the scale factor of the Universe). For this reason, the dominant DM component in our Universe is commonly referred to as ``cold dark matter (CDM).''
\item As mentioned above, its energy density accounts for about $26\ \%$ of the total energy density of the Universe. In our Galaxy, the DM density is expected to be enhanced due to gravitational clustering. In the vicinity of the Solar System, the local DM energy density is typically estimated to be $\rho_{\rm DM}^{\rm (Solar)}\sim O(0.1)\ {\rm GeV/cm^3}$; for example, a reference value given in Ref.\ \cite{Cirelli:2024ssz} is 
\begin{align}
  \rho_{\rm DM}^{\rm (Solar)} \simeq
  0.4\ {\rm GeV/cm^3},
  \label{rhoDM(solar)}
\end{align}
with an uncertainty at the level of a factor of two. Notice that $\rho_{\rm DM}^{\rm (Solar)}$ is a very important parameter in considering DM direct detection experiments because the event rates are sensitive to it.
\end{itemize}

One important parameter characterizing DM is its mass. If the mass is sufficiently large, the occupation number of the DM particles within the de Broglie wavelength becomes much smaller than $1$, and DM behaves as a collection of particles. For smaller masses, by contrast, DM exhibits wave-like behavior, which is the case of interest in this review.

When the occupation number is much larger than unity, DM can often be treated as a classical wave-like field. We first briefly explain how such a classical field can behave as CDM. To see this, let us consider a simple example, a real scalar field $\sigma$ described by the Lagrangian
\begin{align}
  \mathcal{L} = \frac{1}{2} g^{\mu\nu} \partial_\mu \sigma \partial_\nu \sigma - V(\sigma),
\end{align}
where $g^{\mu\nu}$ is the metric and $V(\sigma)$ is the potential. We are interested in the situation in which $\sigma$ oscillates around the minimum of $V(\sigma)$. For simplicity, we assume that $V(\sigma)$ is well approximated by a power-law potential of the form
\begin{align}
  V = \frac{1}{P} \lambda_P \sigma^P,
\end{align}
where $P$ is a positive even integer and $\lambda_P$ is a positive constant.

We first consider the simplest case in which $\sigma$ depends only on the cosmic time $t$, so that the distribution of the DM is fully homogeneous and isotropic. Then, in an expanding Universe with the Robertson-Walker metric $g_{\mu\nu}=\mathrm{diag}(1,-a^2,-a^2,-a^2)$, the equation of motion for $\sigma$ is given by
\begin{align}
  \ddot{\sigma} + 3 H \dot{\sigma} + \partial_\sigma V = 0,
  \label{EoM(phi)}
\end{align}
with
\begin{align}
  H = \frac{\dot{a}}{a}.
\end{align}
The system described by Eq.\ \eqref{EoM(phi)} is analogous to a mechanical system in the potential $V(\sigma)$ with a friction term proportional to $H$; the second term on the left-hand side of Eq.\ \eqref{EoM(phi)} is often referred to as ``Hubble friction.''

The scalar field $\sigma$ may behave as CDM when $H$ becomes much smaller than the oscillation frequency of the field. In this regime, $\sigma$ oscillates around $\sigma=0$ with an amplitude $\bar{\sigma}$, and the oscillation frequency is of order $O(m_{\rm eff})$. Here the ``effective mass'' is defined by $m_{\rm eff}^2\equiv\partial_\sigma^2 V|_{\sigma=\bar{\sigma}}=(P-1)\lambda_P \bar{\sigma}^{P-2}$. When $H\ll m_{\rm eff}$, the variation of $\bar{\sigma}$ over a time scale $m_{\rm eff}^{-1}$ is negligible. Consequently, the oscillation average of $\partial_t f(\dot{\sigma}, \sigma)$ (where $f(\dot{\sigma}, \sigma)$ is any smooth function of $\dot{\sigma}$ and $\sigma$) should vanish. In particular, considering the case of $f=\dot{\sigma}\sigma$, we obtain
\begin{align}
  \ev{ \dot{\sigma}^2 }_{\rm osc} \simeq
  \ev{ \sigma \partial_\sigma V }_{\rm osc} \simeq
  P \ev{ V }_{\rm osc},
\end{align}
with $\ev{\cdots}_{\rm osc}$ denoting the oscillation average. Then, using
\begin{align}
  \rho \simeq \ev{ \frac{1}{2} \dot{\sigma}^2 + V }_{\rm osc},
\end{align}
we find
\begin{align}
  \partial_t \rho \simeq
  - \frac{6P}{P+2} H \rho,
\end{align}
which implies
\begin{align}
  \rho \propto a^{-6P/(P+2)}.
\end{align}
Thus, in the case $P=2$, i.e., in the case of a parabolic potential, we have $\rho \propto a^{-3}$, and $\sigma$ can play the role of CDM.

In what follows, we consider scalar fields (or, more generally, bosonic fields) with a parabolic potential,
\begin{align}
  V = \frac{1}{2} m^2 \sigma^2.
\end{align}
On time scales relevant to DM detection experiments, the effect of cosmic expansion can be neglected. In this case, the scalar-field configuration evaluated at the location of the experimental apparatus (which we take to be $\vec{x}=0$) can be well approximated by the solution of the equation of motion in Minkowski spacetime,
\begin{align}
  \sigma (t) \simeq \bar{\sigma} \cos (m t - \varphi),
  \label{phi(Minkowski)_app}
\end{align}
where $\bar{\sigma}$ and $\varphi$ are the amplitude and the phase parameters, respectively. Thus, the wave-like DM field is oscillating with the angular frequency equal to its mass; numerically, the DM mass and the oscillation frequency $f=\frac{\omega}{2\pi}$ are related as
\begin{align}
  f \simeq 0.24\ {\rm GHz} \times
  \left( \frac{m}{1\ \mu{\rm eV}} \right).
\end{align}
Using Eq.\ \eqref{phi(Minkowski)_app}, the local energy density is
\begin{align}
  \rho = &\, \frac{1}{2} \dot{\sigma}^2 + \frac{1}{2} m^2 \sigma^2
  \simeq \frac{1}{2} m^2 \bar{\sigma}^2.
\end{align}
If $\sigma$ constitutes CDM, the amplitude $\bar{\sigma}$ can therefore be related to the local DM energy density through the above relation. Moreover, Eq.\ \eqref{phi(Minkowski)_app} allows one to show explicitly that the oscillation-average of the pressure, $p=\frac{1}{2}\dot{\sigma}^2 - V$, vanishes. This is consistent with the property of CDM, namely that its energy density redshifts as $\rho \propto a^{-3}$.

\subsection{Decoherence of Wave-Like DM}
\label{subsec:decoherence}

So far, we have considered the case in which the classical equation of motion admits a homogeneous and isotropic solution with a single frequency (see Eq.\ \eqref{phi(Minkowski)_app}). However, in the context of designing direct detection experiments for DM, it is essential to recognize that the DM present in our Galaxy is generally not in a single-frequency state, but rather exists as a superposition of waves with a range of velocities. An important consequence for DM detection experiments is that the coherence of the DM oscillation is preserved only within a finite time interval, as we explain below.

It is well established from the perspective of cosmic structure formation that DM particles are gravitationally bound within galactic halos, resulting in a local overdensity of DM in the Milky Way. In this environment, the DM fluid is expected to be composed of a distribution of various velocity components. Specifically, when DM is sufficiently light and exhibits wave-like properties, it is convenient to describe the field configuration as a superposition of plane waves. Focusing on non-relativistic DM, the scalar field may be expressed as \cite{Foster:2017hbq, Foster:2020fln}
\begin{align}
  \sigma (t, \vec{x}) =
  \sum_{\vec{v}}
  \tilde{\sigma}_{\vec{v}} \cos \left(
    E_v t - m \vec{v} \vec{x} - \varphi_{\vec{v}}
    \right),
    \label{sigma(multifreq)}
\end{align}
where the summation is over various velocity components. Here, $\varphi_{\vec{v}}$ is an unknown, generally random, phase associated with each velocity mode, and 
\begin{align}
  E_v \equiv \frac{m}{\sqrt{1-v^2}} \simeq m + \frac{1}{2} m v^2,
\end{align}
with $v \equiv |\vec{v}|$. In addition, $\tilde{\sigma}_{\vec{v}}$ represents the (real) amplitude of the mode with velocity $\vec{v}$. The characteristic velocity dispersion of DM is determined by the motion of the DM within the Milky Way’s gravitational potential. In the vicinity of the Solar system, $\tilde{\sigma}_{\vec{v}}$ is appreciable for $v \lesssim O(v_{\rm DM})$, and strongly suppressed at larger velocities (where $v_{\rm DM}\sim 10^{-3}$ is the mean velocity of the DM in the Solar system and is typically of the same order as the velocity dispersion).

Using Eq.\ \eqref{sigma(multifreq)}, the energy density of the DM field can be written as the sum of contributions from different velocity modes. In the case of interest, modes with different $\vec{v}$ are assumed to be statistically independent, and their phase factors $\varphi_{\vec{v}}$ are random and mutually uncorrelated. Consequently, in the expression for the energy density, cross terms between different velocities are expected to interfere destructively. When contributions from a large number of velocity components are included, these cross terms vanish upon ensemble averaging (or averaging over a large volume), as well as upon time averaging, and are therefore negligible. The averaged DM energy density is then estimated as
\begin{align}
  \rho_\sigma \simeq \frac{1}{2} m^2 \sum_{\vec{v}} \tilde{\sigma}_{\vec{v}}^2,
\end{align}
and, at each moment, the actual local energy density fluctuates around $\rho_\sigma$ with a magnitude of $O(1)$. The effective amplitude of the DM oscillation (i.e., the parameter $\bar{\sigma}$ in Eq.\ \eqref{phi(Minkowski)_app}) can be interpreted as
\begin{align}
  \bar{\sigma} = \sqrt{ \sum_{\vec{v}} \tilde{\sigma}_{\vec{v}}^2 }.
\end{align}

Focusing on the spatial position of a laboratory-based experiment (i.e., $\vec{x} = 0$), the field value at the apparatus can be written as
\begin{align}
  \sigma (t) = \mathrm{Re}\, \Sigma (t),
\end{align}
where we have introduced
\begin{align}
  \Sigma (t) \equiv
  \sum_{\vec{v}}
  \tilde{\sigma}_{\vec{v}} \, e^{i (E_v t - \varphi_{\vec{v}})}.
\end{align}
For later convenience, we parameterize $\Sigma(t)$ as
\begin{align}
  \Sigma (t) = \bar{\sigma}_\Sigma (t)
  \exp \left[ -i \varphi_\Sigma (t) \right],
\end{align}
where both $\bar{\sigma}_\Sigma(t)$ and $\varphi_\Sigma(t)$ are real, time-dependent functions corresponding to the instantaneous amplitude and phase of the field, respectively.

To examine the time evolution of $\sigma(t)$, we can use the following expression of $\Sigma(t)$:
\begin{align}
  \Sigma (t) = e^{i m (t-t_0)}
  \sum_{\vec{v}}
  \tilde{\sigma}_{\vec{v}} \,
  e^{i (E_v t_0 - \varphi_{\vec{v}})}
  \left[ 1 + \frac{i}{2} m v^2 (t-t_0) + \cdots \right],
\end{align}
where $t_0$ is an arbitrary reference time. For time intervals such that $|t-t_0|\ll O(2\pi/m v_{\rm DM}^2)$, the contribution from higher-order terms in $(t-t_0)$ in the square bracket is small because modes with large $v$ are suppressed by the mode amplitude $\tilde{\sigma}_{\vec{v}}$. Thus, over such time scales the field can be approximated as
\begin{align}
  \sigma (t) \simeq \bar{\sigma}_\Sigma(t_0)
  \cos \left[ m (t-t_0) - \varphi_\Sigma (t_0) \right].
\end{align}
This implies that, for sufficiently short time scales, the field evolution is well approximated by a simple harmonic oscillation with nearly constant amplitude $\bar{\sigma}_\Sigma(t_0)$ and phase $\varphi_\Sigma(t_0)$. However, this description becomes inaccurate when $|t-t_0|\gg 2\pi/m v_{\rm DM}^2$, as the cumulative effects of the velocity dispersion become relevant.

To quantify this, we introduce the coherence time scale of the wave-like DM oscillation (see also Section~\ref{sec:evolution}):
\begin{align}
  \tau_{\rm DM} \equiv \frac{2\pi}{m v_{\rm DM}^2}.
\end{align}
On time scales shorter than $\sim \tau_{\rm DM}$, the phase $\varphi_\Sigma(t)$ evolves approximately linearly in time and the amplitude $\bar{\sigma}_\Sigma(t)$ remains nearly constant. Conversely, on time scales longer than $\sim \tau_{\rm DM}$, both the amplitude and the phase undergo stochastic evolution, reflecting the fact that the local DM field is a random superposition of many velocity modes. This implies that the DM oscillation loses its phase coherence, and the variation of $\bar{\sigma}_\Sigma(t)$ becomes of $O(1)$ relative to its mean value \cite{Centers:2019dyn, Lisanti:2021vij}; the characteristic time scale governing this decoherence is $\sim \tau_{\rm DM}$. On the time scale much longer than $\tau_{\rm DM}$, the time-averaged value of $\bar{\sigma}_\Sigma^2$ is expected to approach $\bar{\sigma}^2$. 

\subsection{Axion and Axion-Like Particles}
\label{subsec:axion}

A leading candidate for wave-like DM is the axion (denoted as $a$) \cite{Weinberg:1977ma, Wilczek:1977pj}, which is a pseudo-Nambu-Goldstone boson arising from the spontaneous breaking of the PQ symmetry \cite{Peccei:1977hh, Peccei:1977ur}. The PQ symmetry was originally introduced as an elegant solution to the strong CP problem, i.e., the empirical fact that the $\theta$ parameter of QCD, the gauge theory of the strong interaction based on $SU(3)_C$, is extremely small. There is no natural explanation for the suppression of the strong CP phase within the Standard Model of particle physics. A comprehensive discussion of the strong CP problem is beyond the scope of this article; interested readers are referred to dedicated reviews (see, e.g., Refs.\ \cite{ParticleDataGroup:2024cfk, Kawasaki:2013ae}).

Here, we merely summarize the properties of the axion that are relevant to our analysis. Before QCD phase transition, the axion interacts with both gluons and photons, and the relevant terms in the Lagrangian are given by
\begin{align}
    \mathcal{L} = \frac{g_3^2}{32 \pi^2 f_a} a\, G_{\mu\nu}^{(a)} \tilde{G}^{(a)\mu\nu}
    + \frac{1}{4} g_{a\gamma\gamma}^{\rm (UV)} a\, F_{\mu\nu} \tilde{F}^{\mu\nu},
\end{align}
where $G_{\mu\nu}^{(a)}$ and $F_{\mu\nu}$ denote the field-strength tensors of the strong and EM interactions, respectively, and their duals are $\tilde{G}^{(a)\mu\nu}\equiv\frac{1}{2}\epsilon^{\mu\nu\rho\sigma}G_{\rho\sigma}^{(a)}$ and $\tilde{F}^{\mu\nu}\equiv\frac{1}{2}\epsilon^{\mu\nu\rho\sigma}F_{\rho\sigma}$. Here, $g_3$ is the $SU(3)_C$ gauge coupling and $f_a$ is the axion decay constant (the scale of PQ symmetry breaking).

The axion-photon-photon coupling at the ultraviolet (UV) scale is expressed as
\begin{align}
    g_{a\gamma\gamma}^{\rm (UV)} \equiv \frac{e^2}{8 \pi^2 f_a} \frac{E}{N_{\rm DW}},
\end{align}
where $N_{\rm DW}$ and $E$ are anomaly coefficients defined by
\begin{align}
    N_{\rm DW} \equiv &\, 2\sum_{i:\, {\rm fermions}} d^{(SU(2)_L)}_i Q^{\rm (PQ)}_{i} T(R_i),
    \\
    E \equiv &\, 2\sum_{i:\, {\rm fermions}} d^{(SU(3)_C)}_i d^{(SU(2)_L)}_i Q^{\rm (PQ)}_i (Q^{\rm (EM)}_i)^2,
\end{align}
where the sums run over all fermions in the theory (expressed in terms of left-handed fields). Here, $d^{(SU(3)_C)}_i$ and $d^{(SU(2)_L)}_i$ are the dimensions of, respectively, the $SU(3)_C$ and $SU(2)_L$ representations for the $i$-th fermion. Moreover, $Q^{\rm (PQ)}_i$ denotes the PQ charge, $Q^{\rm (EM)}_i$ the $U(1)_{\rm EM}$ electric charge, and $T(R_i)$ is the Dynkin index of the $SU(3)_C$ representation $R_i$ and is equal to $\frac{1}{2}$ for the fundamental representation. (We take the convention such that $N_{\rm DW}$ is equal to the number of axionic domain walls.)

After the QCD phase transition, the axion acquires a potential via nonperturbative QCD effects and becomes massive. While the full axion potential is periodic (with periodicity $\mathcal{O}(f_a)$), in the present Universe the axion field amplitude is typically much smaller than $f_a$, justifying an expansion near its minimum. In this regime, the quadratic (mass) term dominates,
\begin{align}
    V \simeq \frac{1}{2} m_a^2 a^2.
\end{align}
The axion mass is evaluated as \cite{Georgi:1986df}
\begin{align}
    m_a^2 \simeq \frac{m_u m_d}{(m_u + m_d)^2} \frac{f_\pi^2}{f_a^2} m_\pi^2,
\end{align}
where $m_u$ and $m_d$ are the up- and down-quark masses, $m_\pi$ is the pion mass, and $f_\pi \simeq 93~{\rm MeV}$ is the pion decay constant. Furthermore, after the QCD phase transition, the axion-photon-photon coupling is given by
\begin{align}
    \mathcal{L}_{a\gamma\gamma} =
    \frac{1}{4} g_{a\gamma\gamma} a F_{\mu\nu} \tilde{F}^{\mu\nu}
    = g_{a\gamma\gamma} a\, \vec{E}\vec{B},
  \label{L_agamma}
\end{align}
where
\begin{align}
    g_{a\gamma\gamma} \simeq
    g_{a\gamma\gamma}^{\rm (UV)} -
    \frac{e^2}{4\pi^2 f_a}
    \left[
      \frac{m_u + 4m_d}{3(m_u + m_d)}
    \right].
    \label{L_agammagamma}
\end{align}

Using precise next-to-leading order QCD calculations and taking into account running effects, the axion mass and axion-photon-photon coupling are given by as \cite{GrillidiCortona:2015jxo}
\begin{align}
    m_a = 5.70\, (7)\ \mu{\rm eV}
    \times
    \left( \frac{f_a}{10^{12}\ {\rm GeV}} \right)^{-1},
\end{align}
and
\begin{align}
    g_{a\gamma\gamma} =
    \frac{\alpha_{\rm EM}}{2 \pi f_a}
    \left( \frac{E}{N_{\rm DW}} - 1.92 \right),
\end{align}
where $\alpha_{\rm EM}$ is the fine-structure constant. In the above expressions, the up-to-down quark mass ratio is taken to be $\frac{m_u}{m_d}=0.48\,(3)$.

In order to evade constraints from the cooling of astrophysical objects, particularly from horizontal branch stars and observations of SN1987A, the allowed values of the axion decay constant are much larger than the electroweak scale (see, e.g., Ref.\ \cite{ParticleDataGroup:2024cfk}). Two prominent axion models that realize such high-scale PQ symmetry breaking are the Kim-Shifman-Vainshtein-Zakharov (KSVZ) model \cite{Kim:1979if, Shifman:1979if} and the Dine-Fischler-Srednicki-Zhitnitsky (DFSZ) model \cite{Zhitnitsky:1980tq, Dine:1981rt}.

The relic abundance of the QCD axion depends on the axion decay constant $f_a$ as well as on the thermal history of the Universe. A key contribution to the present-day cold axion abundance arises from the so-called misalignment mechanism. Whether the PQ symmetry breaking occurs before or after inflation crucially affects the resulting axion abundance. If the PQ symmetry is broken during inflation, the initial value of the axion field is homogenized over our observable Universe by inflation. For the axion density parameter from the misalignment mechanism of this case, we quote the number given in Ref.\ \cite{ParticleDataGroup:2024cfk}:
\begin{align}
  \Omega_a^{\rm (mis)} h^2 &\, \simeq 0.12
  \left( \frac{m_a}{6\ \mu{\rm eV}} \right)^{-1.165}
  \Theta_{\rm i}^2 F,
\end{align}
where $\Theta_{\rm i}$ is the initial misalignment angle of the axion, essentially fixed by the inflationary epoch, and $F \sim O(1)$ denotes a correction factor that accounts for anharmonic effects in the axion potential. Thus, in this case, the axion relic abundance depends on the initial misalignment angle, which is not known a priori. On the other hand, if the PQ symmetry is broken after inflation, the initial misalignment angle varies randomly in different causally disconnected regions (patches) of the Universe. In this scenario, the relic density is typically calculated using the root-mean-square average of the misalignment angle, leading to \cite{ParticleDataGroup:2024cfk}
\begin{align}
  \Omega_a^{\rm (mis)} h^2 &\, \simeq 0.12
  \left( \frac{m_a}{30\ \mu{\rm eV}} \right)^{-1.165}.
\end{align}
In addition to the misalignment mechanism, the decay of axionic cosmic strings and domain walls provides further contributions to the axion abundance. Estimates of the contributions from these topological defects are subject to significant theoretical uncertainties, owing both to the complexity of their dynamics and to the challenges involved in numerical simulations. A detailed discussion of these calculations is beyond the scope of this article; here we just quote a plausible range of axion mass giving the correct DM abundance in post-inflationary PQ breaking scenarios given in Ref.\ \cite{ParticleDataGroup:2024cfk}, which is based on Refs.\ \cite{Klaer:2017ond, Gorghetto:2020qws}:
\begin{align}
  m_a &\, \sim 26\ \mu{\rm eV} - 0.5\ {\rm meV}.
\end{align}

The possible existence of axion-like particles (ALPs) has also been discussed, especially in string-inspired scenarios \cite{Svrcek:2006yi, Arvanitaki:2009fg, Cicoli:2012sz}. An ALP, which we also denote by $a$, is a particle (or field) described by the following effective Lagrangian:
\begin{align}
  \mathcal{L} = \frac{1}{2} \partial_\mu a \partial^\mu a -\frac{1}{2} m_a^2 a^2
  + \frac{1}{4} g_{agg} a\, G_{\mu\nu}^{(a)} \tilde{G}^{(a)\mu\nu}
  + \frac{1}{4} g_{a\gamma\gamma} a\, F_{\mu\nu} \tilde{F}^{\mu\nu} + \cdots,
\end{align}
where we have omitted additional interactions that are not relevant to the present discussion. This Lagrangian is similar in form to that of the QCD axion. However, unlike the QCD axion, for which the mass and couplings are tied to a single scale $f_a$, the ALP mass $m_a$ and the coupling constants are typically treated as independent free parameters. In particular, the ALP mass can arise from various sources and, in general, is not determined by QCD dynamics, implying that ALP models span a much broader parameter space than QCD axion models. Oscillating ALPs are also viable candidates for wave-like DM. Both the QCD axion and ALPs are hereafter referred to as ``axions.''

For the detection of axion DM, the axion-photon-photon coupling is frequently used. In particular, through this coupling, an axion oscillation can be converted into an oscillating electric field in the presence of an external magnetic field. Neglecting the spatial dependence of the axion field, the modified Maxwell equation in the presence of the axion background is given by
\begin{align}
  \Box \vec{A} = g_{a\gamma\gamma} \dot{a} \vec{B}^{\rm (ext)},
\end{align}
where $\vec{B}^{\rm (ext)}$ denotes the external magnetic field. For a static and spatially uniform $\vec{B}^{\rm (ext)}$, the induced electric field is then
\begin{align}
  \vec{E} = - g_{a\gamma\gamma} a \vec{B}^{\rm (ext)},
  \label{EfieldAxion}
\end{align}
where we have used the axion equation of motion $\ddot{a}=-m_a^2 a$.

\subsection{Dark Photon}
\label{subsec:darkphoton}

Another well-known candidate for wave-like DM is the dark photon, which is a massive vector field that kinetically mixes with the ordinary photon. (A detailed discussion of dark photon DM can be found in, e.g., Ref.\ \cite{Caputo:2021eaa}.)

In this section, we summarize the properties of the dark photon. We focus on the regime where the relevant energies are much smaller than the electron mass; thus, we work in the corresponding low-energy effective theory.

The Lagrangian for a theory containing a dark photon (denoted by $X_\mu$ hereafter) is given by
\begin{align}
  \mathcal{L} = -\frac{1}{4} F_{\mu\nu} F^{\mu\nu}
  -\frac{1}{4} X_{\mu\nu} X^{\mu\nu}  + \frac{1}{2} \epsilon F_{\mu\nu} X^{\mu\nu}
  + \frac{1}{2} m_X^2 X_\mu X^\mu +
  e J_\mu^{\rm (EM)} A^\mu,
  \label{L_darkphoton}
\end{align}
where $X_{\mu\nu}\equiv\partial_\mu X_\nu-\partial_\nu X_\mu$ is the field-strength tensor of the dark photon field. Here, $J_\mu^{\rm (EM)}$ denotes the EM current. The parameter $m_X$ is the dark photon mass, and $\epsilon$ is the kinetic-mixing parameter, which is typically assumed to satisfy $\epsilon \ll 1$. It is important to note that, in this basis (referred to as the ``kinetic-mixing basis''), the dark photon $X_\mu$ does not have a direct coupling to the EM current. However, due to its kinetic mixing with the ordinary photon field $A_\mu$, the dark photon effectively interacts with $J_\mu^{\rm (EM)}$.

The nature of this interaction becomes more transparent upon changing to a different field basis. By redefining the photon field as
\begin{align}
  A'_\mu \equiv A_\mu - \epsilon X_\mu,
\end{align}
the kinetic-mixing term is eliminated. The Lagrangian then takes the form
\begin{align}
  \mathcal{L} =
  -\frac{1}{4} F'_{\mu\nu} F'^{\mu\nu}
  -\frac{1}{4} X_{\mu\nu} X^{\mu\nu}
  +\frac{1}{2} m_X^2 X_\mu X^\mu
  + e J_\mu^{\rm (EM)} (A'^\mu + \epsilon X^\mu)
  + \mathcal{O}(\epsilon^2),
\end{align}
where $F'_{\mu\nu}\equiv\partial_\mu A'_\nu-\partial_\nu A'_\mu$. Since $\epsilon \ll 1$, terms of order $\mathcal{O}(\epsilon^2)$ can be neglected. In this basis (often called the ``mass-eigenstate basis''), the dark photon field $X_\mu$ couples directly to the EM current with a strength suppressed by $\epsilon$.

Alternatively, one may perform a further field redefinition:
\begin{align}
  X_\mu''\equiv X_\mu - \epsilon A_\mu.
\end{align}
In this basis (which is sometimes called the ``interaction basis'' \cite{Jaeckel:2012mjv}), the Lagrangian becomes
\begin{align}
  \mathcal{L} &=
  -\frac{1}{4} F_{\mu\nu} F^{\mu\nu}
  -\frac{1}{4} X''_{\mu\nu} X''^{\mu\nu}
  +\frac{1}{2} m_X^2 (X''_\mu X''^\mu + 2\epsilon A_\mu X''^\mu)
  + e J_\mu^{\rm (EM)} A^\mu
  + \mathcal{O}(\epsilon^2),
\end{align}
with $X''_{\mu\nu}\equiv\partial_\mu X''_\nu-\partial_\nu X''_\mu$. In this description, the kinetic mixing term is removed, but there is a mass-mixing term between the $A_\mu$ and $X''_\mu$ fields.

It should be emphasized that these three descriptions, the kinetic-mixing basis, the mass-eigenstate basis, and the interaction basis, are physically equivalent, and the same physics is reproduced in all of them. In practice, however, a particular basis may be more convenient for discussing specific situations or calculations.

Let us now consider the classical dynamics of the dark photon field. Because the kinetic mixing parameter is expected to be extremely small, its effect on the motion of the dark photon is generally negligible, allowing us to focus on the dynamics in the limit $\epsilon \rightarrow 0$. In the flat Minkowski background (which well approximates the environment around the experimental apparatus), the equation of motion for the dark photon field then reads
\begin{align}
  \Box X_\mu + m_X^2 X_\mu = 0, ~~~ \partial_\mu X^\mu = 0.
  \label{darkphotonEoM}
\end{align}
In addition, the energy density is given by
\begin{align}
  \rho = \frac{1}{2} \vec{E}_X^2 + \frac{1}{2} \vec{B}_X^2 
  + \frac{1}{2} m_X^2 \left( \vec{X}^2 + X_0^2 \right),
\end{align}
where $\vec{X}\equiv (X^1, X^2, X^3)$ and
\begin{align}
  \vec{E}_X \equiv -\vec{\nabla} X_0 - \partial_t \vec{X},~~~
  \vec{B}_X \equiv \vec{\nabla} \times \vec{X}.
\end{align}
Since the dark photon is a vector field, its equation of motion is more complicated than that of a (free) scalar field. Nevertheless, in the non-relativistic regime of primary interest here, the dynamics of the dark photon is qualitatively similar to that of a scalar field.

Taking into account that the dark photon DM has a velocity distribution, we can write the ``vector potential'' of the dark photon in the non-relativistic limit as
\begin{align}
  \vec{X} (t, \vec{x}) = 
  \sum_{\vec{v}}  
  \tilde{X}_{\vec{v}} \, \vec{n}_{\vec{v}}\, 
  \cos \left( E_v t - m_X \vec{v} \vec{x} - \varphi_{\vec{v}} \right),
  \label{vec(X)full}
\end{align}
where $\tilde{X}_{\vec{v}}$ is the Fourier amplitude, $\vec{n}_{\vec{v}}$ is the polarization vector (with $|\vec{n}_{\vec{v}}|=1$), $\varphi_{\vec{v}}$ is a random phase, and $E_v \simeq m_X + \frac{1}{2} m_X v^2$. Around the Solar system, we expect $\tilde{X}_{\vec{v}}$ to be significant for $v\lesssim O(v_{\rm DM})$ and suppressed for larger $v$.

From Eq.~\eqref{vec(X)full} and the equation of motion~\eqref{darkphotonEoM}, one finds that
\begin{align}
  X_0 (t, \vec{x}) = - 
  \sum_{\vec{v}}
  \tilde{X}_{\vec{v}}\, \frac{m_X\, (\vec{v} \vec{n}_{\vec{v}})}{E_v}\, 
  \cos \left( E_v t - m_X \vec{v} \vec{x} - \varphi_{\vec{v}} \right).
  \label{X0}
\end{align}
As evident, $X_0$ is suppressed by the DM velocity in the non-relativistic limit compared to $\vec{X}$.

For simplicity, we will henceforth consider the case in which $\vec{n}_{\vec{v}}$ and $\varphi_{\vec{v}}$ are randomly distributed as functions of $\vec{v}$, so that the dark photon DM behaves as an isotropic fluid. For time scales shorter than the DM coherence time $\tau_{\rm DM}$, the dark photon vector potential at the experimental apparatus can be approximated as
\begin{align}
  \vec{X} (t) \simeq \bar{X} \vec{n}_X \cos \left( m_X t - \varphi \right),
  \label{vec(X)}
\end{align}
where $\bar{X}$ is the overall amplitude, $\vec{n}_X$ is a polarization vector with $|\vec{n}_X|=1$, and $\varphi$ is an effective random phase; $\vec{n}_X$ and $\varphi$ are reset randomly on the time scale of the DM coherence time. Neglecting terms suppressed by the DM velocity, the energy density of the dark photon is then given by
\begin{align}
  \rho \simeq \frac{1}{2} m_X^2 \bar{X}^2.
\end{align}
As in the case of scalar DM, the oscillation-averaged pressure is negligible, meaning that the dark photon behaves as CDM. The amplitude $\bar{X}$ of the dark photon oscillation is directly related to the DM energy density via
\begin{align}
  \bar{X} \simeq \frac{\sqrt{2 \rho_{\rm DM}}}{m_X}.
\end{align}
In addition, the effective electric field induced by the dark photon oscillation is given by
\begin{align}
  \vec{E} = - \epsilon \dot{\vec{X}},
  \label{EfieldDPH}
\end{align}
and hence the strength of the DM-induced electric field is $\epsilon \sqrt{2\rho_{\rm DM}}$. The dark photon mass relevant to explaining the present DM density depends on its production mechanism (see, e.g., Refs.\ \cite{Graham:2015rva, Tang:2017hvq, Agrawal:2018vin, Dror:2018pdh, Co:2018lka, Bastero-Gil:2018uel, Long:2019lwl, Nakayama:2019rhg, Kolb:2020fwh, Nakayama:2021avl, Wang:2022ojc, Sato:2022jya, Kasamaki:2026pri}). 

\section{Superconducting Qubit}
\label{app:superconducting}
\setcounter{equation}{0}
\setcounter{figure}{0}
\setcounter{footnote}{1}

In this Appendix, we briefly summarize the basic properties of superconducting qubits that can be used in DM search experiments. (For more details on superconducting qubits, see, e.g., \cite{Clarke:2008ufm}.) We explain the equivalent circuit of a superconducting qubit with a single Josephson junction, as well as that of a SQUID-based qubit, in which the resonance frequency of the qubit is tunable. We also discuss a procedure for calculating the coupling strength between the qubit and an external electric field, which may arise from DM oscillations.

\subsection{Simplest Example of Superconducting Qubit}

Basic properties of superconducting qubits (in particular, the transmon qubit) can be understood by considering an equivalent circuit consisting of a capacitor and a Josephson junction (JJ).

\begin{figure}[t]
  \centering
  \includegraphics[height=0.15\textheight]{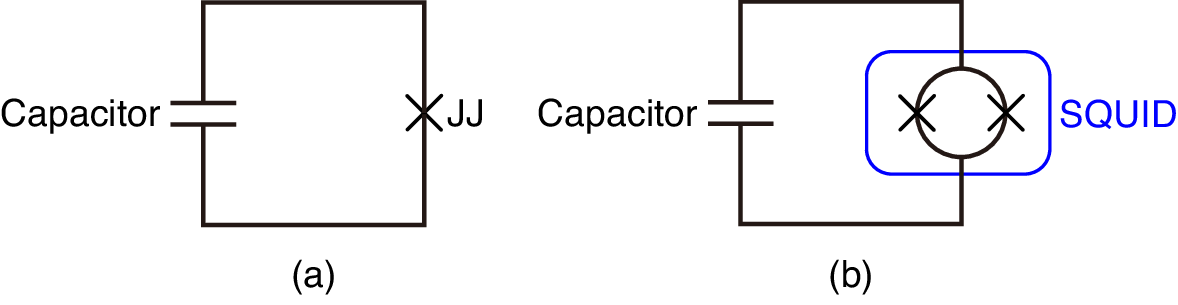}
  \caption{Equivalent circuits of (a) superconducting qubit with single JJ and (b) SQUID-based qubit.}
  \label{fig:qubitcircuit}
\end{figure}

Let us consider a circuit composed of a capacitor with capacitance $C$ and a JJ with Josephson energy $E_{\rm J}$ (see Fig.\ \ref{fig:qubitcircuit}(a)), describing the simplest superconducting qubit. Since a JJ behaves as an inductor, one classically expects the circuit to sustain an oscillating (AC) current in the absence of dissipation. An important difference from a conventional linear LC circuit is that the effective inductance is nonlinear; the JJ acts as an anharmonic inductor, and therefore the oscillation frequency depends on the amplitude, in contrast to the linear LC case.

\begin{figure}[t]
  \centering
  \includegraphics[height=0.15\textheight]{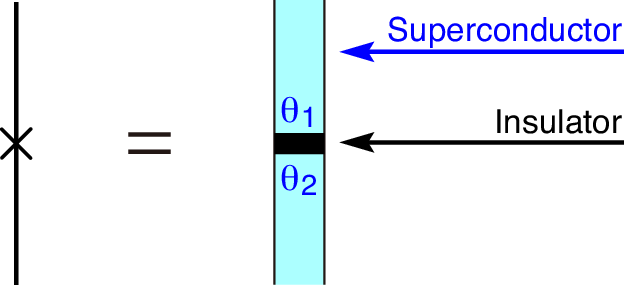}
  \caption{Schematic pictures of the Josephson Junction.}
  \label{fig:JosephsonJunction}
\end{figure}

More quantitatively, the dynamics can be analyzed using the Hamiltonian description. A JJ consists of two superconductors separated by a thin insulating barrier, as shown in Fig.\ \ref{fig:JosephsonJunction} (for more detail, see, e.g., Ref.\ \cite{KittelBook}). Denoting the condensate phases of Cooper pairs on the two sides of the junction by $\theta_1$ and $\theta_2$, respectively, the dynamics are described by the physical (gauge-invariant) phase difference
\begin{align}
  \theta \equiv \theta_2 - \theta_1,
\end{align}
in terms of which the potential energy of the JJ is
\begin{align}
  H_{\rm JJ} = -E_{\rm J}\cos\theta.
\end{align}
Denoting by $Q$ the charge stored in the capacitor, the total Hamiltonian of the circuit is
\begin{align}
  H = \frac{1}{2C} Q^2 - E_{\rm J}\cos\theta.
\end{align}

Using the Josephson relation, which connects the time derivative of $\theta$ to the voltage difference $V$ across the junction,
\begin{align}
  \partial_t \theta = 2 e V,
\end{align}
we obtain
\begin{align}
  Q = \frac{C}{2e} \dot{\theta}.
\end{align}
The Hamiltonian can then be rewritten as
\begin{align}
  H = \frac{1}{2} \frac{C}{(2e)^2} \dot{\theta}^2 - E_{\rm J}\cos\theta =
  \frac{1}{2Z} n^2 - E_{\rm J}\cos\theta,
  \label{H_circuit}
\end{align}
where, in the second equality, we introduced
\begin{align}
  n \equiv \frac{Q}{2e},
  \label{n=Q/2e}
\end{align}
which serves as the canonical momentum conjugate to $\theta$, and
\begin{align}
  Z \equiv \frac{C}{(2e)^2}.
\end{align}
In Eq.\ \eqref{H_circuit}, the first and second terms represent the electrostatic energy of the capacitor and the Josephson potential energy, respectively. Interpreting the first and second terms as kinetic and potential contributions, one sees that the circuit is equivalent to a mechanical system in a cosine potential.

We quantize the system by promoting $\theta$ and $n$ to operators and imposing the canonical commutation relation
\begin{align}
  [\theta, n] = i.
\end{align}

In the following, we focus on the so-called transmon limit, $CE_{\rm J}\gg (2e)^2$ \cite{Koch:2007hay}; in such a limit, quantum fluctuations of $\theta$ are suppressed. In this regime, we may expand the potential around $\theta=0$ and keep the leading nontrivial term. Dropping an irrelevant constant energy shift, we obtain
\begin{align}
  H^{(0)} \equiv \frac{1}{2Z} n^2 + \frac{1}{2} E_{\rm J} \theta^2.
\end{align}
This is the Hamiltonian of a harmonic oscillator. Its spectrum is conveniently described by introducing annihilation and creation operators,
\begin{align}
  a \equiv \frac{1}{\sqrt{2\omega Z}} (n - i \omega Z \theta),~~~
  a^\dagger \equiv \frac{1}{\sqrt{2\omega Z}} (n + i \omega Z \theta),
\end{align}
and hence
\begin{align}
  \theta = &\, i \sqrt{\frac{1}{2\omega Z}} (a - a^\dagger),
  \\
  n = &\, \sqrt{\frac{\omega Z}{2}} (a + a^\dagger).
  \label{theta&n}
\end{align}
Here, 
\begin{align}
  \omega \equiv \sqrt{\frac{E_{\rm J}}{Z}},
  \label{omega_qubit}
\end{align}
which will be regarded as the qubit frequency in the following. One can verify that
\begin{align}
  [a, a^\dagger] = 1.
\end{align}
The Hamiltonian can be written as
\begin{align}
  H^{(0)} = \omega a^\dagger a,
\end{align}
where we again neglect a constant term. The $I$-th energy eigenstate is then
\begin{align}
  \ket{I} \equiv \frac{1}{\sqrt{I!}} (a^\dagger)^I \ket{0}~~~
  (I=0, 1, 2, \cdots),
\end{align}
with $\ket{0}$ the ground state, and the corresponding eigenenergy is
\begin{align}
  E^{(0)}_I = I \omega.
  \label{Energy_I(0)}
\end{align}
At this harmonic level, the level spacings are uniform, as in Eq.\ \eqref{Energy_I(0)}. However, because the JJ is intrinsically anharmonic, this universality is broken once higher-order terms in the expansion of $\cos\theta$ are included. In the transmon limit, these effects can be incorporated perturbatively. Keeping the next-to-leading correction, the $I$-th energy level is approximately
\begin{align}
  E_I \simeq I \omega -\frac{1}{32 Z} (2I^2 + 2I + 1),
\end{align}
so that the adjacent level spacing becomes
\begin{align}
  E_{I+1} - E_I \simeq \omega - \frac{1}{8 Z} (I + 1).
\end{align}

In many applications, the ground state $\ket{0}$ and the first excited state $\ket{1}$ of the superconducting circuit serve as the logical states of a qubit. For this purpose, the anharmonicity induced by the JJ is essential. In actual qubit control, one typically applies microwave radiation whose frequency is tuned to the transition energy $E_1-E_0$. In the present circuit, due to anharmonicity, this resonance condition is satisfied predominantly for the $\ket{0}\leftrightarrow\ket{1}$ transition, and the system can be treated (effectively) as a two-level system, i.e., a qubit. By contrast, in a perfectly harmonic system such as a linear LC circuit, all level spacings are identical; therefore, a drive resonant with $E_1-E_0$ would also resonantly excite higher transitions, and the system cannot be well approximated as a two-level system.

Hereafter, we truncate the Hilbert space to the two-dimensional subspace spanned by $\{\ket{0},\ket{1}\}$. In this subspace, the annihilation and creation operators may be reinterpreted as
\begin{align}
  a = \ketbra{0}{1}, ~~~
  a^\dagger = \ketbra{1}{0}.
\end{align}
Furthermore, neglecting higher-order corrections (or, equivalently, redefining $\omega\equiv E_1-E_0$), the Hamiltonian reduces to
\begin{align}
  H = \omega a^\dagger a = \omega \ketbra{1}{1}.
\end{align}

\subsection{SQUID-Based Qubit}

So far, we have considered a superconducting qubit that contains a single JJ. In that case, the qubit frequency $\omega$ is determined by circuit parameters, in particular the capacitance $C$ and the Josephson energy $E_{\rm J}$. In many practical situations, including applications to DM searches, it is advantageous to employ a SQUID-based qubit, where the single JJ is replaced by a superconducting quantum interference device (SQUID), i.e., a superconducting loop interrupted by one or more JJs. The main advantage of the SQUID-based qubit is that the qubit frequency becomes tunable by varying the external magnetic field applied to the qubit. 

\begin{figure}[t]
  \centering
  \includegraphics[height=0.15\textheight]{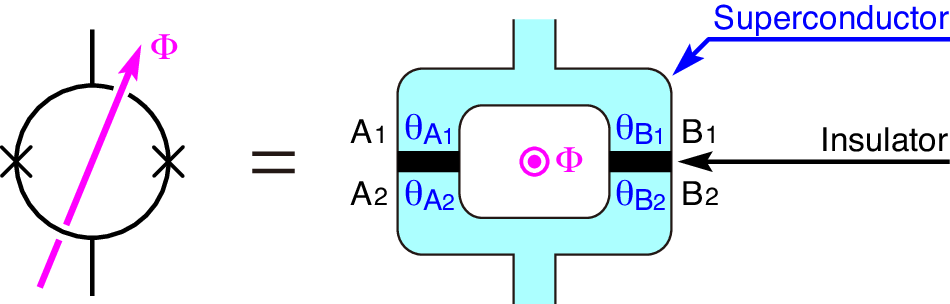}
  \caption{Schematic pictures of the SQUID.}
  \label{fig:SQUID}
\end{figure}

Hereafter, we consider a SQUID consisting of two JJs, as shown in Fig.\ \ref{fig:SQUID}, and a qubit circuit incorporating such a SQUID. We begin by discussing the basic properties of the SQUID depicted in Fig.\ \ref{fig:SQUID}. The SQUID contains two JJs, labeled $A$ and $B$. The points immediately above (below) JJs $A$ and $B$ are denoted by $A_1$ and $B_1$ ($A_2$ and $B_2$), respectively. The Josephson energies of JJs $A$ and $B$ are denoted by $E_{{\rm J},A}$ and $E_{{\rm J},B}$, respectively. Let $\theta_{A_1}$ and $\theta_{B_1}$ ($\theta_{A_2}$ and $\theta_{B_2}$) be the condensate phases of the Cooper-pair wave function at $A_1$ and $B_1$ ($A_2$ and $B_2$). Then, neglecting charging effects for the moment and focusing on the Josephson potential energy, the SQUID Hamiltonian is given by
\begin{align}
  H_{\rm SQUID} = - E_{{\rm J},A} \cos \left( \theta_{A_2} - \theta_{A_1} \right)
  - E_{{\rm J},B} \cos \left( \theta_{B_2} - \theta_{B_1} \right).
\end{align}

Next, we consider the case in which a non-vanishing magnetic flux $\Phi$ threads the SQUID loop, 
\begin{align}
  \Phi \equiv \oint d\vec{x} \vec{A} (\vec{x}),
\end{align}
where $\vec{A}$ is the vector potential of the EM field, and the integration is along the SQUID loop. In the presence of such a flux, the superconducting phases around the loop must satisfy the gauge-invariant relations:
\begin{align}
  \theta_{B_1} - \theta_{A_1} = &\, 2 e \int_{A_1\rightarrow B_1}
  d\vec{x} \vec{A} (\vec{x}),
  \\
  \theta_{A_2} - \theta_{B_2} = &\, 2 e \int_{B_2\rightarrow A_2}
  d\vec{x} \vec{A} (\vec{x}).
\end{align}
These relations express the fact that phase differences between two points depend on the EM gauge field along a chosen path, such that physical quantities depend only on gauge-invariant combinations.

Defining the gauge-invariant phase drops across the two junctions as
\begin{align}
  \theta_A \equiv \theta_{A_2}- \theta_{A_1},~~~
  \theta_B \equiv \theta_{B_2}- \theta_{B_1},
\end{align}
we obtain the flux constraint
\begin{align}
  \theta_B-\theta_A = 2 e \oint d\vec{x} \vec{A} (\vec{x})
  = 2 e \Phi = \frac{2\pi}{\Phi_0} \Phi,
\end{align}
where $\Phi_0$ is the magnetic flux quantum. This relation shows that the difference between the phase drops across the two junctions is fixed by the magnetic flux $\Phi$.

It is convenient to parameterize $\theta_A$ and $\theta_B$ as
\begin{align}
  \theta_A = \theta + \bar{\theta}_A,~~~\theta_B = \theta + \bar{\theta}_B,
\end{align}
where $\theta$ represents the dynamical degree of freedom of the SQUID (which will become the relevant phase variable of the effective qubit), while $\bar{\theta}_A$ and $\bar{\theta}_B$ are constants satisfying $\bar{\theta}_B-\bar{\theta}_A=2e\Phi$. We choose $\bar{\theta}_A$ and $\bar{\theta}_B$ so that $\theta=0$ corresponds to the minimum of the Josephson potential. With this choice, one finds
\begin{align}
  \tan \bar{\theta}_A = &\, -\frac{E_{{\rm J},B} \sin 2e\Phi}{E_{{\rm J},A}+E_{{\rm J},B} \cos 2e\Phi},
  \\
  \tan \bar{\theta}_B = &\, \frac{E_{{\rm J},A} \sin 2e\Phi}{E_{{\rm J},A} \cos 2e\Phi+E_{{\rm J},B}},
\end{align}
and
\begin{align}
  \cos \bar{\theta}_A = &\, \frac{E_{{\rm J},A}+E_{{\rm J},B} \cos 2e\Phi}
  {\sqrt{E_{{\rm J},A}^2+E_{{\rm J},B}^2+2E_{{\rm J},A}E_{{\rm J},B}\cos 2e\Phi}},
  \\
  \cos \bar{\theta}_B = &\, \frac{E_{{\rm J},A} \cos 2e\Phi+E_{{\rm J},B}}
  {\sqrt{E_{{\rm J},A}^2+E_{{\rm J},B}^2+2E_{{\rm J},A}E_{{\rm J},B}\cos 2e\Phi}}.
\end{align}
Substituting these relations back into the SQUID Hamiltonian, the two-junction system can be written in the form of an effective single junction with a flux-dependent Josephson energy:
\begin{align}
  H_{\rm SQUID} = - E_{\rm J}^{\rm (eff)} \cos\theta,
\end{align}
where
\begin{align}
  E_{\rm J}^{\rm (eff)} \equiv
  \sqrt{E_{{\rm J},A}^2 + E_{{\rm J},B}^2 + 2E_{{\rm J},A} E_{{\rm J},B} \cos 2e\Phi}.
  \label{EJ(eff)}
\end{align}
Therefore, the SQUID behaves as a tunable Josephson element; its effective Josephson energy depends on the magnetic flux going through the SQUID loop. 

Now, we consider the qubit consisting of a capacitor as well as a SQUID, as shown in Fig.\ \ref{fig:qubitcircuit}(b). The dynamics of the SQUID-based qubit can be understood by simply replacing the Josephson energy $E_{\rm J}$ by the effective Josephson energy given in Eq.\ \eqref{EJ(eff)}. Then, based on Eq.\ \eqref{omega_qubit}, we can find that, by varying the external magnetic field and hence $\Phi$, one can tune the qubit frequency. This tunability is particularly useful, for example, in qubit-based DM search experiments, where scanning the qubit frequency allows one to probe different candidate DM masses.

\subsection{Interaction of Superconducting Qubit with Electric Field}

So far, we have discussed the behavior of a superconducting qubit that is isolated from external EM fields. In practice, however, quantum operations on the qubit are typically implemented by applying an EM drive in the microwave frequency range. In this section, we describe how to model the interaction between a superconducting qubit and an external EM field.

Because a superconducting qubit includes a capacitive element, it couples capacitively to an external electric field. Let $\vec{d}$ denote the (effective) separation vector between the two plates (or pads) of the capacitor. We parameterize $\vec{d}$ as
\begin{align}
  \vec{d} = d\, \vec{n}_{\rm q},
  \label{dipolevec}
\end{align}
where $d \equiv |\vec{d}|$ denotes the magnitude of the separation, and $\vec{n}_{\rm q}$ is a unit vector parallel to $\vec{d}$ which indicates the direction of maximum sensitivity of the qubit to external electric fields. Then the electric dipole moment associated with the capacitor is given by $Q\vec{d}$, where $Q$ is the charge stored on the capacitor. The interaction Hamiltonian with an external electric field $\vec{E}$ is therefore written as
\begin{align}
  H_{\rm int} = - Q \vec{d} \vec{E}.
\end{align}
In the quantum description of the circuit, $Q$ is promoted to an operator. Using the relations between $Q$ and the ladder operators (see Eqs.\ \eqref{n=Q/2e} and \eqref{theta&n}), one finds that $Q$ can be expressed as
\begin{align}
  Q = \sqrt{\frac{C\omega}{2}} (a + a^\dagger),
\end{align}
and hence the interaction Hamiltonian becomes
\begin{align}
  H_{\rm int} = - \vec{d} \vec{E} \sqrt{\frac{C\omega}{2}}
  (\ketbra{0}{1} + \ketbra{1}{0}).
\end{align}

A particularly important case is a monochromatic drive, where the electric field oscillates in time with a fixed angular frequency $m$. In that case, we may write
\begin{align}
  \vec{E} = \bar{E} \vec{n}_E \cos (mt -\varphi),
\end{align}
where $\bar{E}$ is the field amplitude, $\vec{n}_E$ is the polarization unit vector, and $\varphi$ is the drive phase. In such a case, the effective Hamiltonian describing the qubit driven by the external electric field can be written as
\begin{align}
  H = \omega\ketbra{1}{1} - 2 \eta (\ketbra{0}{1} + \ketbra{1}{0}) \cos (mt -\varphi),
\end{align}
with the drive strength
\begin{align}
  \eta \equiv \frac{1}{2\sqrt{2}} d \sqrt{C\omega} \bar{E} \, (\vec{n}_{\rm q} \vec{n}_E).
  \label{eta_DM}
\end{align}
This is nothing but the Hamiltonian used in the main text (see Eqs.\ \eqref{Hamiltonian_tot} -- \eqref{Hamiltonian_1}).

We note here that, for the case of DM-induced electric field, the phase $\varphi$ cannot be treated as a time-independent variable; in such a case, $\varphi$ is approximately a constant on a time scale shorter than $\tau_{\rm DM}$ while it effectively experiences random resets at a time scale $\tau_{\rm DM}$, where $\tau_{\rm DM}$ is the coherence time of DM (see Section \ref{sec:evolution} as well as Appendix \ref{app:dm} for more details).
 
\subsection{Superconducting Qubit for a Real Experiment}

\begin{figure}[t]
  \centering
  \includegraphics[width=0.5\textwidth]{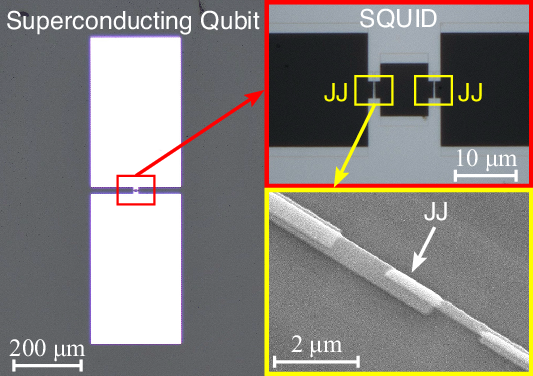}
  \caption{Micrograph of a qubit fabricated for the DarQ-Direct Experiment \cite{DarQ}.}
  \label{fig:qubit_KWatanabe}
\end{figure}

Here, we present a realization of a superconducting qubit, using the device developed for the DarQ-Direct experiment as a representative example. Fig.\ \ref{fig:qubit_KWatanabe} shows a micrograph of the qubit fabricated for the DarQ-Direct experiment.\footnote
{Figure courtesy of K.~Watanabe (DarQ-Direct collaboration).}
Two Nb thin-film electrodes on a sapphire substrate form a capacitor, and an Al/AlO$_x$/Al Josephson junction is integrated (shown in the left panel). The rectangular regions correspond to the Nb electrodes, which are electrically isolated from each other and therefore act as the two plates of a capacitor. These electrodes are interconnected via a superconducting quantum interference device (SQUID), which allows tunability of the Josephson energy; a magnified view of the SQUID loop is shown in the top-right panel. The bottom-right panel displays a close-up image of the Josephson junction.

When the device is cooled, the Nb films transition into the superconducting state, enabling the system to function as a superconducting qubit. In the measurement setup, the qubit chip is mounted inside a Cu microwave cavity for the control and readout. To allow tuning of the qubit transition frequency, a magnetic field is applied to the SQUID by means of a superconducting coil wound around the cavity. This experimental arrangement permits precise adjustment of the qubit characteristics via external magnetic flux.

\section{Coherent State}
\label{app:coherent}
\setcounter{equation}{0}
\setcounter{figure}{0}
\setcounter{footnote}{1}

In this Appendix, we summarize the basic properties of coherent states, which are useful, for example, in analyzing the EM field inside a cavity.

\subsection{Coherent State of a Single Mode}

We focus on a single mode of the cavity field. The annihilation and creation operators for the mode of interest are denoted by $c$ and $c^\dagger$, respectively. They satisfy the canonical commutation relation
\begin{align}
  [c, c^\dagger] = 1.
\end{align}
We also introduce the so-called quadrature operators
\begin{align}
  x_c \equiv &\, \frac{c+c^\dagger}{\sqrt{2}},
  \\
  p_c \equiv &\, \frac{c-c^\dagger}{\sqrt{2}i},
\end{align}
which obey
\begin{align}
  [x_c, p_c] = i.
\end{align}

A coherent state is defined by
\begin{align}
  \ket{\alpha} \equiv e^{-|\alpha|^2/2} e^{\alpha c^\dagger}\ket{0},
\end{align}
where $\alpha$ is a complex number characterizing the coherent state and $\ket{0}$ is the vacuum of this mode, i.e., $c\ket{0}=0$. The coherent state is an eigenstate of the annihilation operator,
\begin{align}
  c \ket{\alpha} = \alpha \ket{\alpha},
\end{align}
and the overlap between two coherent states is given by
\begin{align}
  \langle \beta | \alpha \rangle = \exp\left(\beta^* \alpha - \frac{1}{2}|\alpha|^2 - \frac{1}{2} |\beta|^2 \right),
\end{align}
and hence $\braket{\alpha}{\alpha}=1$.

It is also noteworthy that coherent states provide an (overcomplete) resolution of the identity. Specifically, one has
\begin{align}
  \mathds{I} =
  \frac{1}{\pi} \int d^2 \alpha \ketbra{\alpha}{\alpha},
\end{align}
where the integration over the complex variable is defined as
\begin{align}
  \int d^2 \alpha 
  \equiv
  \int_{-\infty}^\infty d \alpha_{\rm R}
  \int_{-\infty}^\infty d \alpha_{\rm I}, 
\end{align}
with
\begin{align}
  \alpha_{\rm R} \equiv \Re \alpha, ~~~\alpha_{\rm I} \equiv \Im \alpha.
\end{align}
Using this relation, the trace of an operator $A$ can be written as
\begin{align}
  \mathrm{tr} (A) = \frac{1}{\pi} \int d^2 \alpha \mel{\alpha}{A}{\alpha}.
\end{align}
In particular, for the density matrix $\rho$, the quasi-probability distribution over coherent states is known as the Husimi $Q$-function \cite{Husimi1940}:
\begin{align}
  Q (\alpha) \equiv \frac{1}{\pi} \mel{\alpha}{\rho}{\alpha}
  = \mathrm{tr} \left( \frac{1}{\pi} \ketbra{\alpha}{\alpha} \rho \right).
\end{align}

We may also consider the eigenstates of $x_c$ and $p_c$, respectively:\footnote
{The eigenstates of $x_c$ and $p_c$ are normalized as $\braket{x}{x'} = \delta (x-x')$ and $\braket{p}{p'} = \delta (p-p')$, respectively.
}
\begin{align}
  x_c \ket{x} = &\, x \ket{x},
  \\
  p_c \ket{p} = &\, p \ket{p}.
\end{align}
Their inner products with a coherent state are given by
\begin{align}
  \braket{x}{\alpha} = &\, \frac{1}{\pi^{1/4}}
  \exp \left[
  -\frac{1}{2}x^2 + \sqrt{2} \alpha x - \frac{1}{2} (|\alpha|^2 + \alpha^2)
  \right],
  \\
  \braket{p}{\alpha} = &\, \frac{1}{\pi^{1/4}}
  \exp \left[
  -\frac{1}{2}p^2 -i \sqrt{2} \alpha p - \frac{1}{2} (|\alpha|^2 - \alpha^2)
  \right].
\end{align}

\subsection{Displacement Operator}

It is often convenient to introduce the displacement operator to study the properties of the coherent state \cite{Glauber:1963tx, Cahill:1969it, Cahill:1969iq}:
\begin{align}
  D (\alpha) \equiv e^{\alpha c^\dagger - \alpha^* c}
  = e^{-|\alpha|^2/2} e^{\alpha c^\dagger} e^{- \alpha^* c}
  = e^{|\alpha|^2/2} e^{- \alpha^* c} e^{\alpha c^\dagger}.
  \label{DisplacementOp}
\end{align}
The displacement operator is unitary and satisfies
\begin{align}
  D^{-1} (\alpha) = D^\dagger (\alpha) = D (-\alpha).
\end{align}
A coherent state can be constructed by applying the displacement operator to the vacuum,
\begin{align}
  \ket{\alpha} = D (\alpha) \ket{0},
\end{align}
and, more generally,
\begin{align}
  D (\alpha) \ket{\beta} = e^{(\alpha \beta^* - \alpha^* \beta)/2} \ket{\alpha+\beta}.
\end{align}

Importantly, the displacement operator satisfies the completeness relation
\begin{align}
  \mathrm{tr} \left[ D(\alpha) D(-\beta) \right] =
  \pi \delta^{(2)} (\alpha - \beta),
\end{align}
and one can represent an arbitrary operator $A$ as
\begin{align}
  A = \frac{1}{\pi} \int d^2 \alpha\, 
  \mathrm{tr} \left[ A D(\alpha) \right] D(-\alpha).
\end{align}
In particular, the projectors onto the eigenstates of $x_c$ and $p_c$ can be expressed as
\begin{align}
  \ketbra{x}{x} = \frac{1}{\sqrt{2}\pi} \int_{-\infty}^\infty
  d \alpha_{\rm I}\, e^{i\sqrt{2} \alpha_{\rm I} x} D (-i\alpha_{\rm I}),
  \label{projection-xx}
\end{align}
and
\begin{align}
  \ketbra{p}{p} = \frac{1}{\sqrt{2}\pi} 
  \int_{-\infty}^\infty
  d \alpha_{\rm R}\, e^{-i\sqrt{2} \alpha_{\rm R} p} D (-\alpha_{\rm R}),
  \label{projection-pp}
\end{align}
respectively. (Note that the integration variables in the expressions above, $\alpha_{\rm I}$ and $\alpha_{\rm R}$, are real.) Furthermore, for a coherent state, it follows that
\begin{align}
  \ketbra{\alpha}{\alpha} = \frac{1}{\pi} 
  \int d^2\beta e^{(-2\beta^* \alpha + 2\beta \alpha^* - |\beta|^2)/2} D(-\beta).
\end{align}
We can also obtain
\begin{align}
  D (\alpha) = \frac{1}{\pi} e^{-|\alpha|^2/2}
  \int d^2 \beta e^{\beta^* \alpha - \beta \alpha^*} \ketbra{\beta}{\beta}.
  \label{Dwrtcst}
\end{align}

For the density matrix $\rho$, we introduce the so-called characteristic function $\chi$ as
\begin{align}
  \chi (\alpha) \equiv \mathrm{tr} \bigl[ \rho D(\alpha) \bigr],
\end{align}
with which 
\begin{align}
  \rho = \frac{1}{\pi} \int d^2 \alpha \chi (\alpha) D(-\alpha).
\end{align}
With the characteristic function $\chi$, the Husimi $Q$-function can be expressed as
\begin{align}
  Q (\alpha) = \frac{1}{\pi^2} 
  \int d^2\beta e^{(2\beta^* \alpha - 2\beta \alpha^* - |\beta|^2)/2} \chi (\beta).
\end{align}
Formally, the density matrix can also be expressed as
\begin{align}
  \rho = \int d^2 \alpha P (\alpha) \ketbra{\alpha}{\alpha},
\end{align}
where 
\begin{align}
  P (\alpha) \equiv \frac{1}{\pi^2} \int d^2 \beta
  e^{\beta^* \alpha - \beta \alpha^*}
  e^{|\beta|^2/2} \chi (\beta),
\end{align}
which is often called the Glauber-Sudarshan $P$-representation \cite{Glauber:1963tx, Sudarshan:1963ts}. Notice that the $P$-function satisfies
\begin{align}
  \int d^2 \alpha P (\alpha) = 1.
\end{align}

\subsection{$I/Q$ Measurement of the Cavity State}
\label{subsec:cavityreadout}

One important application of the coherent-state formulas discussed so far is the measurement of the cavity state, which is relevant in a variety of experimental contexts (such as the dispersive readout of the qubit state). In particular, one is often interested in obtaining information about both $x_c$ and $p_c$. However, since $x_c$ and $p_c$ do not commute, a simultaneous measurement of these two observables is, strictly speaking, not possible. In what follows, we describe a procedure that allows one to simultaneously acquire information about $x_c$ and $p_c$, which provides a simple example of the measurement of the cavity state. 

The first step is to enlarge the Hilbert space:
\begin{align}
  \mathcal{H}_{\rm Cavity} ~~\rightarrow~~
  \mathcal{H}_{\rm tot} \equiv \mathcal{H}_{\rm Cavity} \otimes \mathcal{H}_{\rm Ancilla},
\end{align}
where $\mathcal{H}_{\rm Cavity}$ is the Hilbert space of the cavity mode, while $\mathcal{H}_{\rm Ancilla}$ is that of the newly introduced ancilla mode which is bosonic. The annihilation and creation operators of the ancilla mode are denoted by $a$ and $a^\dagger$, respectively.\footnote
{One should not confuse $a$ and $a^\dagger$ here with the lowering and raising operators of the qubit introduced in the main part of this article.}
They satisfy the usual commutation relation
\begin{align}
  [a, a^\dagger] = 1.
\end{align}
Since the total Hilbert space is constructed as a tensor product, operators acting on different sectors automatically commute.

We introduce a unitary transformation of the operators. For simplicity, we consider an equal-weight (i.e., $50:50$) mixture of the two operators as
\begin{align}
  \begin{pmatrix}
    b_+ \\ b_-
  \end{pmatrix}
  \equiv
  \frac{1}{\sqrt{2}}
  \begin{pmatrix}
    1 & 1 \\
    1 & -1
  \end{pmatrix}
  \begin{pmatrix}
    c \\ a
  \end{pmatrix} ,~~~
  \begin{pmatrix}
    b_+^\dagger \\ b_-^\dagger
  \end{pmatrix}
  \equiv
  \frac{1}{\sqrt{2}}
  \begin{pmatrix}
    1 & 1 \\
    1 & -1
  \end{pmatrix}
  \begin{pmatrix}
    c^\dagger \\ a^\dagger
  \end{pmatrix} ,
\end{align}
where $b_\pm$ and $b_\pm^\dagger$ denote annihilation and creation operators after the mixture, respectively. A device implementing this transformation is commonly referred to as a ``beam splitter,'' which combines two input light fields into two output fields, $b_+$ and $b_-$. When the ancilla mode is prepared in the vacuum state, the beam splitter effectively splits the cavity field into two output paths, each accompanied by vacuum noise from the ancilla. One can verify the following commutation relations:
\begin{align}
  [b_\pm, b_\pm^\dagger] = 1,~~~
  [b_\pm, b_\mp] = [b_\pm^\dagger, b_\mp^\dagger] = [b_\pm, b_\mp^\dagger] = 0.
\end{align}

In addition, we introduce
\begin{align}
  x_\pm \equiv  &\, \frac{b_\pm+b_\pm^\dagger}{\sqrt{2}},
  \\
  p_\pm \equiv  &\, \frac{b_\pm-b_\pm^\dagger}{\sqrt{2}i}.
\end{align}
Notably, $x_+$ and $p_-$ commute,
\begin{align}
  [x_+, p_-] = 0.
\end{align}
The two commuting observables $x_+$ and $p_-$ can be measured simultaneously. The corresponding outcomes are often denoted by $I$ and $Q$ (the in-phase and quadrature components, respectively), and the resulting scheme is referred to as an $I/Q$ (heterodyne) measurement. From $I$ and $Q$, one can infer the cavity quadratures $x_c$ and $p_c$ up to added vacuum noise.

The simultaneous measurement of $x_+$ and $p_-$ can be described, in the continuum limit, by a (generalized) POVM element given by the following projector:
\begin{align}
  \Pi (\bar{x}_+, \bar{p}_-) \equiv
  \ketbra{\bar{x}_+}{\bar{x}_+} \otimes
  \ketbra{\bar{p}_-}{\bar{p}_-},
\end{align}
where $\ket{\bar{x}_+}$ and $\ket{\bar{p}_-}$ are eigenstates of $x_+$ and $p_-$ (with eigenvalues $\bar{x}_+$ and $\bar{p}_-$), respectively. Here the tensor product refers to the $(b_+, b_-)$ modes, which is related to the original cavity and ancilla modes by the beam-splitter unitary transformation. Using the displacement operator for the cavity mode, $D (\alpha) \equiv e^{\alpha c^\dagger - \alpha^* c}$, and that for the ancilla mode, $D_a (\alpha) \equiv e^{\alpha a^\dagger - \alpha^* a}$, the projector can be expressed as
\begin{align}
  \Pi (\bar{x}_+, \bar{p}_-) = \frac{1}{\pi^2} \int d^2\alpha
  e^{\alpha^* (\bar{x}_+ + i\bar{p}_-) - \alpha (\bar{x}_+ - i\bar{p}_-)}
  D_a (-\alpha^*) D (\alpha).
  \label{Pi(x,p)2}
\end{align}
The probability density for obtaining the measurement result $(\bar{x}_+, \bar{p}_-)$ is given by
\begin{align}
  p (\bar{x}_+, \bar{p}_-) = \mathrm{tr}
  \left[ \Pi (\bar{x}_+, \bar{p}_-) \rho_{\rm tot} \right],
  \label{p(x,p)}
\end{align}
where $\rho_{\rm tot}$ is the density matrix of the total system.\footnote
{
For a rigorous POVM description, one may coarse-grain the continuous outcomes into finite bins. Defining $\bar{x}_j \equiv j \Delta$ and $\bar{p}_k \equiv k \Delta$, where $\Delta$ is a small but finite positive constant while $j$ and $k$ are integers, one can introduce POVM elements
\begin{align*}
  E_{j,k} \equiv
  \int_{\bar{x}_j}^{\bar{x}_j+\Delta}\! d\bar{x}
  \int_{\bar{p}_k}^{\bar{p}_k+\Delta}\! d\bar{p}\,
  \Pi(\bar{x},\bar{p}),
\end{align*}
which satisfy
\begin{align*}
  E_{j,k}\geq 0,~~~
  \sum_{j=-\infty}^\infty \sum_{k=-\infty}^\infty E_{j,k} = \mathds{I}.
\end{align*}
Then, the probability density is 
\begin{align*}
  p (\bar{x}_+, \bar{p}_-) =
  \frac{1}{\Delta^2}
  \int_{\bar{x}_+}^{\bar{x}_++\Delta} d\bar{x}
  \int_{\bar{p}_-}^{\bar{p}_-+\Delta} d\bar{p}\,
  \mathrm{tr} \left[ \Pi (\bar{x}, \bar{p}) \rho_{\rm tot} \right]
  \xrightarrow{\Delta\rightarrow 0}
  \mathrm{tr}\left[ \Pi (\bar{x}_+, \bar{p}_-) \rho_{\rm tot} \right],
\end{align*}
where we assume that the integrand is a smooth function. One may choose Kraus operators $K_{j,k}$ such that $E_{j,k}=K_{j,k}^\dagger K_{j,k}$.}

Hereafter, we consider the ideal situation in which the ancilla is prepared in its vacuum state. In this case, the total density matrix before the measurement is given by
\begin{align}
  \rho_{\rm tot} = \rho \otimes \ketbra{0_a}{0_a},
  \label{rhotot_bs}
\end{align}
where $\rho$ is the density matrix of the cavity state, while $\ket{0_a}$ is the ancilla vacuum. After a straightforward (but tedious) calculation, one obtains
\begin{align}
  p (\bar{x}_+, \bar{p}_-) = Q (\zeta) = \frac{1}{\pi} \mel{\zeta}{\rho}{\zeta}
  = \mathrm{tr} \left( \frac{1}{\pi} \ketbra{\zeta}{\zeta} \rho \right),
  \label{HusimiFn}
\end{align}
where $\ket{\zeta}$ is a coherent state of the cavity mode, with
\begin{align}
  \zeta \equiv \bar{x}_+ + i \bar{p}_-.
\end{align}
From the above expression, we can conclude that the effective measurement on the cavity alone is described by the POVM element $\frac{1}{\pi}\ketbra{\zeta}{\zeta}$, labeled by the coherent-state amplitude $\zeta = \bar{x}_++i\bar{p}_-$.

Through this measurement, we can obtain information about $x_c$ and $p_c$. Under the assumption that the ancilla is prepared in its vacuum state, we find
\begin{align}
  \ev{x_+} = &\, \frac{1}{\sqrt{2}} \ev{x_c},
  \\
  \ev{p_-} = &\, \frac{1}{\sqrt{2}} \ev{p_c},
\end{align}
with $\ev{\cdots}$ denoting expectation values. Thus, information about $x_c$ and $p_c$ is imprinted in the measurement outcomes $\bar{x}_+$ and $\bar{p}_-$. We can also evaluate the corresponding variances and obtain
\begin{align}
  \mathrm{Var}[x_+] = &\,
  \frac{1}{2} \left( \mathrm{Var}[x_c] + \frac{1}{2} \right),
  \\
  \mathrm{Var}[p_-] = &\,
  \frac{1}{2} \left( \mathrm{Var}[p_c] + \frac{1}{2} \right),
\end{align}
where we have used the fact that the cavity and ancilla modes are uncorrelated, i.e., $\ev{x_c x_a}=\ev{p_c p_a}=0$, with $x_a\equiv \frac{1}{\sqrt{2}}(a+a^\dagger)$ and $p_a\equiv \frac{1}{\sqrt{2}i}(a-a^\dagger)$. Note that the $\frac{1}{2}$ terms in the parentheses originate from vacuum fluctuations in the ancilla sector, i.e., from $\ev{x_a^2}$ and $\ev{p_a^2}$. This additional noise is an unavoidable trade-off of measuring $x_c$ and $p_c$ simultaneously \cite{Caves:1982zz} (for a review, see also Ref.\ \cite{Clerk:2008tlb}); since these two operators do not commute, no measurement can determine both without introducing extra fluctuations beyond those already present in the cavity state itself. Indeed, defining $\sqrt{2}\,\bar{x}_+$ and $\sqrt{2}\,\bar{p}_-$ as estimators of $x_c$ and $p_c$, one finds that their variances are given by $\mathrm{Var}[x_c] + \frac{1}{2}$ and $\mathrm{Var}[p_c] + \frac{1}{2}$, respectively; that is, exactly one unit of vacuum noise (in these units) is added to each quadrature as a result of the simultaneous measurement.

It is also notable that, in the case of a coherent state, i.e., $\rho=\ketbra{\alpha}{\alpha}$, the probability density for obtaining the measurement outcome $\zeta=\bar{x}_+ + i\bar{p}_-$ is given by $p(\zeta)=\frac{1}{\pi}e^{-|\zeta-\alpha|^2}$. Thus, under the $I/Q$ measurement, one obtains a Gaussian probability density peaked at $\zeta=\alpha$.

\bibliographystyle{RefStyle}
\bibliography{refs}

\end{document}